%% file: main.tex
\documentclass[11pt,notitlepage, nofootinbib]{revtex4-1}

\usepackage{amsmath}
\usepackage{amsfonts}
\usepackage{amsthm}
\usepackage{amssymb}
\usepackage{graphicx}
\usepackage{mathtools}
\usepackage{txfonts}
\usepackage{subfigure}
\graphicspath{{Images/}}
\usepackage{color}
\usepackage[dvipsnames]{xcolor}
\definecolor{lcolor}{rgb}{0.5,0,0}
\definecolor{citcolor}{rgb}{0,0.3,0.0}
\usepackage[breaklinks,colorlinks,urlcolor=RoyalBlue,citecolor=citcolor,linkcolor=OrangeRed]{hyperref}
\usepackage{slashed}
\usepackage{braket}
\usepackage{mathrsfs}
\newcommand*\diff{\mathop{}\!\mathrm{d}}
\usepackage{etoolbox}

\newcommand{\exercise}[1]{{\color{teal} \textbf{[Exercise] } #1 }}

\begin{document}
\bibliographystyle{apsrev4-1}

\title{Mesons on the light front}
\author{Meijian Li}
\email{meijian.li@usc.es}
\affiliation{Instituto Galego de Fisica de Altas Enerxias (IGFAE), Universidade de Santiago de Compostela, E-15782 Galicia, Spain}
\begin{abstract}
This lecture note is written for \href{https://indico.cern.ch/event/1203236/}{``Courses on Light-Cone Techniques applied to QCD", Nov 21-25, IGFAE}. It is intended to provide basic knowledge and selective perspectives on the application of light-front Hamiltonian approach to mesons in two 1.5-hour lectures.
\end{abstract}

\maketitle
\tableofcontents

\input{L1}

\input{L2}

\appendix
\input{Appendix}
\bibliography{LC.bib}

\end{document}

%% file: L1.tex
\section{Lecture I: Mesons as the QCD bound states}
\subsection{Canonical quantization of the QCD Hamiltonian on the light front}

\subsubsection{Light-front dynamics}
From the viewpoint that the quantum field theory is formulated to reconcile quantum mechanics with special relativity, let us first study how symmetries like Lorentz invariance appear in quantum setting. In particular, we would like to combine the principle of relativity with the Hamiltonian formulation of dynamics. 

Einstein's principle of relativity requires that physical laws shall be invariant under transformations from one space-time coordinate system to another, or in other words, invariant in all inertial frames of reference.
The whole group of the transformations is the inhomogeneous Lorentz group, also known as the Poincar\'e group. Quantum theory postulates that physical states are represented by rays
\footnote{A ray is a set of normalized vectors differed by multiplying an arbitrary scalar of unit magnitude~\cite{Weinberg:1995mt}.} 
in Hilbert space. Therefore we need to implement a representation of the Poincar\'e group. 
The Poincar\'e algebra is the Lie algebra of the Poincar\'e group, and it is given by the commutation relations:
\begin{align}
  \begin{split}
    & [P^\mu, P^\nu]=0 \,,\\
    & [P^\mu, M^{\alpha\beta}]=i(g^{\mu\alpha}P^\beta - g^{\mu\beta}P^\alpha) \,,\\
    & [M^{\mu\nu}, M^{\rho\sigma}]=i(g^{\mu\sigma}M^{\nu\rho} - g^{\nu\sigma}M^{\mu\rho} + g^{\nu\rho}M^{\mu\sigma}- g^{\mu\rho}M^{\nu\sigma}  ) \;.
  \end{split}
\end{align}
It has ten generators, four generators of translations $P^\mu=(P^0, P^1, P^2, P^3)$ and six generators of Lorentz transformations $M^{\mu\nu}$. 
The latter can be further split into the three generators of rotations $J^i=1/2 \epsilon^{ijk}M^{jk}$ and 3 generators of boosts $K^i=M^{0i}$. \footnote{The cyclic symbol $\epsilon^{ijk}$ is $1$ if the indices $ijk$ are in cyclic order, and $0$ otherwise.}

In quantum mechanics, and also in the quantum field theory, the dynamical evolution of a quantum state satisfies the Schr\"{o}dinger equation,
\begin{align}\label{eq:time_ShroEq_general}
    i\frac{\partial}{\partial t}\ket{\psi(t)} = H\ket{\psi(t)}
    \;.
\end{align}
For stationary states,
\begin{align}
\ket{\psi(t)} = e^{-iE t}\ket{\psi(0)}
    \;,
\end{align}
and it leads to the bound-state equation
\begin{align}
    H\ket{\psi(0)} = E\ket{\psi(0)}
    \;,
\end{align}
where $E$ is the bound state energy. Though in its original form the time $t$ is the regular time, there are actually multiple choices of the time variable as a foliation of spacetime. 
\footnote{By foliation it means that the manifold of spacetime is decomposed into hypersurfaces and there exists a smooth scalar field (the ``time") which is regular in the sense that its gradient never vanishes, such that each hypersurface is a level surface of this scalar field. }
P. A. M. Dirac brought up three forms of relativistic dynamics, namely the instant form, the point form, and the front form~\cite{Dirac:1949cp}. 

In the instant form, one works with dynamical variables referring to physical conditions at some instant of time, $x^0$.  The Hamiltonian is $P^0$. The transformations of coordinates associated with the momenta $P^1$, $P^2$, $P^3$ and the rotations $J^1$, $J^2$, $J^3$, leave the instant invariant, and are thus kinematic. The energy $P^0$, and the boosts $K^1$, $K^2$, $K^3$ are dynamical. The instant form seems most intuitive since its time variable is the regular time. Although it is the conventional choice for quantizing field theories, it has many disadvantages. The experiment determining the wavefunction $\psi(t,\vec x)$ solved from the evolution equation of Eq.~\eqref{eq:time_ShroEq_general} requires the simultaneous measurement of all positions of the state. A more practical experimental measurement scatters one plane-wave laser beam, and the signal reaches each part of the object at the same light-front time $x^+=t+z/c$ (this is the same with the definition $x^+=x^0 +x^3$ with the unit $c=1$). 

The point form of dynamics describes physical conditions on the three-dimensional surface, $\tau=\sqrt{x^\mu x_\mu-a^2}=\sqrt{(x^0)^2-(x^1)^2-(x^2)^2-(x^3)^2-a^2}$ with $x^0>0$. The energy $P^0$, and the momenta $P^1$, $P^2$, $P^3$ are all dynamical. The kinematic group consists of the boosts $K^1$, $K^2$, $K^3$ and the rotations $J^1$, $J^2$, $J^3$, which leave the origin point invariant. The point form of relativistic quantum mechanics has been advocated as an appropriate framework for calculating the electroweak structure of mesons and baryons within the scope of constituent-quark models~\cite{Biernat:2009my,Biernat:2010tp,GomezRocha:2012zd}. 

The front form considers the three-dimensional surface in space-time formed by a plane wave front advancing with the velocity of light. The theory describes physical conditions at some constant light-front time $x^+=x^0+x^3$. The front form has the largest number(seven) of kinematic generators that leaves the light front invariant. They are, the transverse momentum $P^1$, $P^2$, the longitudinal momentum $P^+=P^0+ P^3$, the transverse boosts $E^1=K^1 + J^2$, $E^2 = K^2-J^1$, the rotation in the x-y plane $J^3$, and the boost in the longitudinal direction $K^3$.\footnote{The longitudinal boost is actually a scale transformation, seeing that $x^{\pm}\to \tilde{x}^{\pm}=e^{\pm \phi}x^\pm$ with the Lorentz factor $\gamma=\cosh\phi$. It therefore leaves the $x^+=0$ plane invariant. } The remaining generators $\{P^-=P^0-P^3, F^1=J^1+K^2, F^2=J^2-K^1\}$ are dynamical. $P^-$ is the light-front Hamiltonian. It is usually convenient to use the light-front coordinates when implementing the light-front dynamics. We include the conventions of the light-front coordinates in Appendix~\ref{app:LF_cor}.

\begin{figure}[htp!]
  \centering
  \subfigure[instant form \label{fig:zt_instant}]
  {\includegraphics[width=.3\textwidth]
  {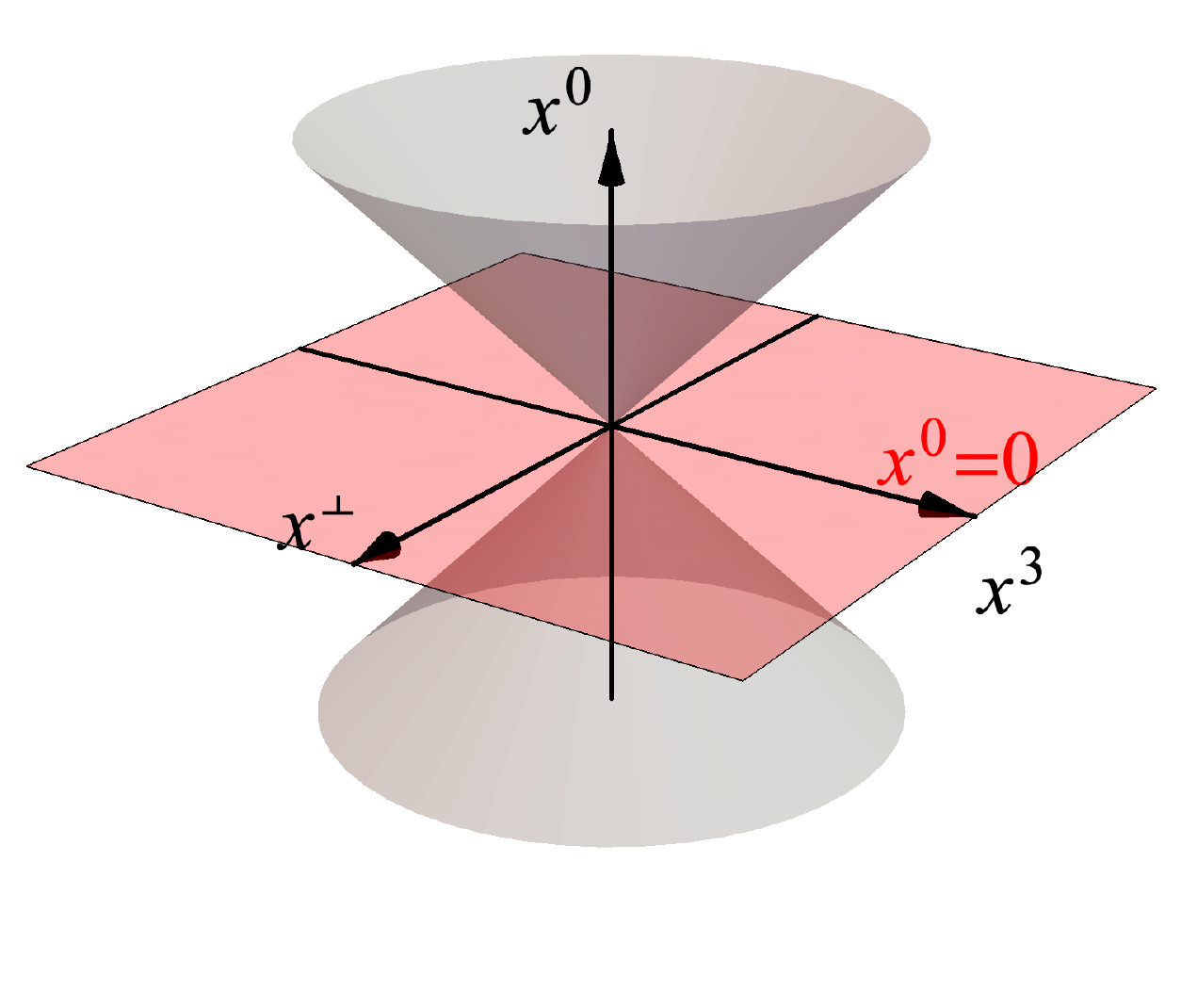}
  }
  \subfigure[front form \label{fig:zt_front_v2}]
  {\includegraphics[width=.3\textwidth]{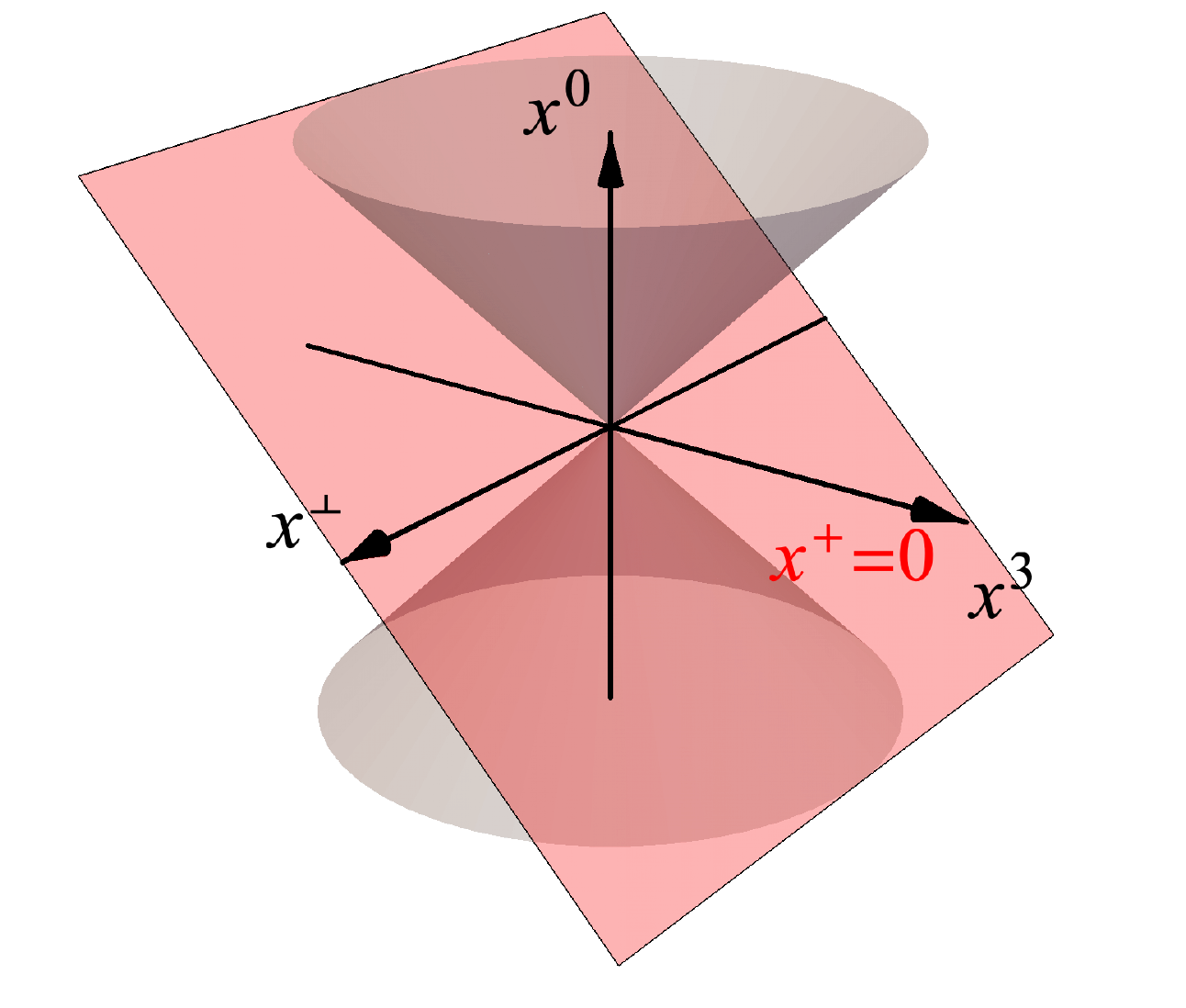}
  }
  \subfigure[point form  \label{fig:zt_point_v2}]
  {\includegraphics[width=.3\textwidth]
  {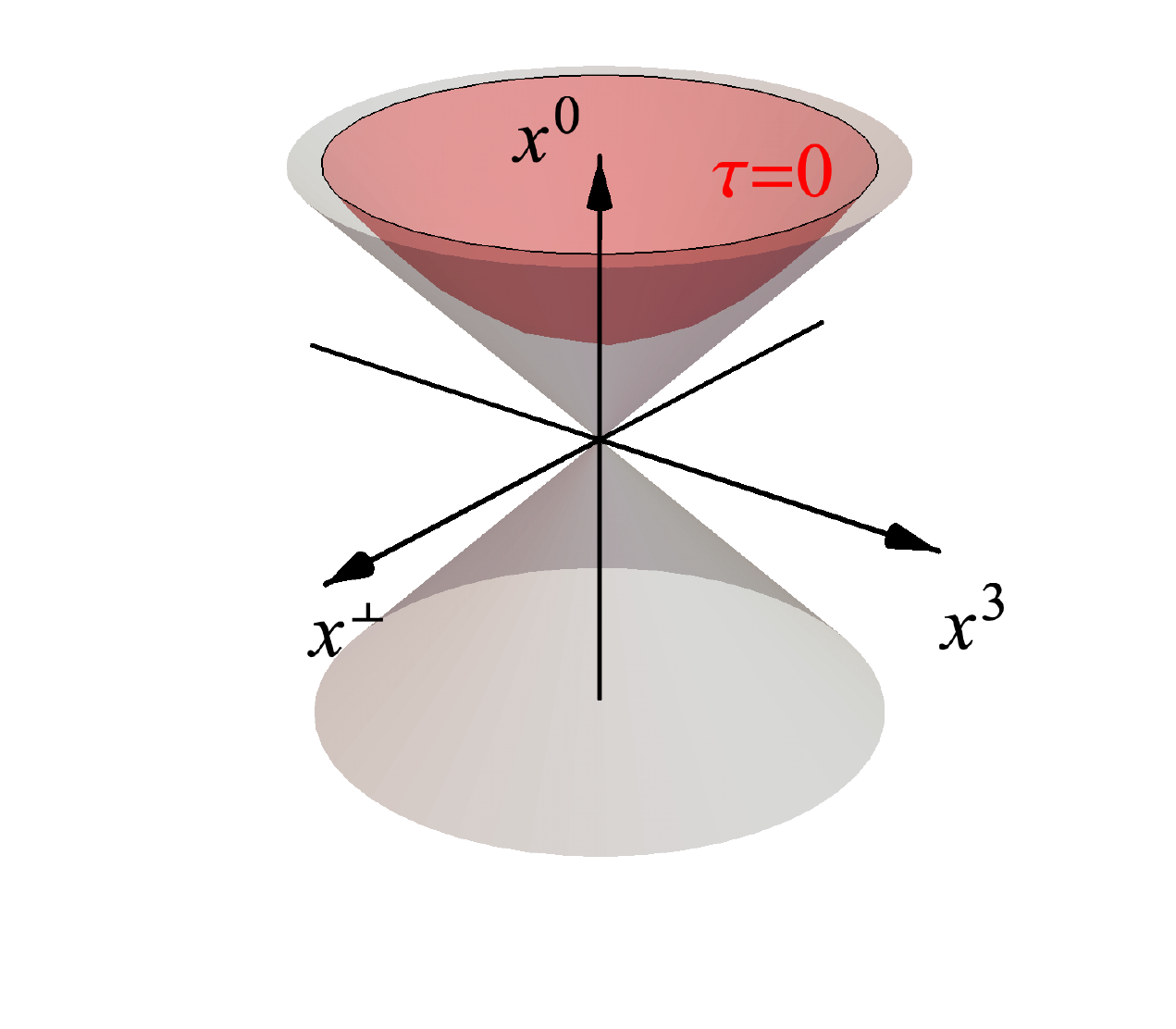}
  }
  \caption{``Time" in the three forms of dynamics. The gray cones are the reference surfaces of the light cones, $t=\sqrt{(x^0)^2+(x^3)^2}$. The equal-``time" surfaces are in red. In (a), the instant form, time is defined as $x^0$ and the shown equal-time surface is $x^0=0$. In (b), the front form, time is defined as $x^+=x^0 +x^3$ and the shown equal-light-front-time surface is $x^+=0$. In (c), the point form, time is defined as $\tau=\sqrt{x^\mu x_\mu-a^2}$ with $x^0>0$ and the shown equal-point-time surface is $\tau=0$. }
  \label{fig:dynamics_Dirac}
\end{figure}

A visualization of the ``time'' in these three forms is presented in Fig.~\ref{fig:dynamics_Dirac}. Be aware that there also exists two other forms of dynamics, with the time defined as $\tau_z =\sqrt{(x^0)^2-(x^3)^2-a^2}$ with $x^0 > 0$ and $\tau_\perp=\sqrt{(x^0)^2-(x^1)^2-(x^2)^2 -a^2}$ with $x^0 > 0$ respectively, though they have a rather small kinematical group and are not commonly used~\cite{Heinzl:2000ht}.

 The quantum field theory quantized on the light-front surface $x^+=0$ is the light front quantum field theory. In the next section, we will carry out the canonical quantization of QCD on the light front.

\subsubsection{The light-front QCD Hamiltonian}
The strong interaction between quarks and gluons is described by the non-Abelian gauge theory with symmetry group SU(3), known as quantum chromodynamics (QCD), and the Lagrangian reads
\begin{align}\label{eq:QCD_L_z}
  \mathcal{L}_{QCD}=-\frac{1}{4}{F^{\mu\nu}}_a F^a_{\mu\nu}+\bar{\Psi}(i\gamma^\mu \bm D_\mu - \bm m)\Psi\;.
\end{align}
$A^\nu_a$ is color vector potential, with the gluon index $a=1,2,\ldots,8$. The quark field
$\Psi_{\alpha,c}$, carries the Dirac index $\alpha=1,2,\ldots,4$ and the color index $c=1,2,3$, which are usually suppressed in expressions like $\bar{\Psi}\gamma^\mu \bm D_\mu\Psi=\bar{\Psi}_c\gamma^\mu (\bm D_\mu)_{cc'}\Psi_{c'}$. $\bm m=m\bm I_3 = m\delta_{cc'}$ is diagonal in color space.
The vector potential can be parameterized as ${( \bm{A}_\mu)}_{cc'}=T^a_{cc'}A^\mu_a$ by the color matrices $T^a_{cc'}$, and its matrix form can be found in Appendix~\ref{app:colorT}. $F^{\mu\nu}_a\equiv\partial^\mu A^\nu_a-\partial^\nu A^\mu_a-gf^{abc}A^\mu_b A^\nu_c$ is the field tensor, and $\bm D^\mu\equiv \partial_\mu \bm I_3+ig \bm A^\mu$ is the covariant derivative. We follow the convention of the covariant derivative from Ref.~\cite{Brodsky:1997de}, such that $g$ is the chromo-electric charge of the anti-fermion. Note that there exists another widely used convention that assigns $g$ to the chromo-electric charge of the fermion instead~\cite{Peskin:1995ev}.
The structure constants $f^{abc}$ are complete anti-symmetric, $f^{abc}=f^{cab}=-f^{acb}$. In the following derivations, we will drop the identity operator in the color space, $\bm I_3$, for simplicity. 
Let us now derive the canonical QCD Hamiltonian according to the procedure in Ref.~\cite{Brodsky:1997de}.

The QCD Lagrangian is a functional of the twelve components $\bm A^\mu$, $\Psi_\alpha$, $\bar{\Psi}_\alpha$ and their space-time derivatives. We can denote them collectively as $\mathcal{L}=\mathcal{L}[\phi_r,\partial_\mu\phi_r]$. The equations of motion are
\begin{align}
  \partial_\kappa\Pi_r^\kappa-\delta\mathcal{L}/\delta\phi_r=0\;,
\end{align}
where the generalized momentum fields are $\Pi_r^\kappa\equiv\delta\mathcal{L}/\delta(\partial_\kappa\phi_r)$. 
Let us write out the equation of motions for each field.
\begin{enumerate}
    \item $A^\mu_a$ and the color-Maxwell equations\\
    The variational derivatives and the generalized momentum fields of the vector field are
\begin{align}
    \frac{\delta\mathcal{L}}{\delta A^s_\kappa}=&-\frac{1}{4}F^{\kappa\mu}_a(-gf^{asc}A^c_\mu)\times
    4+\bar{\Psi}(i\gamma^\kappa(igT^s))\Psi 
    =-gf^{sac}F^{\kappa\mu}_a A^c_\mu-g\bar{\Psi}\gamma^\kappa T^s\Psi,
\qquad
  \Pi_{A^s_\kappa}^\lambda=-F_s^{\lambda\kappa}\;.
\end{align}
The first four equations of motion give the color-Maxwell equations, 
\begin{align}\label{eq:colorM}
  \partial_\lambda F_s^{\lambda\kappa}=gJ_s^\kappa \;,
\end{align}
with the current
density $J_s^\kappa\equiv f^{sac}F^{\kappa\mu}_a A^c_\mu+\bar{\Psi}\gamma^\kappa T^s\Psi$.
In the light-cone gauge of $A_a^+=0$, the $\kappa=+$ component of Eq.~\eqref{eq:colorM} does not contain time derivatives, and can be written as
\begin{align}
  g J_a^+=\partial_\lambda F_a^{\lambda +}=-\partial^+\partial_-A^-_a-\partial^+\partial_i A^i_a
  \;.
\end{align}
By inverting the above equation, we get
\begin{align}\label{eq:Aconstrain}
  \frac{1}{2}A^-_a=-g\frac{1}{{(\partial^+)}^2}J^+_a-\frac{1}{\partial^+}\partial_i A^i_a
  \;.
\end{align}
We define the free solution $\tilde{A}_a^\mu$ such that $\lim_{g\to 0}A_a^\mu=\tilde{A}_a^\mu$. According to Eq.~\eqref{eq:Aconstrain}, the free field reads,
\begin{align}
  \tilde{A}^\mu_a=(0,\tilde{A}^-_a,A^i_a), \quad \text{with}\ \
  \frac{1}{2}\tilde{A}^-_a\equiv\frac{1}{2}A^-_a+g\frac{1}{{(\partial^+)}^2}J^+_a=-\frac{1}{\partial^+}\partial_i A^i_a
  \;.
\end{align}
$\tilde{A}_a^\mu$ is thereby purely transverse.
    \item $\Psi_\alpha$ and the (adjoint) color-Dirac equations\\
The variational derivatives and the generalized momentum fields of the fermion field are
\begin{align}
    \frac{\delta\mathcal{L}}{\delta \Psi}=-g\bar{\Psi}\gamma^\mu A_\mu-m\bar{\Psi}-\frac{i}{2}\bar{\Psi}\gamma^\mu\overleftarrow{\partial}_\mu,
    \quad
  \Pi_{\Psi}^\lambda=\frac{i}{2}\bar{\Psi}\gamma^\lambda
  \;.
\end{align}
Note that the second term of the Lagrangian in Eq.~\eqref{eq:QCD_L_z} written more explicitly is 
\begin{align}
  \frac{1}{2}\left[\bar{\Psi}(i\gamma^\mu \bm D_\mu - \bm m)\Psi + h.c. \right]
  =\frac{1}{2}\left[\bar{\Psi}(i\gamma^\mu \bm D_\mu - \bm m)\Psi + \bar{\Psi}(-i\gamma^\mu \bm \overleftarrow{D}_\mu - \bm m)\Psi\right]
  \;,
\end{align}
in which $\overleftarrow{\bm D}^\mu\equiv \overleftarrow{\partial}_\mu \bm I_3-ig \bm A^\mu$.

The equations of motion for $\Psi$ give the adjoint color-Dirac equation,
\begin{align}\label{eq:cDirac1}
  \bar{\Psi}[i\gamma^\mu(\overleftarrow{\partial}_\mu -ig\bm A_\mu)+m]=0 \;.
\end{align}
Take Hermitian conjugate on the equation and use the relation $\bar{\Psi}=\Psi^\dagger\gamma^0$, we have
\begin{align}
  [-i{\gamma^\mu}^\dagger(\partial_\mu+ig\bm A_\mu)+m]\gamma^0\Psi=0
  \;.
\end{align}
By moving $\gamma^0$ to the left,
we arrive at the color-Dirac equation,
\begin{align}\label{eq:cDirac2}
  [i\gamma^\mu(\partial_\mu +ig\bm A_\mu)-m]\Psi=0
  \;.
\end{align}
Similar to the gluon field, we also want to separate the dynamical components of the fermion field.
Define the projected spinors $\Psi_\pm=\Lambda^\pm \Psi$, with $\Lambda^\pm=\frac{1}{2}\gamma^0\gamma^\pm$, see more definitions of $\Lambda^\pm$ in Appendix~\ref{app:gamma}.
First multiply Eq.~\eqref{eq:cDirac2} by $\gamma^0$ on the left,
\begin{align}
  \begin{split}
    [i(\gamma^0\gamma^+ \bm  D_++\gamma^0\gamma^- \bm D_-+\alpha^i \bm D_i)-m\beta]\Psi=0 \;,\\
    \text{which is,}\ \   [i(2\Lambda^+ \bm D_++2\Lambda^- \bm D_-+\alpha^i \bm D_i)-m\beta]\Psi=0 \;.
  \end{split}
\end{align}
Then multiply the equation by $\Lambda^+ (\Lambda^-)$ on the left, and bring it to the right,
\begin{align}
  [i(2 \bm D_\pm\Lambda^\pm+\alpha^i \bm D_i\Lambda^\mp)-m\beta\Lambda^\mp]\Psi=0 \;.
\end{align}
One thereby obtains a coupled set of spinor equations,
\begin{align}
  2i\partial_+\Psi_+=(-i\alpha^i \bm D_i+m\beta)\Psi_-+2g\bm A_+\Psi_+ \label{eq:psi_1}\;,\\
  2i\partial_-\Psi_-=(-i\alpha^i \bm D_i+m\beta)\Psi_++2g\bm A_-\Psi_- \label{eq:psi_2}\;.
\end{align}
Then, in the light-cone gauge, $2A_-=A^+=0$.
Equation~\eqref{eq:psi_2} does not contain time derivatives, and can be written as a constraint relation, 
\begin{align}\label{eq:psi_constraint}
    \Psi_-=\frac{1}{2i\partial_-}(m\beta-i\alpha^i \bm D_i)\Psi_+
    \;.
  \end{align}
By substituting Eq.~\eqref{eq:psi_constraint} into Eq.~\eqref{eq:psi_1}, we get
\begin{align}\label{eq:d-psi+}
  2iD_+\Psi_+=(m\beta-i\alpha^i \bm D_i)\frac{1}{2i\partial_-}(m\beta-i\alpha^i \bm D_i)\Psi_+
  \;.
\end{align}
In analogy to the free solution $\tilde{A}$, we define the free spinor $\tilde{\Psi}=\tilde{\Psi}_++\tilde{\Psi}_-$ with 
\begin{align}\label{eq:free_spinor}
  \tilde{\Psi}_+=\Psi_+,\quad\tilde{\Psi}_-=\frac{1}{2i\partial_-}(m\beta-i\alpha^i \partial_i)\Psi_+
  \;.
\end{align}
The projection still holds, $\tilde{\Psi}_\pm=\Lambda^\pm \tilde{\Psi}$.

\item $\bar\Psi_\alpha$ and the color-Dirac equations\\
The variational derivatives and the generalized momentum fields of the anti-fermion field are
\begin{align}
    \frac{\delta\mathcal{L}}{\delta \bar\Psi}=-g\gamma^\mu A_\mu \Psi
    -m\Psi+\frac{i}{2}\gamma^\mu \partial_\mu \Psi,\quad \Pi_{\bar{\Psi}}^\lambda=-\frac{i}{2}\gamma^\lambda\Psi
  \;.
\end{align}
The equations of motion for $\bar \Psi$ give the  color-Dirac equation, Eq. \eqref{eq:cDirac2}, which we have already arrived from $\Psi$, not surprisingly.

\end{enumerate}

We now turn to the construction of the canonical Hamiltonian density through a Legendre transformation,
\footnote{Note that we are taking the derivative $\partial^+$, in terms of $x^+$, and the conjugate quantity is $P_+$, or equivalently $P^-/2$.} 
\begin{align}\label{eq:Legendre_H}
  \begin{split}
    \mathcal{P}_+=& (\partial_+ A^s_\kappa)\Pi_{A^s_\kappa}^+ +(\partial_+ \Psi)\Pi_{\Psi}^++(\partial_+ \bar{\Psi})\Pi_{\bar{\Psi}}^+-\mathcal{L}\\
    =&-F_s^{+\kappa}\partial_+ A^s_\kappa
    +\frac{1}{2}[i\bar{\Psi}\gamma^+\partial_+ \Psi+h.c.]
    +\frac{1}{4}{F^{\mu\nu}}_a F^a_{\mu\nu}
    -\frac{1}{2}[\bar{\Psi}(i\gamma^\mu \bm D_\mu-\bm{m})\Psi+
    \bar{\Psi}(-i\gamma^\mu \overleftarrow{\bm D}_\mu-\bm{m})\Psi
    ]\\
    =&-F_s^{+\kappa}\partial_+ A^s_\kappa+\frac{1}{2}[i\bar{\Psi}\gamma^+\partial_+ \Psi+h.c.]
    +\frac{1}{4}{F^{\mu\nu}}_a F^a_{\mu\nu}
    \;,
  \end{split}
\end{align}
where we have used the color-Dirac equations as in Eqs.~\eqref{eq:cDirac1} and~\eqref{eq:cDirac2} in the last line. It is convenient to add a total derivative $-\partial_\kappa(F_s^{\kappa+}A^s_+)$ to the Hamiltonian $P^-=2 P_+$, 
\begin{align}\label{eq:Legendre_Pmn}
  \begin{split}
    P^-=&2 \int\diff x_+\diff^2 x_\perp\ \mathcal{P}_+\\
    =&\int\diff x^-\diff^2 x_\perp\ -F_s^{+\kappa}\partial_+ A^s_\kappa+\frac{1}{2}[i\bar{\Psi}\gamma^+\partial_+ \Psi+h.c.]+\frac{1}{4}{F^{\mu\nu}}_a F^a_{\mu\nu}-\partial_\kappa(F_s^{\kappa+}A^s_+)
    \;.
  \end{split}
\end{align}
We can rewrite the first and the last terms into
\begin{align}
  \begin{split}
    F_s^{\kappa+}\partial_+ A^s_\kappa -\partial_\kappa(F_s^{\kappa+}A^s_+)
    =&F_s^{\kappa+}\partial_+ A^s_\kappa-(\partial_\kappa F_s^{\kappa+})A^s_+-F_s^{\kappa+}\partial_\kappa A^s_+\\
    =&F_s^{\kappa+}(\partial_+ A^s_\kappa-\partial_\kappa A^s_+)-(\partial_\kappa F_s^{\kappa+})A^s_+\\
    =&-F_s^{\kappa+}(F_{\kappa+}^s+gf^{sbc}A^b_\kappa A^c_+)-gJ_s^+ A^s_+\\
    =&-F_s^{\kappa+}F_{\kappa+}^s-g\bar{\Psi}\gamma^+T^s A^s_+\Psi
    \;.
  \end{split}
\end{align}
The Hamiltonian becomes
\begin{align}\label{eq:Hamiltonian_h}
  \begin{split}
    P^-=&\int\diff x^-\diff^2 x_\perp\
    \frac{1}{4}{F^{\mu\nu}}_a F^a_{\mu\nu}
    -F_s^{\kappa+}F_{\kappa+}^s-g\bar{\Psi}\gamma^+T^s A^s_+\Psi
    +\frac{1}{2}[i\bar{\Psi}\gamma^+\partial_+ \Psi+h.c.]\\
    =&\int\diff x^-\diff^2 x_\perp\
    \frac{1}{4}{F^{\mu\nu}}_a F^a_{\mu\nu}
    -F_s^{\kappa+}F_{\kappa+}^s
    +\frac{1}{2}[i\bar{\Psi}\gamma^+\bm{D}_+ \Psi+h.c.]
    \;.
  \end{split}
\end{align}

\begin{enumerate}
    \item 1st part of Eq.~\eqref{eq:Hamiltonian_h}

Let us also rewrite the color-electro-magnetic energy density and separate the longitudinal and the transversal contributions,
\begin{align}\label{eq:fields_first}
  \begin{split}
    \frac{1}{4}F^{\mu\nu}_a F_{\mu\nu}^a-F^{\mu+}_a F_{\mu+}^a
    =&\frac{1}{4}(F^{ij}_a F_{ij}^a+F^{\mu+}_a F_{\mu+}^a+F^{+\nu}_a F_{+\nu}^a+F^{\mu-}_a F_{\mu-}^a+F^{-\nu}_a F_{-\nu}^a\\
    &-F^{+-}_a F_{+-}-F^{-+}_a F_{-+})-F^{\mu+}_a F_{\mu+}\\
    =&\frac{1}{4}F^{ij}_a F_{ij}^a+\frac{1}{2}(F^{\mu+}_a F_{\mu+}^a+F^{\mu-}_a F_{\mu-}^a-F^{+-}_a F_{+-}^a)-F^{\mu+}_a F_{\mu+}^a\\
    =&\frac{1}{4}F^{ij}_a F_{ij}^a-\frac{1}{2}F^{+-}_a F_{+-}^a
    \;.
  \end{split}
\end{align}
Note that $F^{\mu+}_a F_{\mu+}^a=F^{\mu-}_a F_{\mu-}^a$ by $F^{\mu-}_a=g^{\mu\nu}F_{\nu+}^a g^{+-}$. Substituting $A^-_a$ by Eq.~\eqref{eq:Aconstrain}, the color-electric part becomes,
\begin{align}
  \begin{split}
    F^{+-}_a F_{+-}^a
    =&-\partial^+A_a^-\partial_-A_+^a\\
    =&-\frac{1}{4}\partial^+A_a^-\partial^+A^-_a\\
    =&-(-g\frac{1}{\partial^+}J^+_a-\partial_i A^i_a)^2\\
    =&g^2 J^+_a\frac{1}{{(\partial^+)}^2}J^+_a-{(\partial_i A^i_a)}^2-g J^+_a\tilde{A}_a^-
    \;.
  \end{split}
\end{align}
In deriving the last line, we introduced an extra term $g^2 \frac{1}{\partial^+}(J^+_a\frac{1}{\partial^+}J^+_a)$, taking that it should vanish under the integral of $\int \diff x^-$. The color-magnetic part can be written as
\begin{align}\label{eq:FijFij}
  \begin{split}
    F^{ij}_a F_{ij}^a=&2\partial^i A^j_a\partial_i A_j^a-2\partial^i A^j_a\partial_j A_i^a-4g f^{abc}\partial^i A^j_a A_i^b
    A_j^c+g^2f^{abc}A^i_b A^j_c f^{aef}A_i^e A_j^f\\
    =&-2 A^j_a\partial^i\partial_i A_j^a+2A^j_a\partial^i\partial_j A_i^a-4g f^{abc}\partial^i A^j_a A_i^b
    A_j^c+g^2f^{abc}A^i_b A^j_c f^{aef}A_i^e A_j^f\\
    =&2 A^j_a\nabla^2_\perp A_j^a-2( \partial_j A^j_a\partial^i A_i^a)-4g f^{abc}\partial^i A^j_a A_i^b A_j^c+g^2f^{abc}A^i_b A^j_c f^{aef}A_i^e A_j^f
    \;.
  \end{split}
\end{align}
    \item 2nd part of Eq.~\eqref{eq:Hamiltonian_h}
For the spinor terms, 
\begin{align}
  i\bar{\Psi}\gamma^+\bm D_+ \Psi=i\Psi^\dagger\gamma^0\gamma^+\bm D_+ \Psi=2i\Psi^\dagger\Lambda^+\bm D_+ \Psi=2i\Psi^\dagger\Lambda^+\bm D_+ \Lambda^+\Psi=2i\Psi^\dagger_+\bm D_+ \Psi_+
  \;.
\end{align}

Substitution of the time derivative in Eq.~\eqref{eq:d-psi+} and the free spinors defined in Eq.~\eqref{eq:free_spinor} leads to
\begin{align}
  \begin{split}
    &2i\Psi^\dagger_+\bm D_+ \Psi_+\\
    =&\Psi^\dagger_+(m\beta-i\alpha^i \bm D_i)\frac{1}{2i\partial_-}(m\beta-i\alpha^i \bm D_i)\Psi_+\\
    =&\Psi^\dagger_+(m\beta-i\alpha^i \partial_i)\frac{1}{2i\partial_-}(m\beta-i\alpha^i \partial_i)\Psi_+
    +g^2\Psi^\dagger_+\alpha^i \bm A_i\frac{1}{2i\partial_-}\alpha^i \bm A_i\Psi_+\\
    &+g\Psi^\dagger_+\alpha^i \bm A_i\frac{1}{2i\partial_-}(m\beta-i\alpha^i \partial_i)\Psi_+
    +g\Psi^\dagger_+(m\beta-i\alpha^i \partial_i)\frac{1}{2i\partial_-}\alpha^i \bm A_i\Psi_+\\
    =&\Psi^\dagger_+(m\beta-i\alpha^i \partial_i)\frac{1}{2i\partial_-}(m\beta-i\alpha^i \partial_i)\Psi_+
    +g^2\Psi^\dagger_+\alpha^i \bm A_i\frac{1}{2i\partial_-}\alpha^i \bm A_i\Psi_+
    +g\Psi^\dagger_+\alpha^i \bm A_i\tilde{\Psi}_-
    +g\tilde{\Psi}^\dagger_-\alpha^i \bm A_i\Psi_+
    \;.
  \end{split}
\end{align}
The first term reads, with recalling that $\tilde{\Psi}_+ =\Psi_+$,
\begin{align}
  \begin{split}
    \tilde{\Psi}^\dagger_+(m\beta-i\alpha^i \partial_i)\frac{1}{2i\partial_-}(m\beta-i\alpha^j \partial_j)\tilde{\Psi}_+
    =&\tilde{\Psi}^\dagger\Lambda_+(m\beta-i\alpha^i \partial_i)\frac{1}{2i\partial_-}(m\beta-i\alpha^j \partial_j)\Lambda_+\tilde{\Psi}\\
    =&\frac{1}{2}\bar{\tilde{\Psi}}\gamma^+(m\beta-i\alpha^i \partial_i)\frac{1}{2i\partial_-}(m\beta-i\alpha^j \partial_j)\tilde{\Psi}\\
    =&\frac{1}{2}\bar{\tilde{\Psi}}\gamma^+(m+i\gamma^i \partial_i)\frac{{(\gamma^0)}^2}{2i\partial_-}(m-i\gamma^j \partial_j)\tilde{\Psi}\\
    =&\frac{1}{2}\bar{\tilde{\Psi}}\gamma^+\frac{m^2-\nabla_\perp^2}{2i\partial_-}\tilde{\Psi}
    \;.
  \end{split}
\end{align}
From the first to the second equation, the $\Lambda_+$ on the right is brought to the left, using the relations in Eq.\eqref{eq:lambda_pro}. 
The second term reads,
\begin{align}
  \begin{split}
    g^2\tilde{\Psi}^\dagger_+\alpha^i \bm A_i\frac{1}{2i\partial_-}\alpha^j \bm A_j\tilde{\Psi}_+
    =\frac{g^2}{2}\bar{\tilde{\Psi}}\gamma^+\gamma^0\gamma^i \bm A_i\frac{1}{2i\partial_-}\gamma^0\gamma^j \bm A_j\tilde{\Psi}=\frac{g^2}{2}\bar{\tilde{\Psi}}\gamma^i \bm A_i\frac{\gamma^+}{2i\partial_-}\gamma^j \bm A_j\tilde{\Psi}
    \;.
  \end{split}
\end{align}
The last two terms combine into
\begin{align}\label{eq:LFH_spinor_end}
  \begin{split}
    g\tilde{\Psi}^\dagger_+\alpha^i \bm A_i\tilde{\Psi}_-
    +g\tilde{\Psi}^\dagger_-\alpha^i \bm A_i\tilde{\Psi}_+
    =g(\tilde{\Psi}^\dagger_++\tilde{\Psi}_-)\alpha^i \bm A_i(\tilde{\Psi}^\dagger_++\tilde{\Psi}_-)
    =g\tilde{\Psi}^\dagger\alpha^i \bm A_i\tilde{\Psi}
    =g\bar{\tilde{\Psi}}\gamma^i \bm A_i\tilde{\Psi}
    \;.
  \end{split}
\end{align}
\end{enumerate}

We can also define the current density of free fields solution $\tilde{J}^\mu_a$ in analogy to $J^\mu_a$, and notice that their "+" components are the same,
\begin{align}\label{eq:fields_last_1}
  \begin{split}
    J_s^+= &f^{sac}F^{+\mu}_a A^c_\mu+\bar{\Psi}\gamma^+T^s\Psi 
    = f^{sac}\partial^+A^\mu_a A^c_\mu+\bar{\Psi}\gamma^+T^s\Psi \\
    = &f^{sac}\partial^+A^i_a A^c_i+\bar{\Psi}\gamma^+T^s\Psi 
    = f^{sac}\partial^+\tilde{A}^i_a \tilde{A}^c_i+\bar{\tilde{\Psi}}\gamma^+T^s\tilde{\Psi}
    =\tilde{J}^+_s
    \;.
  \end{split}
\end{align}
Let us also introduce the fermion current $\tilde{\jmath}^\mu_a\equiv\bar{\tilde{\Psi}}\gamma^\mu T^a\tilde{\Psi}$ as part of the total current $\tilde{J}^\mu_a$. 
By substituting Eqs.~\eqref{eq:fields_first} to~\eqref{eq:fields_last_1} into Eq.~\eqref{eq:Hamiltonian_h}, and with $\tilde{A}^i=A^i$, we finally get the front form Hamiltonian,
\begin{align}\label{eq:Pmn_QCD}
  \begin{split}
    P_{QCD}^-=&\int\diff x^-\diff^2 x_\perp\
    -\frac{1}{2} \tilde{A}^j_a{(i\nabla_\perp)}^2 \tilde{A}_j^a
    +\frac{1}{2}\bar{\tilde{\Psi}}\gamma^+\frac{m^2-\nabla_\perp^2}{i\partial^+}\tilde{\Psi}\\
    &-g f^{abc}\partial^i \tilde{A}^j_a \tilde{A}_i^b
    \tilde{A}_j^c
    +g \tilde{J}^+_a\tilde{A}^a_+
    +g\bar{\tilde{\Psi}}\gamma^i \tilde{\bm{A}}_i\tilde{\Psi}
    \\
    &-\frac{1}{2}g^2 \tilde{J}^+_a\frac{1}{{(\partial^+)}^2}\tilde{J}^+_a
    +\frac{g^2}{4}f^{abc}\tilde{A}^i_b \tilde{A}^j_c f^{aef}\tilde{A}_i^e\tilde{A}_j^f
    \\
    &    +\frac{g^2}{2}\bar{\tilde{\Psi}}\gamma^i \tilde{\bm{A}}_i\frac{\gamma^+}{i\partial^+}\gamma^j \tilde{\bm{A}}_j\tilde{\Psi}
    \;.
  \end{split}
\end{align}

\begin{figure}[htp!]
  \centering
  \subfigure[The three-gluon interaction, $-g f^{abc}\partial^i A^j_a A_i^b
  A_j^c$  \label{fig:ggg}]
  {\includegraphics[width=.17\textwidth]
  {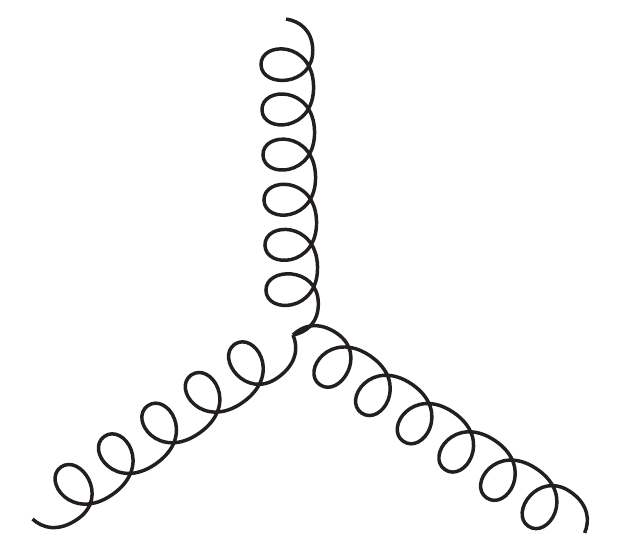}
  }
\quad
  \subfigure[The gluon emission, $g\bar{\Psi}\gamma^\mu \bm A_\mu\Psi$ \label{fig:qqg}]
  {\includegraphics[width=.2\textwidth]{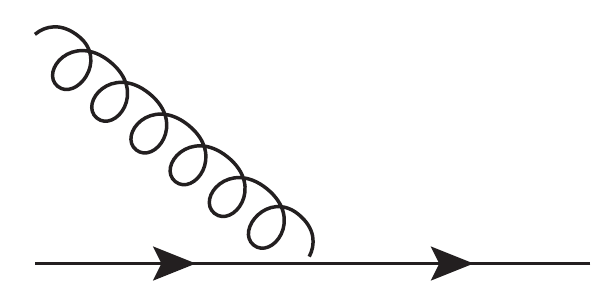}
  }
  \quad
  \subfigure[The four-gluon interaction, $\frac{1}{4} g^2f^{abc}A^i_b A^j_c f^{aef}A_i^eA_j^f$  \label{fig:gggg}]
  {\includegraphics[width=.17\textwidth]
  {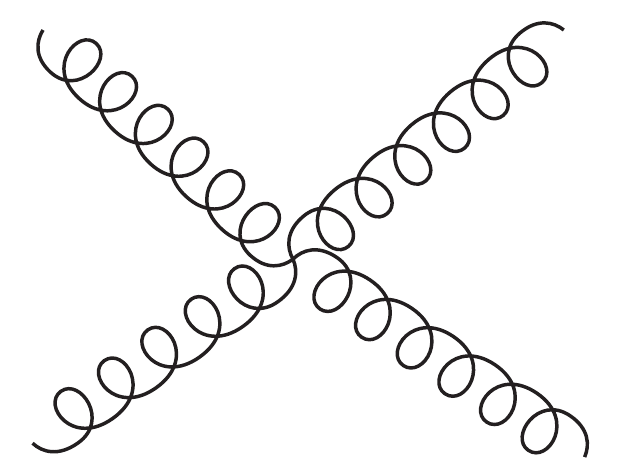}
  }
\quad
  \subfigure[The instantaneous-quark-interaction, $\frac{1}{2} g^2\bar{\Psi}\gamma^i \bm A_i\frac{\gamma^+}{i\partial^+}\gamma^j \bm A_j\Psi$ \label{fig:qg_ins_q}]
  {\includegraphics[width=.2\textwidth]{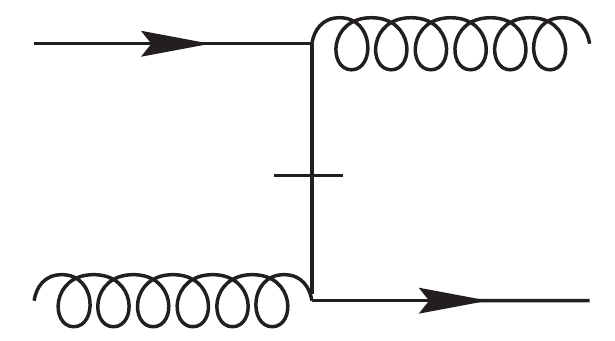}
  }
  \\
  \subfigure[The instantaneous-gluon-interaction,  
  $-\frac{1}{2}g^2 J^+_a\frac{1}{{(\partial^+)}^2}J^+_a$\label{fig:qq_ins_g}]
  {
    \includegraphics[width=.2\textwidth]{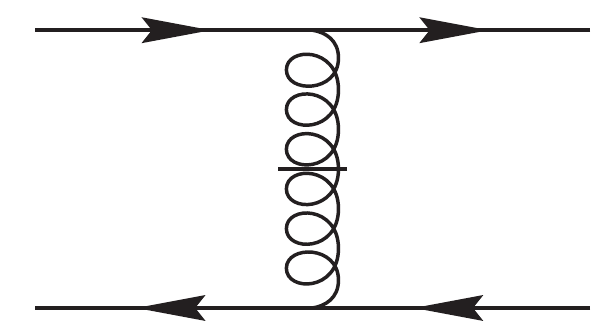}
    \includegraphics[width=.2\textwidth]{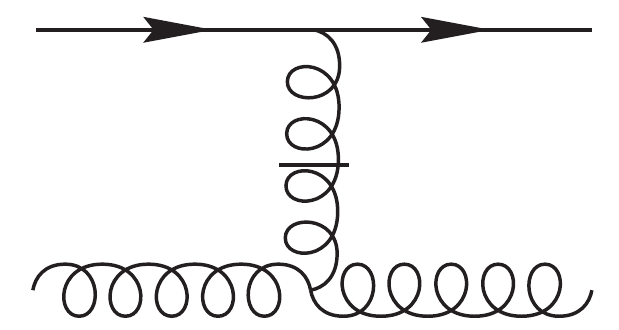}
    \includegraphics[width=.2\textwidth]{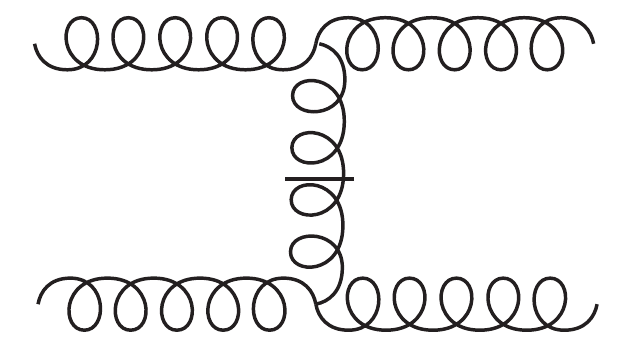}
  }
  \caption{Vertex diagram representation of the light-front QCD Hamiltonian in Eq.~\eqref{eq:Pmn_QCD}. The solid lines represent the quark operators, and the curly lines represent the gluon operators. The instantaneous quark (gluon)
  propagator $1/(i\partial^+)$ [$1/(\partial^+)^2$] is represented by a quark (gluon) line with a bar across it.}
  \label{fig:LFQCD_vertices}
\end{figure}

The two terms in the first line are the kinetic energy for the gauge field and the fermion respectively. The three terms in the second line can be written collectively as $g J^\mu_a A^a_\mu$, which include the three-gluon-interaction, the gluon emission and quark-antiquark-pair-production processes. The two terms in the third line are the instantaneous-gluon-interaction and the four-gluon-interaction respectively. The last line contains the instantaneous-fermion-interaction. The vertex diagrams for these interactions are shown in Fig.~\ref{fig:LFQCD_vertices}

The fields for QCD admit free-field expansions at $x^+ = 0$ ~\cite{Brodsky:1997de},
\begin{align}
    \label{eq:field_expand_psi}
  \Psi_{\alpha cf}(x)=\sum_{\lambda=\pm\frac{1}{2}}
\int\frac{\diff^2 p_\perp \diff p^+}{{(2\pi)}^3 2p^+}
\theta(p^+)
\left[b_q(p)u_\alpha(p,\lambda)e^{-ip\cdot x} +
d^{\dagger}_q(p)v_\alpha(p,\lambda)e^{ip\cdot x}\right]
\;,\\
\label{eq:field_expand_A}
A_{\mu a}(x)=\sum_{\lambda=\pm 1}
  \int\frac{\diff^2p_\perp \diff p^+}{{(2\pi)}^3 2p^+}
  \theta(p^+)
 \left[a_q(p) \epsilon_\mu(p,\lambda) e^{-ip\cdot x}+
  a^\dagger_q(p) \epsilon_\mu^* (p,\lambda) e^{ip\cdot x}\right]
\; ,
\end{align}
where $\theta(p^+)$ is the Heaviside unit step function. $\alpha$ denotes the spinor components of $\Psi$, and $\mu$ denotes the vector components of $A$. $\lambda$ is the light-front helicity of the corresponding field ($\lambda=\pm 1/2$ for quarks and $\lambda=\pm 1$ for gluons). $c = 1,2,3$ and $a=1,2,\ldots,8$ are the color indices of quarks (antiquarks) and gluons respectively. $q$ contains the quantum numbers of single particle state, for fermion $q= \{\lambda, c, f (\text{flavor})\}$ and for gluons $q= \{\lambda, a\}$. The creation and annihilation operators obey the commutation and anti-commutation relations. For gluons,
\begin{align}
\begin{split}
 & [a_{\lambda a}(p),a^{\dagger}_{\lambda' a'}(p')]=2p^+\theta(p^+){(2\pi)}^3\delta^3
 (p-p')\delta_{\lambda \lambda'}\delta_{aa'}\;,
\end{split}
\end{align}
where $\delta^3(p-p')=\delta(p^+-{p'}^+)\delta^2(\vec p_\perp- \vec p_\perp')$.
For quarks and antiquarks,
\begin{align}
  \begin{split}
  &\{b_{\lambda c f}(p),b^{\dagger}_{\lambda' c' f'}(p')\}=2p^+\theta(p^+){(2\pi)}^3\delta^3 (p-p')\delta_{\lambda \lambda'}\delta_{cc'}\delta_{ff'}\\
 & \{d_{\lambda c f}(p),d^{\dagger}_{\lambda' c' f'}(p')\}=2p^+\theta(p^+){(2\pi)}^3\delta^3
 (p-p')\delta_{\lambda \lambda'}\delta_{cc'}\delta_{ff'}\;.
\end{split}
\end{align}
All the other commutation and anti-commutation relations vanish,
\begin{align}
  [a_{\lambda a}(p),a_{\lambda' a'}(p')]=
 \{b_{\lambda c f}(p),b_{\lambda' c' f'}(p')\}=
 \{d_{\lambda c f}(p),d_{\lambda' c' f'}(p')\}=
  \{b_{\lambda c f}(p),d^{\dagger}_{\lambda' c' f'}(p')\}=\cdots=0
  \;.
\end{align}

The fields obey the standard equal-light-front-time commutation relations, and here we write it out for the dynamical components (suppressing the flavor indices):
\begin{align}
     \{\Psi_{+, c}(x),\Psi_{+,c'}^\dagger(y)\}_{x^+=y^+}=
\Lambda_+ \delta(x^- - y^-) \delta^2(\vec x_\perp - \vec y_\perp)\delta_{c,c'}\;, 
\end{align}
in which recall that $ \Lambda_+ =\gamma^0\gamma^+/2$ is the light-front projector , and
\begin{align}
   [A_{i, a}(x),A_{j, a'}^\dagger(y)]_{x^+=y^+}
    =-\frac{i}{4} \epsilon(x^- - y^-) \delta^2(\vec x_\perp - \vec y_\perp)\delta_{i,j} \delta_{a,a'}
    \;,
  \end{align}
with $i,j=1,2$, and $\epsilon(x)$ is the sign function.

\subsubsection{Fock space representation}\label{sec:Fock_rep}
The Hilbert space for the single-particle creation and destruction operators is the Fock space. The Fock space can be decomposed
into sectors with $n$ Fock particles, in which the number of quarks, antiquarks and gluons, $N$, $\bar{N}$ and $\tilde N$,
respectively and $n = N + \bar N + \tilde N$. Fock states
can be defined in terms of the eigenstates of the free-field Hamiltonian, i.e., the light-front kinetic operator, and can be obtained by applying the creation operators on the Fock vacuum $\ket{0}$.
The hadron state vector $ \ket{\psi_h(P,j,m_j)}$ can be expanded in the Fock space. We use $j$ as the total spin
of meson and $m_j$ as its magnetic projection. In the single-particle coordinates, it reads
\begin{align}\label{eq:hadron_Fockrep}
  \begin{split}
    \ket{\psi_h(P,j,m_j)}
    =&\sum_{n=0}^\infty\int \prod_{i=1}^n \frac{\diff \kappa^+_i \diff^2
      \kappa_{i \perp}}{{(2\pi)}^32 \kappa^+_i}\theta(\kappa_i^+)
    2P^+\theta(P^+){(2\pi)}^3 \delta^3(\kappa_1+\kappa_2+\cdots +\kappa_n-P)\\
    &\times
 \sum_{\{l_i,s_i\}}\psi_{n/h}^{(m_j)}(\{\kappa_i, s_i, l_i\})
    c^\dagger_{s_1 l_1}(\kappa_1)\ldots c^\dagger_{s_n l_n}(\kappa_n)\ket{0}
    \; ,
  \end{split}
\end{align}
where $i$ is the index of the Fock particle, and it takes values of $i=1,\ldots,n$ for the n-particle sector. $c^\dagger_{s_i l_i}(\kappa_i)$ is the creation operator for the corresponding constituent (quark, antiquark or gluon). $\kappa_i$ is the
momentum, and each particle is on its mass-shell $\kappa_i^2=m_i^2$.
$l$ is the color index, and $s$ is the spin projection of the particle. 
$\sum_{\{l_i,s_i\}}$ means the sum of all color and spin arrangements in the string of the creation operators resulting in a sum over a unique set of creation operators with the restriction of producing color-singlet projected states. The construction of the global color singlets for multi-particle states
can be found in Ref.~\cite{1stBLFQ}. We suppress flavor indices but they can be included in a straightforward manner. $\psi_{n/h}^{(m_j)}(\{\kappa_i, s_i, l_i\})$ are the projection of the physical states to the Fock states, called the light-front
wavefunctions (LFWFs).

In the relative particle coordinates, we define
\begin{align}\label{eq:xk}
  x_i\equiv\frac{\kappa_i^+}{P^+},\qquad 
  \vec k_{i\perp}\equiv \vec \kappa_{i\perp}-x_i \vec P_\perp
  \; .
\end{align}
$x_i$ are known as the longitudinal momentum fractions; $ \vec k_{i\perp}$ are the relative transverse momenta. They are independent
of the total momentum of the bound state, and satisfy $0<  x_i < 1$, $\sum^n x_i = 1$ and $\sum\vec k_{i\perp} = \vec 0$. 
\exercise{Show that the quantities defined in Eq.~\eqref{eq:xk} are invariant under Lorentz boosts 
  \begin{align}
    &v^+ \to v^+,\quad \vec{v}_\perp \to \vec{v}_\perp +v^+\vec{\beta}_\perp \;,\\
    & v^+ \to c_\omega v^+,\quad \vec{v}_\perp \to \vec{v}_\perp  \;,
  \end{align}
    with c-numbers $\vec \beta_\perp$ and $c_\omega$.
    }
    
The hadron state vector now reads, 
    \begin{align}
      \begin{split}
        \ket{\psi_h(P,j,m_j)}
        =&\sum_{n=0}^\infty\int\prod_{i=1}^n
        \frac{\diff x_i \diff^2
      k_{i\perp}}{{(2\pi)}^32 x_i}
    2{(2\pi)}^3
\delta(x_1 + x_2 +\cdots+x_n - 1)
    \delta^2(\vec k_{1\perp}+\vec k_{2\perp}+\cdots +\vec k_{n\perp})\\
    &\times
    \sum_{\{l_i,s_i\}}\psi_{n/h}^{(m_j)}(\{x_i,\vec k_{i\perp}, s_i,l_i\})
    c^\dagger_{s_1 l_1}(x_1 P^+, \vec k_{1\perp}+x_1\vec P_\perp)\cdots c^\dagger_{s_n l_n}(x_n P^+, \vec k_{n\perp}+x_n\vec P_\perp)\ket{0}
    \; ,
  \end{split}
    \end{align}
    with the LFWFs $\psi_{n/h}^{(m_j)}(\{x_i,\vec k_{i\perp}, s_i,l_i\})$ in the relative coordinates.
    
The hadron state vector is normalized as,
\begin{align}\label{eq:hadron_norm}
  \braket{\psi_h(P,j,m_j)|\psi_{h'}(P', j', m_j')}
  =2P^+\theta(P^+) {(2\pi)}^3\delta^3(P-P')\delta_{m_j, m_j'}\delta_{j,j'}\delta_{h, h'}
  \; .
\end{align}
Then the normalization of the LFWFs reads,
\begin{align}\label{eq:hadron_LFWF_norm}
  \begin{split}
\sum_{n=0}^\infty\int\prod_{i=1}^n
    \frac{\diff x_i \diff^2
  k_{i\perp}}{{(2\pi)}^3 2 x_i}
2{(2\pi)}^3
\delta(x_1 + \cdots+x_n - 1)
&\delta^2(\vec k_{1\perp}+\cdots +\vec k_{n\perp})
\sum_{\{l_i,s_i\}} \left|\psi_{n/h}^{(m_j)}(\{x_i,\vec k_{i\perp}, s_i,l_i\})\right|^2
=1
\; .
\end{split}
\end{align}
\exercise{
For practical calculations, the infinite Fock space needs to be truncated. 
Consider a meson state in the $\ket{q \bar q}$ Fock sector, write out its light-front wavefunction representation in terms of the relative momenta,
\begin{align}
  x\equiv\frac{k_q^+}{P^+},\qquad \vec k_\perp\equiv \vec k_{q \perp}-x \vec P_\perp
  \; .
\end{align}
You will get,
\begin{align}
  \begin{split}
    \ket{h_{q\bar{q}}(P,j,m_j)}
    =&\sum_{s, \bar s}
    \int_0^1\frac{\diff x}{2x(1-x)}
    \int\frac{\diff^2  k_\perp}{{(2\pi)}^3}
    \psi_{s\bar s/h}^{(m_j)}(\vec k_\perp, x)\\
    &\times\frac{1}{\sqrt{N_c}}\sum_{i=1}^{N_c}
    b^\dagger_{s i}(xP^+, \vec k_\perp+x \vec P_\perp)d^\dagger_{\bar s i}((1-x)P^+,-\vec k_\perp+(1-x) \vec P_\perp)\ket{0}
    \; .
  \end{split}
\end{align}
Here we write the color singlet configuration of the $q\bar q$ state, $1/\sqrt{3}(r\bar r+ g\bar g+b\bar b)$, explicitly with color index $i$ and $N_c=3$ in the above equation.
The normalization relation of the valence LFWF $\psi_{s\bar s/h}^{(m_j)}(\vec k_\perp, x)$ is
\begin{align}
  \sum_{s, \bar s}\int_0^1\frac{\diff x}{2x(1-x)}
  \int\frac{\diff^2k_\perp}{{(2\pi)}^3}
  \psi_{s\bar s/h'}^{(m_j')*}(\vec k_\perp, x)
  \psi_{s\bar s/h}^{(m_j)}(\vec k_\perp, x)
  =\delta_{hh'}\delta_{m_j, m_j'}\delta_{h, h'}
  \; .
\end{align}
}

\subsubsection{The eigenvalue equation}

The quarkonium state $ \ket{\psi_h}$ is an eigenstate of the light-front Hamiltonian, and satisfies
\begin{align}\label{eq:H_lf}
  H_{LF}\ket{\psi_h} = M_h^2 \ket{\psi_h} 
  \;,
\end{align}
where $H_{LF}=P^+ P^- + \vec P_\perp^2$ is the light-front Hamiltonian and $M_h$ is the mass of the bound state. Each eigenstate $ \ket{\psi_h} $ can be labeled with six eigenvalues, $M_h$, $P^+$, $\vec P_\perp$, the total spin $j$ and its longitudinal projection $m_j$. 

Projecting the Hamiltonian eigenvalue
equation of Eq.~\eqref{eq:H_lf} onto the Fock space results in an infinite number of coupled
integral eigenvalue equations. The solutions of these equations consist of the spectrum and the corresponding wavefunctions, which could
fully describe the bound state system. Fock states
can be defined in terms of the eigenstates of the free-field Hamiltonian, i.e., the light-front kinetic operator, and can be obtained by applying the creation operators on the Fock vacuum $\ket{0}$:
\begin{align}\label{eq:Fock_space}
\begin{split}
  Q_0 &\equiv \ket{q \bar q:~ k_i^+, \vec k_{i \perp}, \lambda_i} = b_{\lambda_1}^\dagger(k_1) d_{\lambda_2}^\dagger(k_2) \ket{0}\\
  Q_1 &\equiv \ket{q \bar q g:~ k_i^+, \vec k_{i \perp}, \lambda_i} = b_{\lambda_1}^\dagger(k_1) d_{\lambda_2}^\dagger(k_2) a_{\lambda_3}^\dagger(k_3)\ket{0}\\
    Q_2 &\equiv \ket{q \bar q q \bar q:~ k_i^+, \vec k_{i \perp}, \lambda_i} = b_{\lambda_1}^\dagger(k_1) d_{\lambda_2}^\dagger(k_2)b_{\lambda_3}^\dagger(k_3) d_{\lambda_4}^\dagger(k_4) \ket{0}\\
    \ldots
  \end{split}
\end{align}
For convenience, we have labeled the various Fock states
with index $n = 1,2,\ldots$. Each Fock state $Q_n$ is an eigenstate of $P^+$ and $\vec P_\perp$, satisfying $P^+ = \sum_i k_i^+$ and $\vec P_\perp = \sum_i \vec k_\perp$. 

In practical calculations, only a finite number of the leading Fock sectors are considered. The eigenvalue equation, Eq.~\eqref{eq:H_lf}, 
can be written explicitly on the finite Fock basis truncated as, 
\begin{align}\label{eq:H_lf_finiteF}
  \sum_{j = 1}^N H_{ij} \ket{\psi_j}= M_h^2 \ket{\psi_i}
  \qquad
  \text{for all } i = 1,2,\ldots,N
  \;.
\end{align}
We define the block matrices $H_{ij} \equiv Q_i H_{LF} Q_j$, and the projected eigenstates $\ket{\psi_i} \equiv Q_i \ket{\psi_h}$. One could then proceed to solve the coupled matrix equations in Eq.~\eqref{eq:H_lf_finiteF}. The resulting eigenstate can be written as
\(\ket{\psi_h} = \sum_{n=1}^N \int \diff [k_i] Q_n \ket{\psi_n}\).

Even with a finite truncation scheme, solving the Hamiltonian matrix becomes a major challenge in numerical calculations with increasing number of Fock sectors. Could we include the physics from higher Fock sectors while carrying out the calculation at a smaller feasible Fock space? A well known and widely used method is the effective interactions. In field theories, it was first introduced by I.Tamm~\cite{Tamm:1945qv} and rediscovered by S.M.Dancoff~\cite{Dancoff:1950ud} to describe the two nucleon forces. It reduces and solves the field equations according to the number of Fock particles. 

Although the Tamm-Dancoff approach was applied originally in the instant form, we can derive it analogously in the front form. The Fock space could be arbitrarily divided into two parts, namely the P-space and the Q-space. By choosing a specific partition, we wish to formulate an effective potential acting only in the P-space but including the effects generated by the Q-space. The Hamiltonian matrix equation, Eq.~\eqref{eq:H_lf_finiteF}, can then be rewritten as a coupled matrix equation involving the block matrices \(H_{\alpha \beta} \equiv \braket{\alpha|H_{LF}|\beta}\) and the projected eigenfunctions \( \ket{\psi_h}_\alpha = \braket{\alpha|\psi_h}\) with \((\alpha,\beta = P, Q) \):
\begin{subequations}\label{eq:PQ_H}
\begin{align}
H_{PP} \ket{\psi}_P + H_{PQ} \ket{\psi}_Q = \omega \ket{\psi}_P \;,\label{eq:PQ_H_a}\\
H_{QP} \ket{\psi}_P + H_{QQ} \ket{\psi}_Q = \omega \ket{\psi}_Q \;.\label{eq:PQ_H_b}
\end{align}
\end{subequations}
The mass eigenvalue is unknown at this point, and it is written as $\omega = M_h^2$ in the above equations. One can express the Q-space wavefunction $\ket{\psi_h}_Q$ in terms of the P-space wavefunction $\ket{\psi_h}_P$ from Eq.~\eqref{eq:PQ_H_b} as,
\begin{align}
\ket{\psi}_Q = \frac{1}{\omega - H_{QQ} }H_{QP} \ket{\psi}_P\;.
\end{align}
Plugging it into Eq.~\eqref{eq:PQ_H_a}, we arrive at an eigenvalue equation with an ``effective Hamiltonian" acting only in the P-space:
\begin{align}\label{eq:P_Heff_eq}
H_{\text{eff}}\ket{\psi}_P = \omega \ket{\psi}_P 
\;,
\end{align}
with 
\begin{align}\label{eq:P_Heff}
H_{\text{eff}} = H_{PP} + H_{PQ} \frac{1}{\omega - H_{QQ} }H_{QP} 
\;.
\end{align}
We can see that the effective interaction contains two parts: the original block matrix $H_{PP}$, and a contribution where the system is scattered virtually into the Q-space and then scattered back to the P-space. 

One key problem now is to compute the energy denominator $(\omega - H_{QQ})^{-1}$, since the value of $\omega$ is unknown before solving the equations. One could start with some fixed value of $\omega$ as the ``starting point energy" and calculate $M_h^2(\omega)$ from the eigenvalue equation. The true eigenvalues are determined by varying $\omega$ until $\omega = M_h^2(\omega)$ ~\cite{Pauli:1981zik,Zheng:1993wr}. This procedure, involving inverting a Q-space matrix, however, does not seem to reduce the numerical work of diagonalizing the (P+Q)-space matrix directly. An alternative way is to substitute the eigenvalue $\omega$ by $T^*$, the average kinetic energy of the initial and final P-space states~\cite{Krautgartner:1991xz}. The idea is to reduce the matrix $\omega - H_{QQ}$ to its dominant term as a c-number.
The Q-space matrix $H_{QQ}$ splits into a diagonal kinetic term $T_{QQ}$ and an off-diagonal interaction term $U_{QQ}$. The inverse matrix could then be written as 
\begin{align}\label{eq:ED}
 \frac{1}{\omega - H_{QQ}} = \frac{1}{T^* - T_{QQ} -\delta U(\omega)}
,\qquad
\delta U(\omega) = \omega - T^* - U_{QQ}
\;.
\end{align}
In the case of a sufficiently small $\delta U(\omega)$, the energy denominator can be approximated by the kinetic energy $T^* - T_{QQ}$, which no longer depends on the energy eigenvalue. 

\subsubsection{Mesons in the valence Fock sector}
\begin{figure}[h!tb] \centering
  \includegraphics[width=.8\textwidth]{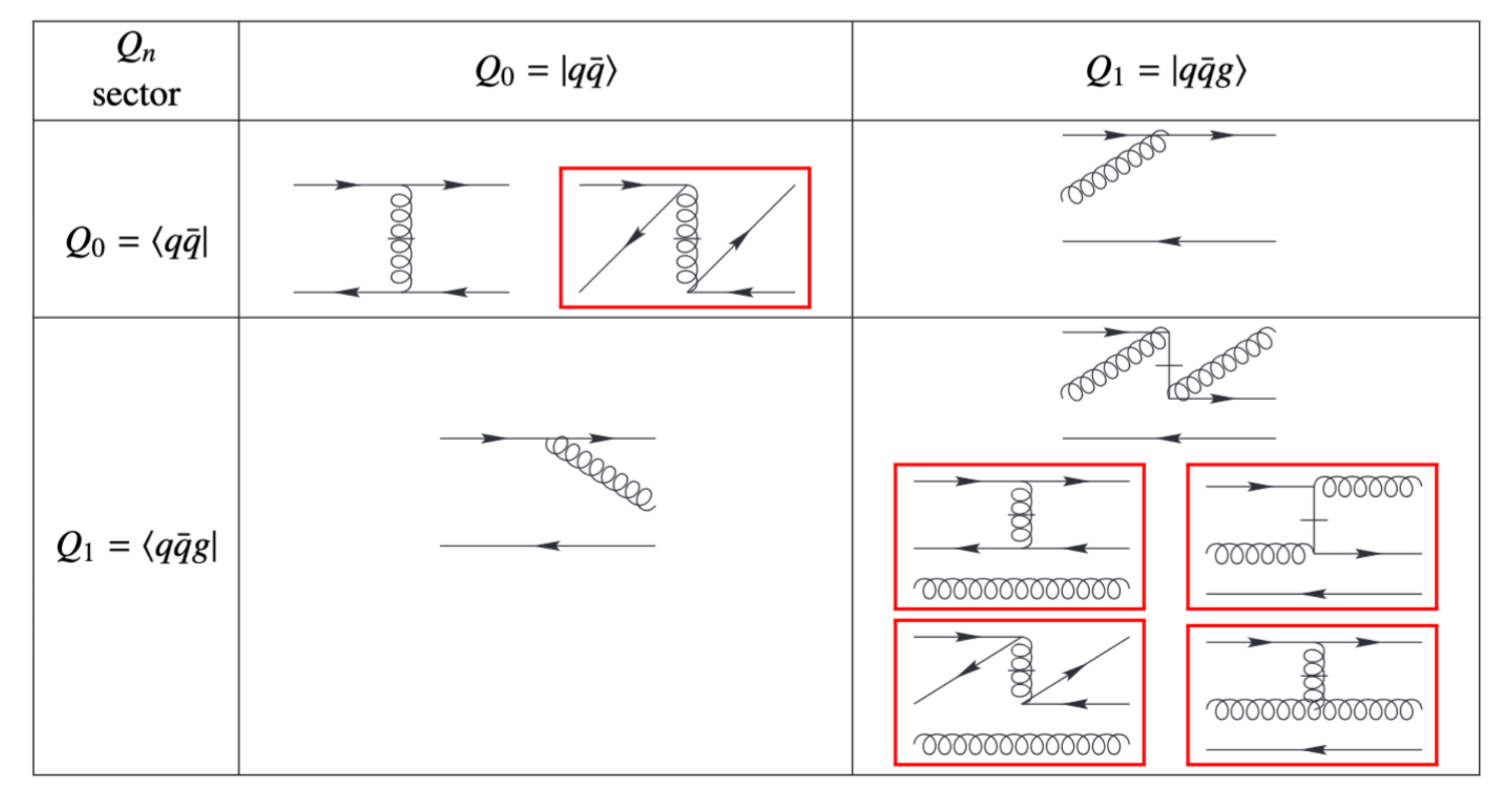}
  \caption{The interaction matrix $U$ for a meson in the Fock space $\ket{q\bar q}+ \ket{q\bar q g}$. The matrix elements are
  represented by diagrams. For each diagram where the gluon couples to the quark, there also exists a corresponding diagram with the gluon
  coupling to the antiquark.
  Diagrams in the red frames are excluded by color factor or gauge cutoff, see details in the text. }
 \label{tab:H01}
\end{figure}

In solving bound state systems with effective Hamiltonian approaches, the simplest P-space one can choose is the valence Fock sector. For heavy quarkonium, constituent quark models have shown reasonable first approximations in non-relativistic potential models \cite{Appelquist:1978aq, Godfrey:1985xj}. In the following, we illustrate the formulation of the effective Hamiltonian in the valence Fock sector by choosing $Q_0=\ket{q\bar q }$ as the P-space and $Q_1=\ket{q\bar q g}$ as the Q-space. The eigenvalue equation now reads (signifying the $Q_i$ by its index "$i$" in the following),
\begin{align}\label{eq:H_01}
\bigg(H_{00} + H_{01} \frac{1}{\omega - H_{11}} H_{10} \bigg)\psi_0 = \omega \psi_0\;.
\end{align}
We can write the Hamiltonian as a summation of the kinetic energy and the interaction operator, $H=T+U$. The diagonal block $H_{ii}$ contains $T_{ii}$ and $U_{ii}$, and the off-diagonal block is $H_{ij}= U_{ij}, ~(i\neq j)$.
The interaction matrix $U$ is illustrated in Table~\ref{tab:H01}.

We first focus on the denominator of the second term in Eq.~\eqref{eq:H_01}. To maintain the gauge
invariance in the truncated Fock space, we implement the ``gauge cutoff" formulated by Tang, Brodsky, and Pauli~\cite{Tang:1991rc}; that is, the instantaneous parton graph is only retained if the corresponding propagating parton graph contributes in the truncated theory. As
a consequence, some instantaneous interactions in $U_{00}$ and $U_{11}$ are excluded. In the $U_{11}$ block, the diagrams in the red frames should not be considered since the corresponding $\ket{q\bar q gg}$ sector is absent in the truncated space. 
The second diagram in $U_{00}$ block vanishes for another reason: zero color factor.
We further adopt the approximation $\delta U(\omega)\approx 0$ in Eq.~\eqref{eq:ED},
i.e. $U_{11}\to 0$. In principle, this approximation can be improved systematically by performing an expansion in $\delta U(\omega)$ and retaining terms order-by-order in that expansion. The energy denominator now reduces to $T^* - T_{11}$. 

\begin{figure}[h!tb] \centering
  \includegraphics[width=.2\textwidth]{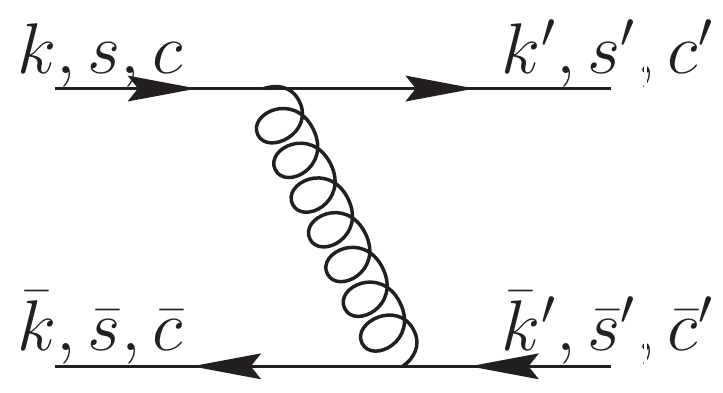}
  \qquad
  \includegraphics[width=.2\textwidth]{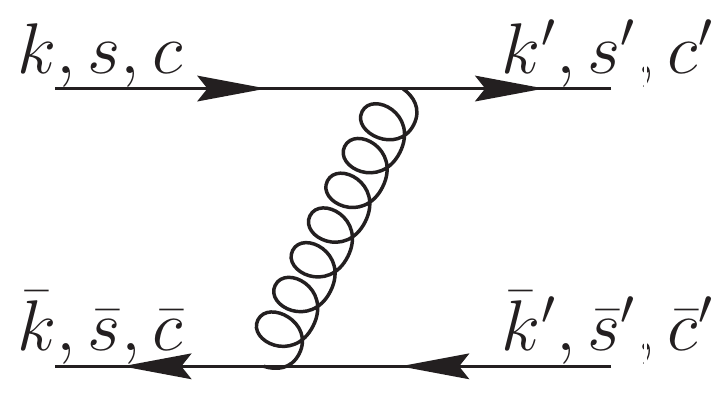}\\
~\\
  \includegraphics[width=.2\textwidth]{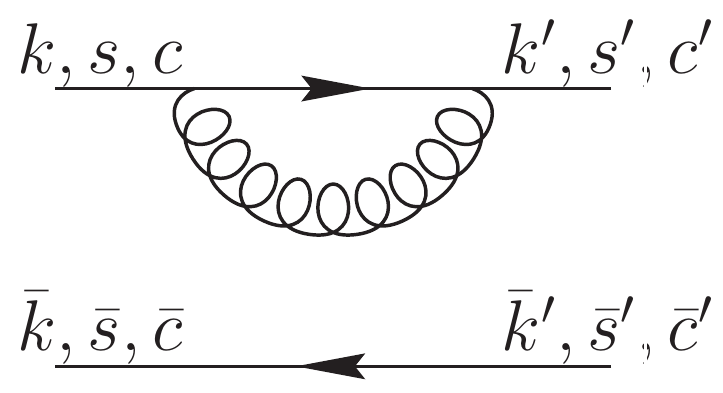}
  \qquad
  \includegraphics[width=.2\textwidth]{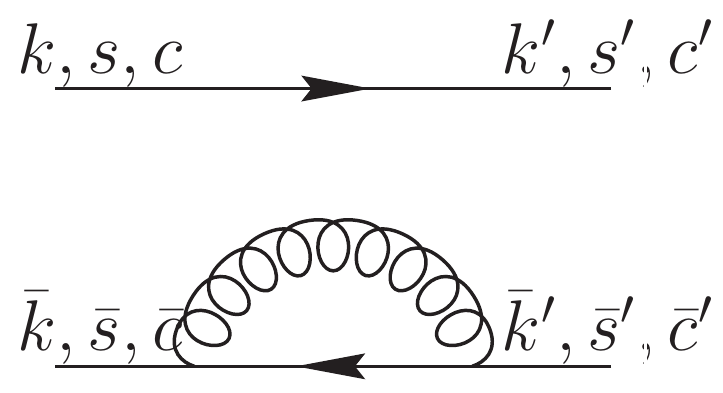}
  \caption{Iterated interactions generated in the two-body effective interaction. The top two panels are the gluon-exchange diagrams. The bottom two panels are the fermion-self-energy contributions. Each fermion line is labeled by its momentum ($k$), spin ($s$), and color ($c$).}
 \label{fig:H_qqg}
\end{figure}
The first term in Eq.~\eqref{eq:H_01}, $H_{00}$, contains an instantaneous gluon-exchange interaction, $U_{00}$. The second term, by stitching $U_{01}$ and
$U_{10}$, generates both fermion-self-energy loops and exchanges of gluons between the quark and the antiquark as shown in
Fig.~\ref{fig:H_qqg}.
We simplify the interaction by neglecting the self-energy terms in these investigations and we will adopt the strategy of using quark masses as adjustable parameters (called ``constituent quarks"). The remaining one-gluon exchange can be combined together with
the instantaneous contributions from $U_{00}$ into one term, namely $V_{\text{OGE}}$. In the Basis Light-Front Quantization (BLFQ) formalism of ref.~\cite{Yang_run}, the one-gluon exchange term reads,
\begin{align}\label{eq:OGE}
V_{\text{OGE}} =  - \frac{C_F 4\pi\alpha_s(q^2)}{q^2}\bar u_{s'}(k')\gamma_\mu u_s(k) \bar v_{\bar s}(\bar k) \gamma^\mu v_{\bar s'}(\bar k')
\;.
\end{align}
The energy denominator can now be interpreted as the average 4-momentum squared carried by the exchanged gluon, $q^2 = -(1/2) (k'-k)^2- (1/2) (\bar k'-\bar k)^2$. $C_F$ is the color factor of the one-gluon exchange diagram, and its calculation follows the corresponding QCD vertices~\cite{Griffiths:2008zz}. Here the initial and final quark-antiquark pairs are both in the color singlet configuration, thereby
$C_F = 1/4(1/\sqrt{3}{c'}^\dagger T^\alpha c)(1/\sqrt{3}{c}^\dagger T^\alpha c')=4/3$, where $T^\alpha ~(\alpha=1,\ldots,8)$ are the Gell-Mann matrices and $c,c'=$ red, blue, green are the color vectors, their expressions can be found in Appendix~\ref{app:colorT}. The overall``-'' sign in Eq.~\eqref{eq:OGE} results from the anti-communitation relation of the fermion fields in calculating the vertices, in analogy to the Coulomb potential between two opposite charges in electrodynamics.
This term
implements the short-distance physics between the quark and the antiquark, and 
determines the spin structure of the mesons.
The eigenvalue equation of Eq.~\eqref{eq:H_01} then reduces to
\begin{align}\label{eq:H_01_reduce}
(T_{00} + V_{\text{OGE}} )\psi_0 = \omega_0 \psi_0
\;.
\end{align}
The one-gluon exchange interaction $V_{\text{OGE}}$ is identical to the one-photon exchange in quantum electrodynamics (QED), except for the color factor. The eigenvalue equation of Eq.~\eqref{eq:H_01_reduce} with the one-photon exchange has been applied to the positronium system in the basis function approach by Ref.~\cite{positronium}.
In the relative coordinate presentation, the kinetic term can be written as $T_{00}= (\vec k_\perp^2 + m_q^2)/x
 + (\vec k_\perp^2 + m_{\bar q}^2)/(1-x)$.
 \exercise{
 Derive this expression of $T_{00}$ from the light-front QCD Hamiltonian in  Eq.~\eqref{eq:Pmn_QCD}. Recall that $x=p^+_q/P^+$ is the longitudinal momentum fraction of the quark and $\vec k_\perp = \vec k_{q\perp} - x \vec P_\perp$ is the relative transverse momentum.
 }

One can imagine that expanding the Q-space directly would introduce more  interaction terms. Apart from the standard way of including interactions from a finite Q-space, there are also phenomenological approaches.
Light-front holography constructs an effective Hamiltonian based on inspirations from string theory. It addresses confinement, an essential feature of QCD, by holographic mapping gravity in a higher-dimensional anti-de Sitter(AdS) space to light-front dynamics \cite{holography}. In the soft-wall model, a 2-dimensional soft-wall confinement originates from the gravitational background field \cite{Karch:2006pv}. Y. Li et al. further improved the confinement by including the longitudinal degree of freedom \cite{Yang_fix,Yang_run}, 
\begin{align}\label{eq:confinement}
V_{\text{confinement}} = \kappa^4 x(1-x) r_\perp^2
-\frac{\kappa^4}{(m_q + m_{\bar q})^2}\partial_x(x(1-x)\partial_x)
\;.
\end{align}
 $\kappa$ is the strength of the confinement, $r_\perp = |\vec r_{q\perp} - \vec r_{\bar q \perp}|$ is the transverse separation of the partons. This phenomenological confinement takes into account long-distance physics, and provides another approximation to QCD. The eigenvalue equation provides a more extensive model of QCD by absorbing the confining potential,
\begin{align}\label{eq:H_01_reduce_conf}
(T_{00} + V_{\text{OGE}} + V_{\text{confinement}})\psi_0 = \omega \psi_0
\;.
\end{align}
Conventionally, all the contributions in the Hamiltonian excluding the kinetic energy are combined and the resulting interaction is referred to as the effective interaction, $V_{\text{eff}} = V_{\text{OGE}} + V_{\text{confinement}}$. 
  The mass spectrum and LFWFs are the direct solutions of the eigenvalue equation, and
could be obtained as in BLFQ by diagonalizing the Hamiltonian in a basis representation. 
\subsection{Two studies of meson light-front wavefunctions }
\subsubsection{Basis Light-Front Quantization (BLFQ)}
In the Hamiltonian approach to studying the meson states, the central task is to diagonalize the QCD Hamiltonian to solve the eigenvalue equation.
The basis light-front quantization (BLFQ) has been developed as a flexible computational platform for such purpose, dealing with relativistic strong interaction many-body bound-state problems \cite{1stBLFQ}.
Based on the Hamiltonian formalism in light-front dynamics, which we have discussed previously, BLFQ adopts basis function representation. This key feature provides us considerable freedom in the choice of the orthonormal and complete set of basis functions with convenience and convergence rates.

The BLFQ approach is first applied to the heavy quarkonium system in Ref.~\cite{Yang_fix, Yang_run}. As we have already introduced in the proceeding section, the effective light-front Hamiltonian is constructed in the $\ket{q\bar q}$ space, consisting of the holographic QCD Hamiltonian and the one-gluon exchange, as in Eq.~\eqref{eq:H_01_reduce_conf}. 
There are two model parameters, $\kappa$ and $m_q$, and they are determined by fitting the mass spectrum of the quarkonium system to experiments.
In solving the eigenvalue equation of  Eq.~\eqref{eq:H_01_reduce_conf}, the eigenfunctions of part of the Hamiltonian, $T_{00} + V_{\text{confinement}}$, are taken as the basis functions, which largely brings in numerical efficiency.
The heavy quarkonium system is then solved in such basis representation, giving the spectrum and the meson LFWFs. 
Following the applications in the heavy meson system, BLFQ is further developed and applied to the heavy-light system \cite{Tang:2019gvn}, the light mesons \cite{Jia:2018ary,Qian:2020utg}, and the nucleon \cite{Mondal:2020mpv, Xu:2021wwj}.

A significant and more challenging step is including higher Fock sectors to understand and explain the meson systems from QCD first principles. The BLFQ study, Ref. \cite{Lan:2021wok}, addresses the light meson systems in the $\ket{q\bar q} +\ket{q\bar q g}$ space. 

\subsubsection{Small-basis Light-Front Wavefunction (sLFWF) by design}
In a standard light-front Hamiltonian formalism, the meson LFWFs are solved from the Schr\"{o}dinger(-like) eigenvalue equations.
A different path to obtain the meson LFWF is to model it directly or determine it from other formalisms. Works in this category include the widely used boosted Gaussian~\cite{Kowalski:2006hc, Nemchik:1994fp, Nemchik:1996cw}, the LFWFs determined from the Dyson-Schwinger and Bethe-Salpeter approach~\cite{Shi:2018zqd, Mezrag:2016hnp,dePaula:2020qna, Shi:2021taf}, and LFWFs boosted from the NRQCD solution in the rest frame~\cite{Krelina:2019egg, Lappi:2020ufv}.

As a complementary study to the existing modeled LFWFs, Ref. \cite{Li:2021cwv} proposed a method of designing the LFWFs of meson bound states with a simple-functional form. Such ``by design'' approach is apparently not first-principle, in which one chooses by hand to apply the constraints being considered for phenomenological applications of the wavefunctions. This involves a certain amount of judgment on how many basis functions to include and which constraints to impose. One primary advantage of this approach is that the resulting LFWFs are analytically tractable and can be used to calculate a wide variety of physical observables. 

The following briefs the basic idea of this method.
Consider a meson state $h$ consisting of a quark and an antiquark, with momentum $(P^+, \vec P_\perp)$, and expand its wavefunction on an orthonormal basis $\{\beta_1, \beta_2, \ldots, \beta_{N_\beta}\}$,
\begin{equation}
  \psi_h(\vec k_\perp, x) = \sum_{i=1}^{N_\beta} C_{h,i} \beta_i(\vec k_\perp, x)\;,
\end{equation}
where $C_{h,i}$ are the basis coefficients for $h$ and $N_\beta$ is the number of basis states.
Here we are writing the wavefunction in a relative coordinate, where $x=p^+_q/P^+$ is the longitudinal momentum fraction of the quark and $\vec k_\perp = \vec k_{q\perp} - x \vec P_\perp$ is the relative transverse momentum. 

The wavefunctions should satisfy the orthonormalization relation
\begin{equation}\label{eq:norm}
\sum_{i=1}^{N_\beta} C_{h,i}C^*_{h',i} = \delta_{h,h'}\;.
\end{equation}
Physical quantities and observables ($O$) such as decay widths and charge radius are functions ($f_O$) of the basis coefficients,
\begin{equation}\label{eq:Oh}
  O_h = f_O (C_{h,i})\;.
\end{equation}
The constraints Eqs.~\eqref{eq:norm} and \eqref{eq:Oh} form a system of equations, and the unknowns are the basis coefficients $C_{h,i}$ and could also include parameters in the basis functions. 
The procedure of designing LFWFs is, in essence, solving such a system of equations. 

In work \cite{Li:2021cwv}, we modeled the LFWFs for four charmonium states, $\eta_c$, $J/\psi$, $\psi'$, and $\psi(3770)$ as superpositions of orthonormal basis functions.
We choose the basis functions as eigenfunctions of an effective Hamiltonian, which has a longitudinal confining potential in addition to the transverse confining potential from light-front holographic QCD, the same basis functions as in BLFQ \cite{Yang_run}.
We determine the basis function parameters and superposition coefficients by employing both guidance from the nonrelativistic description of the meson states and the experimental measurements of the meson decay widths. With the obtained wavefunctions, we study the features of those meson states, including charge radii and parton distribution functions. The obtained LFWFs have simple-functional forms and can be readily used to predict additional experimental observables.

%% file: L2.tex
\section{Lecture II: Electromagnetic transitions and form factors}

In quantum field theory, the electromagnetic (EM) elastic form factors (EFFs) characterize the structure of a bound state system, which generalize the multipole expansion of the charge and current densities in the nonrelativistic quantum mechanics. The physical process that
determines the EFFs is $\psi_h(P) + \gamma^{(*)}(q=P'-P)\to \psi_h(P')$. The form factors are defined as the Lorentz invariants arising in the Lorentz structure decomposition of the hadron matrix element
$\braket{\psi_h(P')|J^\mu(x)|\psi_h(P)}$. For a spin-j particle, assuming charge conjugation, parity and time reversal symmetries, there are $2j+1$ independent Lorentz invariant form factors. 
Similarly, the EM transition form factors (TFFs) arise in the transition between two meson states via emission of a photon, $\psi_A(P')\to \psi_B(P) + \gamma(q=P'-P)$, and the corresponding hadron matrix element is $\braket{\psi_B(P)|J^\mu(x)|\psi_A(P')}$.
Both EFFs and TFFs could help us in understanding the internal structure of mesons.

There is a connection between the experimentally measured decay width and the hadron matrix elements.
In the physical process of $A \to B+\gamma$, the photon is on shell ($q^2=0$). The transition amplitude is
\begin{align}\label{eq:amplitude}
  \mathcal{M}^\lambda_{m_j, m_j'}= I^\mu_{m_j, m_j'}\epsilon_{\mu,\lambda}^*(q) |_{q^2=0}\;,
  \qquad
    I^\mu_{m_j, m_j'}\equiv \bra{B(P, j_B, m_j)} J^\mu(0) \ket{A (P', j_A, m_j')}\;,
\end{align}
with the hadron matrix element $I^\mu_{m_j, m_j'}$ written out with the magnetic projections,
and $\epsilon_{\mu,\lambda}$ the polarization vector of the final-state photon with its spin projection $\lambda=\pm 1$.
The decay width is usually measured in the rest frame of the initial particle $A$, as such, the momenta of the initial meson, final meson, and the photon read (see Appendix~\ref{app:LF_cor} for the convention of ordering the 4-vector components in light-front coordinates), 
  $P'=(m_{A },m_{A  }, 0, 0)$, 
  $P=(\sqrt{m_{B }^2+k^2}, \sqrt{m_{\mathcal{P} }^2+k^2}, k, 0)$, and 
  $q = P' - P = (k, k, -k, 0)$.
The momentum of the photon is determined by energy-momentum conservation, \(|\vec{q}|=k=( m_{A}^2-
m_{B}^2)/2 m_{A }\). 
The decay width of $A \to B+\gamma$ follows by averaging over the initial polarization and summing over the final
polarization. 
\begin{align}\label{eq:ABwidth}
  \begin{split}
    \Gamma(A \to  B+\gamma)
    =&\int\diff \Omega_q
    \frac{1}{32\pi^2}
    \frac{|\vec{q}|}{m_{A }^2}
    \frac{1}{2 j_{A }+1}
    \sum_{m_j,m_j',\lambda}|\mathcal{M}^\lambda_{m_j, m_j'}|^2
    \;.
  \end{split}
\end{align}

We proceed by first examining the Lorentz vector decomposition of the hadron matrix elements in Sec.~\ref{sec:Lorentz_decompose}, then writing out the matrix elements in the light-front wavefunction representation in Sec.~\ref{sec:hadronMatrix}, and finally we calculate the EFFs and TFFs in Secs.~\ref{sec:EFF_J0} and \ref{sec:TFF_M1}.

\subsection{Lorentz structure decomposition}\label{sec:Lorentz_decompose}
According to spacetime translational invariance, the matrix element of the EM current operator satisfies
\begin{align}
  \bra{\psi_h'(P',j',m_j')} J^\mu (x) \ket{\psi_h(P,j,m_j)}
  =   \bra{\psi_h'(P',j',m_j')} J^\mu (0) \ket{\psi_h(P,j,m_j)}
  e^{i(P-P')\cdot x}
  \;.
\end{align}
\exercise{Check the above equation. You will see it again in Eq.~\eqref{eq:I_trans}.}

The current conservation condition $\partial_\mu J^\mu=0$ leads to
\begin{align}\label{eq:Qcon}
  {(P'-P)}_\mu \bra{\psi_h'(P',j',m_j')} J^\mu (0) \ket{\psi_h(P,j,m_j)}=0\;.
\end{align}
The charge operator on the light front is defined as
\begin{align}
  Q\equiv\int\diff x_+\diff^2 \bm{x}_\perp  J^+ (x)
  \;.
\end{align}
The eigenvalue of $Q$ on for a particle state is interpreted as the charge of that particle,
\begin{align}
  Q\ket{\psi_h(P,j,m_j)}=e_h\ket{\psi_h(P,j,m_j)}
  \;.
\end{align}
The evaluation of the charge operator on a particle state leads to a normalization relation at zero momentum transfer, i.e., $P'=P$,
\begin{align}
  \bra{\psi_h(P',j,m_j)} Q\ket{\psi_h(P,j,m_j)}
  = \int\diff x_+\diff^2  \bm{x}_\perp   \bra{\psi_h(P',j,m_j)} J^+(x) \ket{\psi_h(P,j,m_j)}
  \;.
\end{align}
That is,
\begin{align}
  2e_h P^+ {(2\pi)}^3\delta^3(P-P')
  ={(2\pi)}^3\delta^3(P-P')
  \bra{\psi_h(P',j,m_j)} J^+(0) \ket{\psi_h(P,j,m_j)}
  \;,
\end{align}
thus
\begin{align}\label{eq:j+0}
  \bra{\psi_h(P,j,m_j)} J^+(0) \ket{\psi_h(P,j,m_j)}
  =    2e_h P^+
  \;.
\end{align}
The form factors of particle transitions are those coefficients $F_i$ of vectors $V_i$ obtained by decomposing the hadron matrix element,
\begin{align}
  \bra{\psi_h'(P',j',m_j')} J^\mu (0) \ket{\psi_h(P,j,m_j)}
  =\sum_i^n F_i V_i^\mu
  \;.
\end{align}
\subsubsection{Spin $0$ mesons}\label{app:LD_spin0}
In the quark model, a spin-$0$ ($J=0$) meson could be either a scalar $0^{+}$ or a pseudo-scalar $0^{-}$. And $\mathrm{C}=1$ for quarkonium.
The matrix element of the current reads
\begin{align}
  \bra{h'_{q\bar{q}}(P',j'=m_i'=0)} J^\mu(0) \ket{h_{q\bar{q}}(P,j=m_i=0)}
  = e_q \mathscr{J}^\mu \;,
\end{align}
where $\mathscr{J}^\mu$ is a four-vector function of ${P'}^\mu$ and $P^\mu$. Relevant scalars are $|P'|$,  $|P|$ and $P'\cdot
P$. The first two are fixed by the on shell conditions, 
\begin{align}
  {P'}^\mu{P'}_\mu=m_{h'}^2,\qquad P^\mu P_\mu=m_h^2\;.
\end{align}
Therefore the coefficients of vectors should only depend on $P'\cdot P$.
Define
\begin{align}
  q^\mu\equiv {P'}^\mu-P^\mu,\qquad \bar{p}^\mu\equiv {P'}^\mu+P^\mu \;,
\end{align}
and decompose $\mathscr{J}^\mu$ into the form of
\begin{align}
  \mathscr{J}^\mu=  q^\mu H(q^2)+\bar{p}^\mu F(q^2)
  \;,
\end{align}
The condition of current conservation in Equation~\eqref{eq:Qcon} requires,
\begin{align}
  q_\mu \cdot q^\mu H(q^2)+  q_\mu \cdot \bar{p}^\mu F(q^2)=0
  \;.
\end{align}
This means there is only one independent form factor,
\begin{align}
  H(q^2)=-\frac{ q_\mu \cdot  \bar{p}^\mu}{ q_\mu \cdot q^\mu } F(q^2)
  =-\frac{ m_{h'}^2-m_h^2}{ q^2} F(q^2)
  \;.
\end{align}
It follows that 
\begin{align}\label{eq:spin0F}
  \bra{h'_{q\bar{q}}(P')} J^\mu(0) \ket{h_{q\bar{q}}(P)}
  = e_q [ \bar{p}^\mu -\frac{ m_{h'}^2-m_h^2}{ q^2}  q^\mu ]F(q^2)
  \;,
\end{align}
and $F(q^2)$ is the electromagnetic form factor.
To satisfy hermiticity, 
\begin{align}
  \bra{h'_{q\bar{q}}(P')} J^\mu(0) \ket{h_{q\bar{q}}(P)}
  =\bra{h_{q\bar{q}}(P)} J^\mu(0) \ket{h'_{q\bar{q}}(P')}^*\;,
\end{align}
$F(q^2)$ must be real. 

For the elastic scattering, $h'=h$, thus $m_{h'}=m_h$,
\begin{align}\label{eq:EFF_J0_e}
  \bra{h_{q\bar{q}}(P')} J^\mu(0) \ket{h_{q\bar{q}}(P)}
  = e_q \bar{p}^\mu F(q^2)
  \;.
\end{align}
Compare with the normalization relation in Eq.~\eqref{eq:j+0}, we get $F(0)=1$.

Let us now analyze the symmetries of parity and charge conjugation (see Appendix~\ref{app:symmetry}), and find out what kind of transitions are allowed. We first insert two complete sets of the parity operator to the matrix element,
\begin{align}\label{eq:PP}
  \begin{split}
    \bra{h'_{q\bar{q}}(P',\mathrm{P}_2)} J^\mu(0)\ket{h_{q\bar{q}}(P,\mathrm{P}_1)}
    =&
    \bra{h'_{q\bar{q}}(P',\mathrm{P}_2)}\varmathbb{P}   \varmathbb{P}^{-1} J^\mu(0) \varmathbb{P}  \varmathbb{P}^{-1} \ket{h_{q\bar{q}}(P,\mathrm{P}_1)}\\
    =&\mathrm{P}_2 \mathrm{P}_1 \mathcal{P}^\mu_\nu \bra{h'_{q\bar{q}}( \mathcal{P}\cdot P',\mathrm{P}_2)} J^\nu(0) \ket{h_{q\bar{q}}(
      \mathcal{P}\cdot P,\mathrm{P}_1)}\\
    =&e_q \mathrm{P}_2 \mathrm{P}_1 \mathcal{P}^\mu_\nu 
    \mathcal{P}^\nu_\rho [ \bar{p}^\rho -\frac{ m_{h'}^2-m_h^2}{ q^2}  q^\rho ]F(q^2)\\
    =&e_q \mathrm{P}_2 \mathrm{P}_1  [ \bar{p}^\mu -\frac{ m_{h'}^2-m_h^2}{ q^2}  q^\mu ]F(q^2)
    \;.
  \end{split}
\end{align}
Compare with Eq~\eqref{eq:spin0F}, we arrive at
\begin{align}
  \mathrm{P}_2 \mathrm{P}_1=+1\;.
\end{align}
This means the electromagnetic transitions of spin $0$ particles preserves the parity. The allowed transition modes are
$0^+\to 0^+$ (scalar-to-scalar) and $0^-\to 0^-$ (pseudoscalar-to-pseudoscalar).

We then consider the charge conjugation of quarkonium.
\begin{align}\label{eq:CC}
  \begin{split}
    \bra{h'_{q\bar{q}}(P',\mathrm{C}_2)} J^\mu(0)\ket{h_{q\bar{q}}(P,\mathrm{C}_1)}
    =&
    \bra{h'_{q\bar{q}}(P',\mathrm{C}_2)}\varmathbb{C}   \varmathbb{C}^{-1} J^\mu(0) \varmathbb{C}  \varmathbb{C}^{-1} \ket{h_{q\bar{q}}(P,\mathrm{C}_1)}\\
    =&-\mathrm{C}_2 \mathrm{C}_1 \bra{h'_{q\bar{q}}( P',\mathrm{C}_2)}  
    J^\mu(0)  \ket{h_{q\bar{q}}(
      P,\mathrm{C}_1)}
    \;.
  \end{split}
\end{align}
Compare with Eq~\eqref{eq:spin0F}, we arrive at
\begin{align}
  \mathrm{C}_2 \mathrm{C}_1=-1\;.
\end{align}
This means the electromagnetic transitions of  quarkonium must change the charge conjugation. However, all the spin $0$
quarkonium have the same parity conjugation $\mathrm{C}=+1$. Therefore the form factors for spin-$0$ quarkonium are zero.

 \subsubsection{Spin-$0\leftrightarrow$ spin-$1$ mesons}\label{app:LD_spin01}
The matrix element of the transition between a spin-0 and a spin-1 meson reads
\begin{align}\label{eq:spin01}
  \bra{h_{q\bar{q}}'(P', j'=1, m_j'=0,\pm 1)} J^\mu(0) \ket{h_{q\bar{q}} (P, j=0, m_j=0)}
  =
  {e^\alpha}^* (P', m'_j) \Gamma^\mu_\alpha 
  \;,
\end{align}
 where $e^*$ is the spin vector defined in Appendix~\ref{sec:spinvector} and $\Gamma^\mu_\alpha$ is a 2nd-order tensor function of 
$P^\mu$, ${P'}^\mu$, $g^{\mu\nu}$ and $\epsilon^{\mu\nu\rho\sigma}$. Note that we did not write out the charge here for simplicity. 
All possible non-vanishing combinations are
\begin{align}
  \begin{split}
    P^\mu, {P'}^\mu: &P^\mu P_\alpha, P^\mu P'_\alpha, 
    {P'}^\mu P'_\alpha , {P'}^\mu P_\alpha \;,\\
    g_{\alpha\beta}:  &g^\mu_\alpha \;,\\
    \epsilon^{\mu\nu\rho\sigma}: &\epsilon^\mu_{\alpha\rho\sigma} P^\rho {P'}^\sigma \;.
  \end{split}
\end{align}
Contracting with the spin vectors in Eq.~\eqref{eq:spin01}, and according to the Proca equation in~\ref{sec:spinvector},
\begin{align}
  P_\beta e^\beta(P,m_j)=0\;,\qquad  P'_\alpha {e^{\alpha}}^*(P', m_j')=0\;,
\end{align}
we get all possible non-vanishing vectors of ${e^{\alpha}}^*(P', m_j')\Gamma^\mu_\alpha$.
\begin{align}
  \begin{split}
    & P^\mu (e^*(P', m_j')\cdot P ),  
    {P'}^\mu (e^*(P', m_j')\cdot P )  \;,\\
&    {e^{\mu}}^*(P', m_j')  \;,\\
 &   \epsilon^\mu_{\alpha\rho\sigma} P^\rho {P'}^\sigma  {e^{\alpha}}^*(P', m_j')\;.
  \end{split}
\end{align}
Their coefficients are functions of $|P'|$,  $|P|$ and $P'\cdot P$. The first two are fixed by on shell conditions, 
\begin{align}
  {P'}^\mu{P'}_\mu=m_{h'}^2,\qquad P^\mu P_\mu=m_h^2\;.
\end{align}
Therefore those coefficients should only depend on $P'\cdot P$. For convenience, we define
\begin{align}
  q^\mu\equiv {P'}^\mu-P^\mu,\qquad \bar{p}^\mu\equiv {P'}^\mu+P^\mu \;.
\end{align}
The on shell condition now reads
\begin{align}
  q_\mu q^\mu=q^2\;, \qquad
  q_\mu \bar{p}^\mu=m_{h'}^2-m_h^2\equiv \Delta_m
  \;.
\end{align}
We thereby write ${e^{\alpha}}^*(P', m_j')\Gamma^\mu_\alpha$ as a linear combination of the vectors we found,
\begin{align}
  \begin{split}
    {e^{\alpha}}^*(P', m_j')\Gamma^\mu_\alpha
    =& \bar{p}^\mu (e^*(P', m_j')\cdot P )F_1
    +q^\mu (e^*(P', m_j')\cdot P ) F_2
    +{e^{\mu}}^*(P', m_j') F_3\\
   &+\epsilon^\mu_{\alpha\rho\sigma} \bar{p}^\rho q^\sigma  {e^{\alpha}}^*(P', m_j')F_4
\;.
  \end{split}
\end{align}
The condition of current conservation in Eq.~\eqref{eq:Qcon} requires,
\begin{align}
  q_\mu  {e^{\alpha}}^*(P', m_j')\Gamma^\mu_\alpha=0
  \;.
\end{align}
That is, 
\begin{align}
  \begin{split}
    0    = &  (e^*(P', m_j')\cdot P )[\Delta_mF_1    +q^2  F_2 - F_3]
    +\epsilon^\mu_{\alpha\rho\sigma} \bar{p}^\rho q^\sigma q_\mu  {e^{\alpha}}^*(P', m_j')F_4\;.
  \end{split}
\end{align}
$F_4$ survives since
\begin{align}
  \epsilon^\mu_{\alpha\rho\sigma} \bar{p}^\rho q^\sigma q_\mu=0 \;
\end{align}
The other terms satisfy,
\begin{align}
    \Delta_mF_1    +q^2  F_2 - F_3=0
\end{align}
We therefore rewrite the vector decomposition with new coefficients,
\begin{align}
  \begin{split}
    {e^{\alpha}}^*(P', m_j')\Gamma^\mu_\alpha
    =& [\bar{p}^\mu (e^*(P', m_j')\cdot P )-q^\mu (e^*(P', m_j')\cdot P ) \Delta_m/q^2]H_1\\
    &+[{e^{\mu}}^*(P', m_j')+ q^\mu (e^*(P', m_j')\cdot P )/q^2]H_2
+\epsilon^\mu_{\alpha\rho\sigma} \bar{p}^\rho q^\sigma  {e^{\alpha}}^*(P', m_j')H_3
    \;.
  \end{split}
\end{align}
Parity invariance requires that
\begin{align}\label{eq:PPspin01}
  \begin{split}
    \bra{h_{q\bar{q}}'(P',j'=1, m'_j,\mathrm{P}_2)} &J^\mu (0)\ket{h_{q\bar{q}}(P,j=0, m_j, \mathrm{P}_1)}\\
    =&
    \bra{h_{q\bar{q}}'(P', j'=1, m'_j, \mathrm{P}_2)}\varmathbb{P}   \varmathbb{P}^{-1} J^\mu(0)\varmathbb{P}
    \varmathbb{P}^{-1} \ket{h_{q\bar{q}}(P, j=0, m_j, \mathrm{P}_1)}\\
    =&\mathrm{P}_2 \mathrm{P}_1 \mathcal{P}^\mu_\nu \bra{h_{q\bar{q}}'( \mathcal{P}\cdot P', j'=1,m'_j, \mathrm{P}_2)}  J^\nu (0) \ket{h_{q\bar{q}}(
      \mathcal{P}\cdot P, j=0,m_j, \mathrm{P}_1)}
    \;.
  \end{split}
\end{align}
The matrix element under the parity transformation reads
\begin{align}
  \begin{split}
    \bra{h_{q\bar{q}}'( \mathcal{P}\cdot P', j'=1,m'_j, \mathrm{P}_2)} & J^\nu(0)  \ket{h_{q\bar{q}}(
      \mathcal{P}\cdot P, j=0, m_j, \mathrm{P}_1)}\\
    =&\mathcal{P}^\nu_\kappa  [\bar{p}^\kappa (e^*( \mathcal{P}\cdot P', m_j')\cdot ( \mathcal{P}\cdot P) )
    -q^\kappa (e^*( \mathcal{P}\cdot P', m_j')\cdot ( \mathcal{P}\cdot P) ) \Delta_m/q^2]H_1\\
    &-\mathcal{P}^\nu_\kappa[{e^\kappa}^*( \mathcal{P}\cdot P', m_j')
    + \mathcal{P}^\nu_\kappa q^\kappa (e^*( \mathcal{P}\cdot P', m_j')\cdot ( \mathcal{P}\cdot P) )/q^2]H_2\\
    &-\mathcal{P}^\rho_{\kappa _1}\mathcal{P}^\sigma_{\kappa _2}\mathcal{P}^\alpha_{\kappa_3}
    \epsilon^\nu_{\alpha\rho\sigma} \bar{p}^{\kappa _1} q^{\kappa _2} {e^{\kappa_ 3}}^*(P', m_j')H_3
    \;.
  \end{split}
\end{align}
By using the following transformation relations,
\begin{align}
  & e^\mu(\mathcal{P}\cdot  P, m_j)=-\mathcal{P}^\mu_\nu e^\nu(P, m_j)\label{eq:ep1}\;,\\
  &  e^*(\mathcal{P}\cdot P', m_j')\cdot (\mathcal{P}\cdot P) 
    =-\mathcal{P}^\nu_\kappa {e^\kappa}^*( P', m_j') \mathcal{P}^\chi_\nu P_\chi
    =- e^*(P', m_j')\cdot P\label{eq:ep2}\;,\\
  &  e^*(\mathcal{P}\cdot P', m_j')\cdot e(\mathcal{P}\cdot P, m_j)
    =\mathcal{P}^\nu_\kappa {e^\kappa}^*( P', m_j') \mathcal{P}^\chi_\nu {e^\chi}^*(P', m_j')
    =e^*( P', m_j')\cdot e(P, m_j)\label{eq:ep3}\;,
\end{align}
 we get
\begin{align}
  \begin{split}
    \bra{h_{q\bar{q}}'( \mathcal{P}\cdot P', j'=1,m'_j, \mathrm{P}_2)} & \bar{\Psi}\gamma^\nu\Psi  \ket{h_{q\bar{q}}(
      \mathcal{P}\cdot P, j=0, m_j, \mathrm{P}_1)}\\
    =&-\mathcal{P}^\nu_\kappa  [\bar{p}^\kappa (e^*(P', m_j')\cdot P )-q^\kappa (e^*(P', m_j')\cdot P ) \Delta_m/q^2]H_1\\
    &-\mathcal{P}^\nu_\kappa[{e^\kappa}^*(P', m_j')+ \mathcal{P}^\nu_\kappa q^\kappa (e^*(P', m_j')\cdot P )/q^2]H_2\\
&+\mathcal{P}^\nu_\kappa
    \epsilon^\kappa_{\kappa_3 \kappa_1 \kappa_2} \bar{p}^{\kappa 1} q^{\kappa 2} {e^{\kappa 3}}^*(P', m_j')H_3
    \;.
  \end{split}
\end{align}
Plugging it back into Eq.~\eqref{eq:PPspin01}, we find
\begin{align}
  \mathrm{P}_2 \mathrm{P}_1=
  \begin{cases}
    +1&\to H_1,H_2=0\\
    -1 &\to H_3=0
  \end{cases}\;.
\end{align}
 $H_1, H_2$ are form factors of parity flipped transition, and $H_3$ are form factors of parity conserved transition. 
 
 To summarize, there are two classes of allowed transitions, (1):
 \[  0^{++} (\text{scalar})\to  1^{--} \text{(vector)}\;,\qquad 0^{-+} \text{( pseudoscalar )}\to 1^{+-}\text{(axial-vector )}.\]
 and the transition form factors are $H_1$ and $H_2$; (2):
 \[  0^{++} (\text{scalar})\to  1^{+-} \text{(axial-vector)}\;,\qquad 0^{-+} \text{( pseudoscalar )}\to 1^{--}\text{(vector )}\]
  and the transition form factor is $H_3$.
   The transitions between
 pseudoscalar $0^{-+}$ and axial-vector $1^{++}$, between 
scalar $0^{++}$ and axial-vector $1^{++}$ are forbidden due to charge conjugation.
 
\subsection{The hadron matrix element}\label{sec:hadronMatrix}
The electromagnetic transition between two hadron states $\psi_A$ and $\psi_B$ is governed by the matrix element 
$\bra{\psi_B(P,j,m_j)} J^\mu (x) \ket{\psi_A(P',j',m'_j)}$. The elastic process is a special case where $\psi_A=\psi_B$. In this section, we derive the light-front wavefunction representation of the hadron matrix element, which we will use later in calculating the elastic form factor and the transition form factor. 

The EM current operator is defined as \( J^\mu=\bar{\Psi}\gamma^\mu \Psi \). In the light-front representation,
\begin{align}
  \begin{split}
    J^\mu (x)=&
    \sum_{\lambda_1,\lambda_2}\sum_{c_1,c_2}
    \int\frac{\diff^2  p_{1 \perp} \diff p_1^+}{{(2\pi)}^3 2p_1^+}
    \int\frac{\diff^2p_{2 \perp} \diff p_2^+}{{(2\pi)}^3 2p_2^+}
    \left[b^\dagger_{\lambda_2 c_2}(p_2)\bar{u}_{\lambda_2 }(p_2)e^{ip_2\cdot x}
    +d_{\lambda_2 c_2}(p_2)\bar{v}_{\lambda_2}(p_2)e^{-ip_2\cdot x}
    \right]\\
    &\gamma^\mu\left[b_{\lambda_1 c_1}(p_1)u_{\lambda_1}(p_1)e^{-ip_1\cdot x} +
    d^{\dagger}_{\lambda_1 c_1}(p_1)v_{\lambda_1}(p_1)e^{ip_1\cdot x}\right]
    \; .
  \end{split}
\end{align}
By spacetime translation invariance, 
\begin{align}\label{eq:I_trans}
  \begin{split}
    \bra{\psi_B(P,j,m_j)} J^\mu (x) \ket{\psi_A(P',j',m_j')}
    =&\bra{\psi_B(P,j,m_j)}  e^{-i\hat{p}x}\bar{\Psi}(0)e^{i\hat{p}x}\gamma^\mu e^{-i\hat{p}x}\Psi (0) e^{i\hat{p}x}\ket{\psi_A(P',j',m_j')}\\
    = &\bra{\psi_B(P,j,m_j)} J^\mu (0) \ket{\psi_A(P',j',m_j')}e^{i(P'-P)x}\;.
  \end{split}
\end{align}
The argument $x$ only results in an overall phase factor, so in the literature one usually takes $J^\mu (0)$ in calculating the matrix element.

\begin{figure}[!ht]
  \centering 
  \subfigure[\ $n\to n$ transition \label{fig:n_to_n}]{\includegraphics[width=0.35\textwidth]{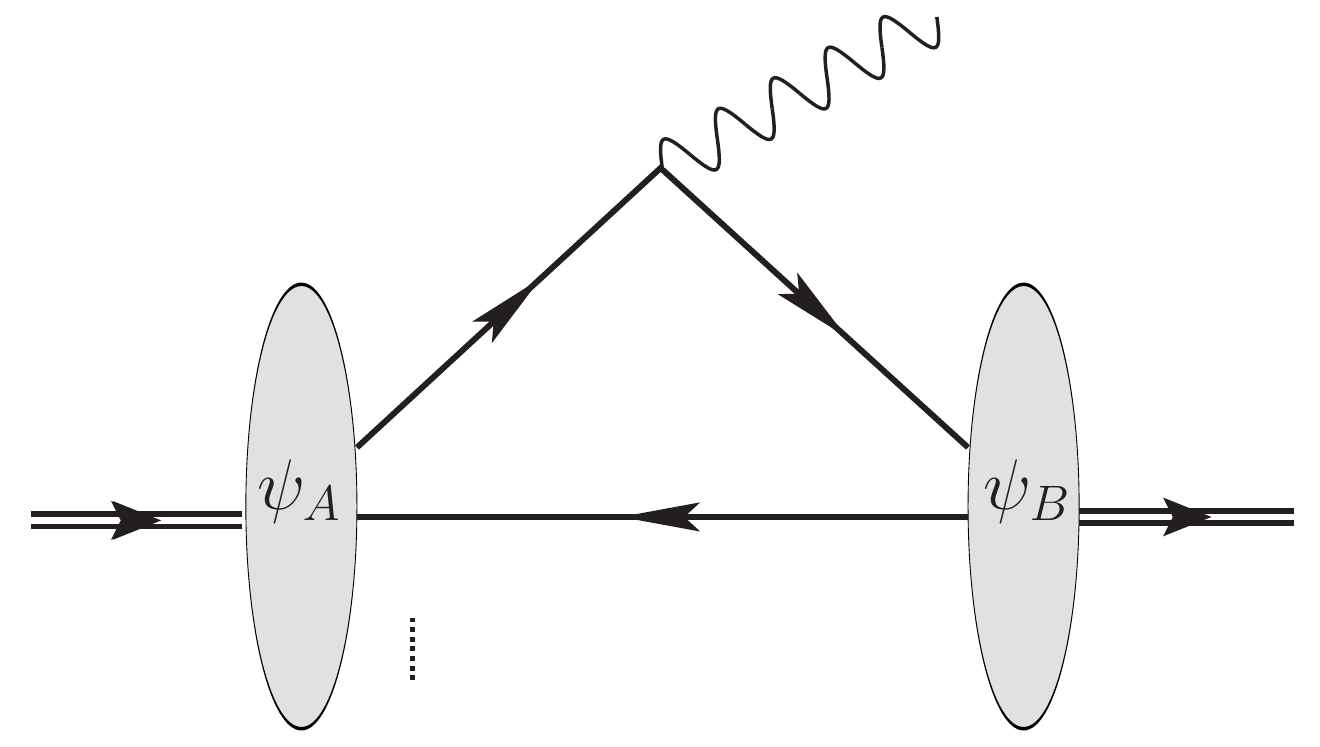}} \qquad
  \subfigure[\ $n+2\to n$ transition \label{fig:np2_to_n}]{\includegraphics[width=0.35\textwidth]{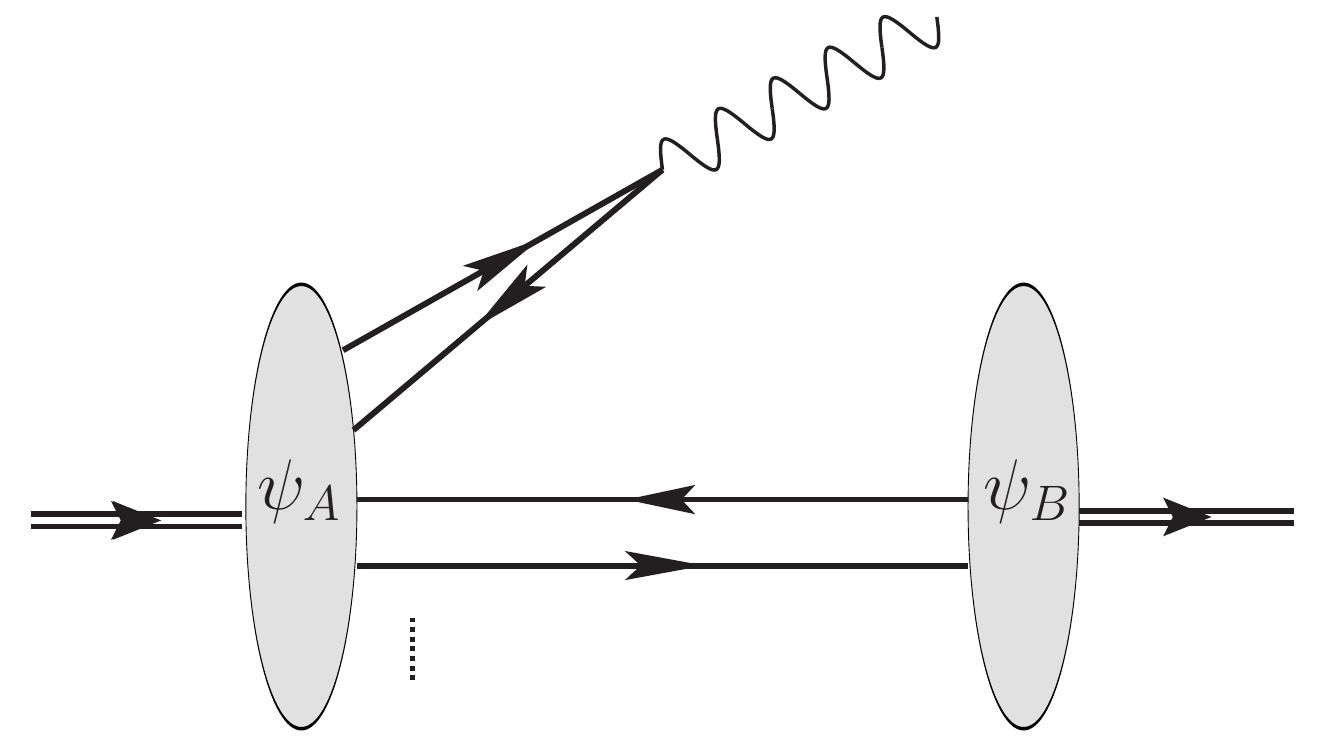}}
  \caption{Light-front wavefunction representation of the hadron matrix element. The double-lines represents the hadrons.
    The solid lines represent the partons. The wavy lines represent the external photon. The shaded areas
    represent the light-front wavefunctions. These diagrams are ordered by light-front time $x^+$, which flows from left to 
    right. In (a), the $n\to n$ transition, parton number is conserved, whereas in (b), the $n+2\to n$ transition, parton number is reduced by 2 due to pair annihilation. (Figure adapted from Ref.~\cite{Li:2019kpr}.)}
  \label{fig:LFWF_rep}
\end{figure}

We have shown in Part 1 that the meson state vector $ \ket{\psi_h(P,j,m_j)}$ can be expanded in the light-front Fock space. The coefficients of the Fock expansion are the complete set
of n-particle light-front wavefunctions, $\{\psi_{n/h}^{(m_j)}(x_i,\vec k_{i\perp}, s_i)\}$. $x_i\equiv
{\kappa_i^+}/{P^+}$ is the longitudinal momentum fraction of the i-th parton, and $\vec k_{i\perp}\equiv \vec \kappa_{i\perp}-x \vec
P_\perp$ is the relative transverse momenta, with $\kappa_i$ being the momenta of the corresponding parton. $s$ is the spin of the parton.
The electromagnetic current matrix element is in general given by the sum of the diagonal $n\to n$ and off-diagonal $n+2\to
n$ transitions, as shown in Fig.~\ref{fig:LFWF_rep}.
\begin{align}
  \begin{split}
    \bra{\psi_{B}} J^\mu \ket{\psi_{A}}
    =&\bra{\psi_{B}} J^\mu  \ket{\psi_{A}}_{n\to n}
    +\bra{\psi_{B}} J^\mu \ket{\psi_{A}}_{n+2\to n}
    \;.
  \end{split}
\end{align}
In the former case, the external photon is coupled to a quark or an antiquark. In the latter case, a
quark-antiquark pair is annihilated into the external photon.

Let us make the derivation of the $2\to 2$ transition explicitly. In the case where the hadrons are solved in the $\ket{q\bar{q}}$ Fock sector, only the $n\to n ~ (n=2)$ term would contribute to the transition,
\begin{align}
  \begin{split}
    \bra{\psi_{q\bar{q}/B}(P,j,m_j)}& J^\mu (0) \ket{\psi_{q\bar{q}/A}(P',j',m_j')}\\
    =
    &   \frac{1}{N_c}\sum_{i, j=1}^{N_c} \bra{0}
    \sum_{s,\bar s}
    \int_0^1\frac{\diff x}{2x(1-x)}
    \int\frac{\diff^2 k_\perp}{{(2\pi)}^3}
    \psi_{s\bar s/B}^{(m_j)*}(\vec k_\perp, x)\\
    &
d_{j\bar s }((1-x)P^+,-\vec k_\perp+(1-x) \vec P_\perp)
    b_{js }(xP^+, \vec k_\perp+x \vec P_\perp)
    \\
    &\sum_{c_1,c_2} \sum_{\lambda_1,\lambda_2}
    \int\frac{\diff^2 \vec p^1_\perp \diff p_1^+}{{(2\pi)}^3 2p_1^+}
    \int\frac{\diff^2 \vec p^2_\perp \diff p_2^+}{{(2\pi)}^3 2p_2^+}
    \left[b^\dagger_{\lambda_2 c_2}(p_2)\bar{u}_{\lambda_2 }(p_2)
    +d_{\lambda_2 c_2}(p_2)\bar{v}_{\lambda_2}(p_2)
    \right]\\
    &\gamma^\mu\left[b_{\lambda_1 c_1}(p_1)u_{\lambda_1}(p_1) +
    d^{\dagger}_{\lambda_1 c_1}(p_1)v_{\lambda_1}(p_1)\right]\\
    &\times 
    \sum_{s',\bar s'}
    \int_0^1\frac{\diff x'}{2x'(1-x')}
    \int\frac{\diff^2\vec k'_\perp }{{(2\pi)}^3}
   \psi_{s'\bar s'/A}^{(m_j')}(\vec k'_\perp  , x')\\
    &
    b^\dagger_{is'}(x'{P'}^+, \vec k'_\perp +x' \vec P'_\perp )
    d^\dagger_{i\bar s '}((1-x')P',-\vec k'_\perp +(1-x') \vec P'_\perp )
    \ket{0}
    \;.
  \end{split}
\end{align}
There are two non-vanishing terms as we pair up the creation and annihilation operators. One is the contribution from the quark
radiation and the other from the antiquark. We will use $J_q^\mu$ ($J_{\bar q}^\mu$) as the operator acting on the quark (antiquark).

\exercise{ Why do the other two terms $ db$ and $d^\dagger b^\dagger$ vanish? What physical process do they describe? See Fig.~\ref{fig:LFWF_rep}}.

    Contracting the creation and annihilation operators, 
    \begin{align}\label{eq:hadron_matrix_unint}
      \begin{split}
        &  \bra{\psi_{q\bar{q}/B}(P,j,m_j)}J_q^\mu (0) \ket{\psi_{q\bar{q}/A}(P',j',m_j')}\\
        =
        &   \frac{1}{N_c}\sum_{i, j=1}^{N_c} \sum_{\lambda_1,\lambda_2}\sum_{c_1,c_2}\sum_{s',\bar s'}\sum_{s,\bar s}
        \int\frac{\diff^2 \vec p^1_\perp \diff p_1^+}{{(2\pi)}^3 2p_1^+}
        \int\frac{\diff^2 \vec p^2_\perp \diff p_2^+}{{(2\pi)}^3 2p_2^+}
        \int_0^1\frac{\diff x'}{2x'(1-x')}\\
        &   \int\frac{\diff^2\vec k'_\perp }{{(2\pi)}^3}
        \int_0^1\frac{\diff x}{2x(1-x)}
        \int\frac{\diff^2 k_\perp}{{(2\pi)}^3}
       \psi_{s' \bar s'/A}^{(m_j') }(\vec k'_\perp  , x')
        \psi_{s\bar s/B}^{(m_j)*}(\vec k_\perp, x)\\
        &
        2 p_2^+\theta(p_2^+){(2\pi)}^3\delta(p_2^+ -x{P}^+ )\delta^2(\vec p^2_\perp-\vec k_\perp -x \vec P_\perp
        )\delta_{j,c_2}\delta_{s,\lambda_2}\\
        &
        2 p_1^+\theta(p_1^+){(2\pi)}^3\delta(p_1^+ -x'{P'}^+ )\delta^2(\vec p^1_\perp-\vec k'_\perp -x'\vec P'_\perp)\delta_{i,c_1}\delta_{s'
          ,\lambda_1}\\
        &
        2 (1-x){P}^+\theta({P}^+){(2\pi)}^3\delta((1-x'){P'}^+-(1-x){P}^+)\\
        &\delta^2(-\vec k'_\perp+(1-x') \vec P'_\perp+\vec k_\perp -(1-x) \vec P_\perp )
        \delta_{j,i}\delta_{\bar s',\bar s}\\
        &
        \bar{u}_{\lambda_2 }(p_2)
        \gamma^\mu u_{\lambda_1}(p_1)
        \;.
      \end{split}
    \end{align}
    We could first integrate over $x$ and $\vec{k}_\perp $ by the last two delta functions and get,
    \begin{align}
      x=1-(1-x'){P'}^+/P^+
           ,\qquad
        \vec k_\perp =\vec k'_\perp-(1-x') \vec P'_\perp+(1-x) \vec P_\perp 
        \;.
    \end{align}
    Integrate over $p_1$, $p_2$, we get
    \begin{align}\label{eq:jmuq}
      \begin{split}
        &\bra{\psi_{q\bar{q}/B}(P,j,m_j)}J_q^\mu (0) \ket{\psi_{q\bar{q}/A}(P',j',m_j')}\\
        =
        &\sum_{s,\bar s}
        \int_{\max(0, 1-P^+/{P'}^+)}^1\frac{\diff x'}{2x'(1-x')}
        \int\frac{\diff^2 k'_\perp}{{(2\pi)}^3}
        \frac{1}{x}
        \sum_{s'}
        \psi_{s' \bar s/A}^{(m_j')}(\vec k'_\perp, x')
        \psi_{s \bar s/B}^{(m_j) *}
        (\vec k_\perp, x)  \\
        &\times \bar{u}_{s  }(x{P}^+ , \vec k_\perp+x\vec P_\perp)
        \gamma^\mu u_{s'}(x'{P'}^+, \vec k'_\perp +x'\vec P'_\perp)
        \;.
      \end{split}
    \end{align}

    Note that the lower bound of the integral over $x$ is not 0 when ${P}^+<{P'}^+$, which results from the condition $x
    \in [0,1] $. However, when evaluating the $x'$-integral numerically, it would be more convenient to have the integral range as
    $[0,1]$. This is actually possible by integrating over $x'$ and $\vec {k}'_\perp $ instead in Eq.~\eqref{eq:hadron_matrix_unint},
    \begin{align}\label{eq:jmuq_v2}
      \begin{split}
        &\bra{\psi_{q\bar{q}/B}(P,j,m_j)}J_q^\mu (0) \ket{\psi_{q\bar{q}/A}(P',j',m_j')}\\
        =
        &\sum_{s,\bar s}
        \int_{\max(0, 1-{P'}^+/P^+)}^1\frac{\diff x}{2x(1-x)}
        \int\frac{\diff^2\vec k_\perp}{{(2\pi)}^3}
        \frac{1}{x'}
        \sum_{s'}
        \psi_{s' \bar s/A}^{(m_j')}(\vec k'_\perp, x')
        \psi_{s\bar s/B}^{(m_j) *}
        (\vec k_\perp, x)  \\
        &\times \bar{u}_{s }(xP^+, \vec k_\perp +x\vec P_\perp)
        \gamma^\mu u_{s'}(x'{P'}^+ , \vec k'_\perp+x'\vec P'_\perp)
        \;,
      \end{split}
    \end{align}
    where
    \begin{align}
      x'=1-(1-x){P}^+/{P'}^+,\qquad
      \vec k'_\perp =\vec k_\perp-(1-x) \vec P_\perp+(1-x') \vec P'_\perp \;.
    \end{align}
    As expected, the lower bound of the integral over $x$ is 0 when ${P}^+<{P'}^+$. Eqs.~\eqref{eq:jmuq} and ~\eqref{eq:jmuq_v2} are
    equivalent, and one could choose the one that facilitates the numerical calculations. If one considers the process $ \psi_A(P')\to \psi_B(P) + \gamma^{(*)}(q=P'-P)$ where ${P}^+ < {P'}^+$, it is more convenient to use the expression in Eq.~\eqref{eq:jmuq_v2}. However, if one considers the process $ \psi_A(P')+ \gamma^{(*)}(q=P-P') \to \psi_B(P)$, which is usually the case in calculating the elastic form factor, where ${P}^+ > {P'}^+$, it would be more convenient to use the expression in Eq.~\eqref{eq:jmuq}.
In analogy, we get the hadron matrix element of the antiquark current,
    \begin{align}\label{eq:2to2_qbar}
      \begin{split}
        &\bra{\psi_{q\bar{q}/B}(P,j,m_j)}J_{\bar{q}}^\mu (0) \ket{\psi_{q\bar{q}/A}(P',j',m_j')}\\
        =& - \sum_{s,\bar s}
        \int_0^{\min(1,{P}^+/{P'}^+)}\frac{\diff x'}{2x'(1-x')}
        \int\frac{\diff^2 k'_\perp}{{(2\pi)}^3}
        \frac{1}{1-x}
        \sum_{\bar s'}
        \psi_{s \bar s/B}^{(m_j)*}(\vec k_\perp, x)
        \psi_{s \bar s'/A}^{(m_j')}
        (\vec k'_\perp, x')  \\
        &\times
        \bar{v}_{\bar s '}((1-x'){P'}^+, -\vec k'_\perp+(1-x') \vec P'_\perp)      
        \gamma^\mu 
        v_{\bar s }((1-x)P^+, -\vec k_\perp+(1-x) \vec P_\perp)
        \;,
      \end{split}
    \end{align}
  where
    \begin{align}
        x=x'{P'}^+/P^+
        \;,\qquad
        \vec k_\perp 
        =\vec k'_\perp+x' ({P}^+\vec P'_\perp-{P'}^+\vec P_\perp)/{P}^+
        \;.
    \end{align}
    \exercise{What does the ``-'' sign in Eq.~\eqref{eq:2to2_qbar} imply physically? Think about the charge.}

To have a more explicit form for the purpose of calculation, let us put in the expressions of spinors as in Appendix. \ref{sec:spinvector}. 

    \exercise{Write out the expressions$\bar u_{s_2}(p_2)\gamma^\mu u_{s_1}(p_1)$ for $\mu=+,-,x,y$, and $s_1,s_2=\pm 1/2$.}

\subsection{Frames and kinematics}\label{sec:frame}
\begin{figure}[htp!]
  \subfigure[\ Regional plot of $q^2(\Delta_\perp, z)$ \label{fig:LFWF_rep_a}]
{\includegraphics[width=0.35\textwidth]{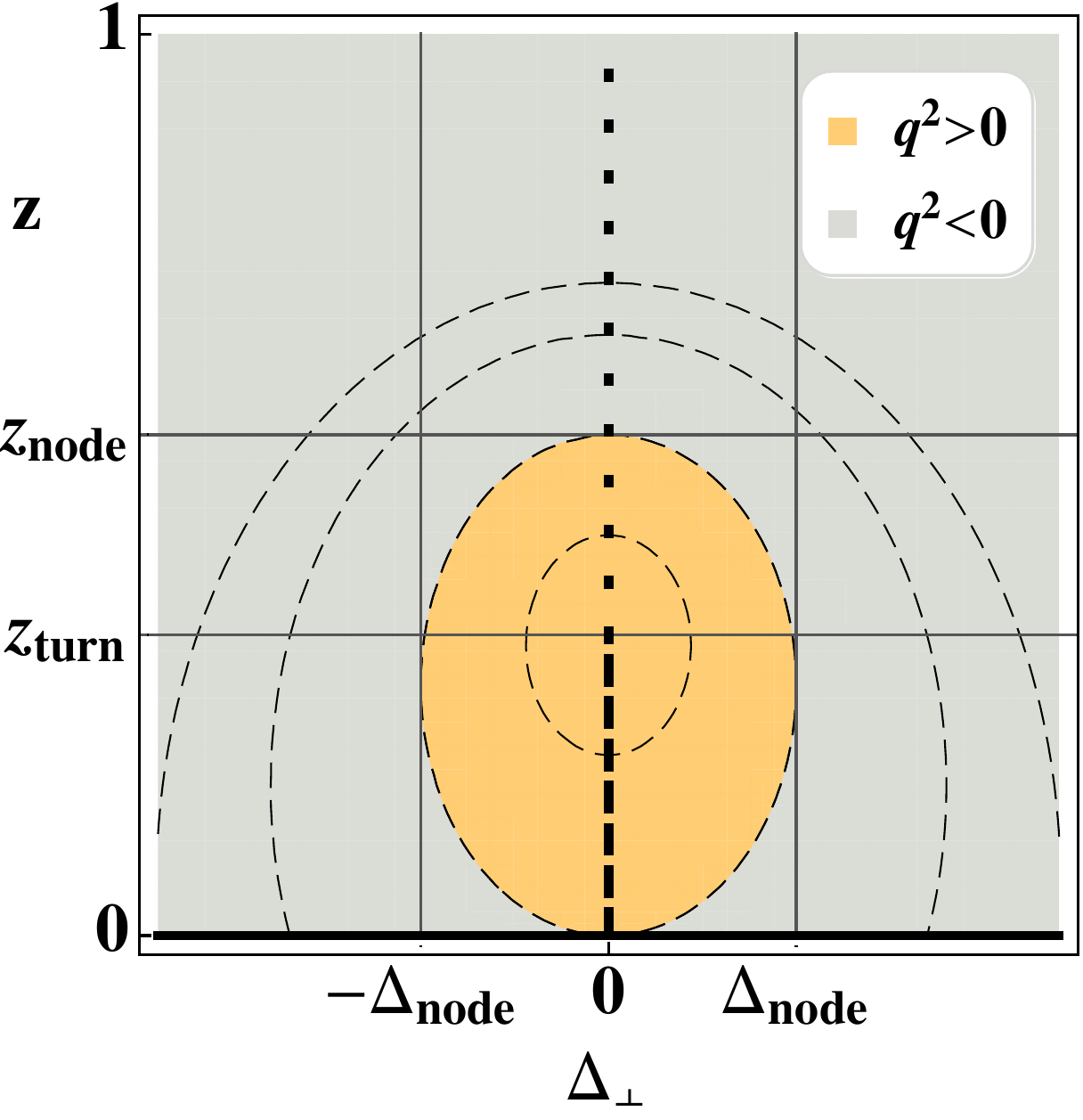}
} 
\qquad
  \subfigure[\ 3D plot of $q^2(\Delta_\perp, z)$ \label{fig:LFWF_rep_b}]
{\includegraphics[width=0.35\textwidth]{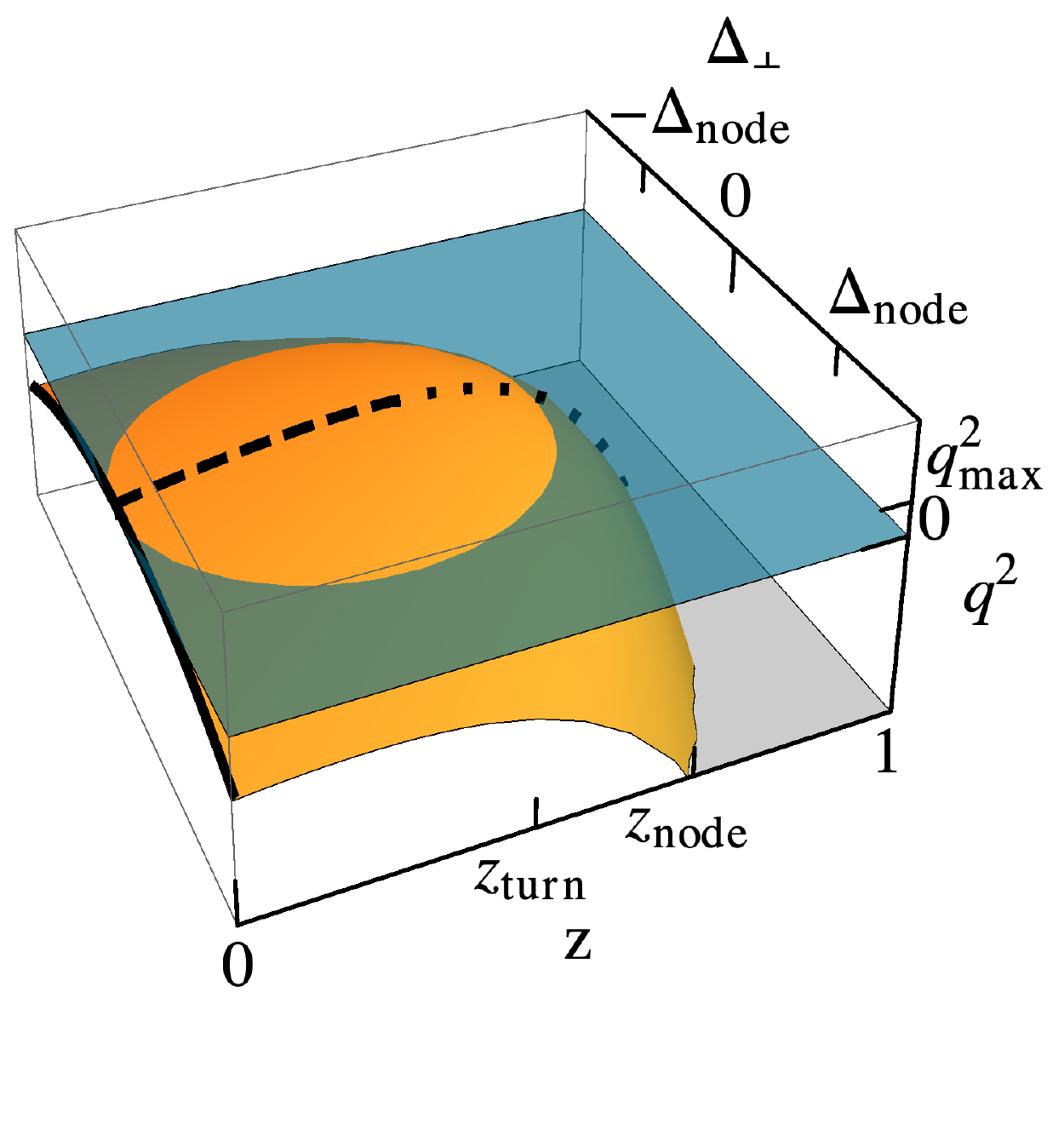}
}   
  \caption{\label{fig:q2_z_delta}Visualization of the Lorentz invariant momentum transfer squared $q^2$ as a function of $z$ and $\vec \Delta_\perp$ at $\arg \vec\Delta_\perp=0,\pi$. (a): regional plot of $q^2$. The time-like region ($q^2>0$) is the orange oval shape, bounded by $\Delta_{\text{node}}=(m^2_A-m^2_B)/2m_A$ and $z_{\text{node}}=1-m^2_B/m^2_A$. The space-like region ($q^2<0$) is in light gray. Contour lines of $q^2$ are indicated with thin dashed curves. The maximal value $q_{\max}^2=(m_A-m_B)^2$ occurs at $(z_{\text{turn}}=1-m_B/m_A, \Delta_\perp=0)$. (b): 3D plot of $q^2$ showing a convex shape in the $(z, \Delta_\perp)$ representation. The blue flat plane is the reference plane of $q^2=0$. In each figure, the Drell-Yan frame is shown as a thick solid line, and the longitudinal I and II frames are shown as thick dotted and thick dashed lines respectively. (Figure adapted from Ref.~\cite{Li:2019kpr}.)}
\end{figure}

Considering the process $\psi_A(P')\to\psi_B(P)+X(q=P'-P)$ or $\psi_B(P)+X(q=P'-P)\to \psi_A(P')$, the Lorentz invariant momentum transfer $q^2$ can be written as a function of two boost invariants~\cite{Li:2017uug, Li:2019kpr} according to the four-momentum conservation $q^2=(P'-P)^2$,
\begin{align}\label{eq:q2_z_delta}
q^2=zm_A^2-\frac{z}{1-z}m_B^2-\frac{1}{1-z}\vec\Delta_\perp^2\;.
\end{align}
where,
\[ z\equiv ({P'}^+-P^+)/{P'}^+, \qquad \vec \Delta_\perp\equiv\vec q_\perp-z\vec P'_\perp\;.\]
Both $z$ and $\vec \Delta_\perp$ are invariant under the transverse Lorentz boost specified by the velocity vector $\vec \beta_\perp$,
  \begin{align}
    v^+ \to v^+,\quad \vec{v}_\perp \to \vec{v}_\perp +v^+\vec{\beta}_\perp \;.
  \end{align}
$z$ can be interpreted as the relative momentum transfer in the longitudinal direction, and $\vec \Delta_\perp$ describes the
momentum transfer in the transverse direction. Note that $z$ is restricted to $0\le z<1$ by definition.
For each possible value of $q^2$, the values of the pair $(z,\vec\Delta_\perp)$ are not unique, and those different choices correspond to different reference
frames (up to longitudinal and transverse light-front boost transformations). Fig.~\ref{fig:q2_z_delta} should help visualize the functional form of $q^2(z,\vec\Delta_\perp)$. Since $q^2$ is relevant to the magnitude of $\vec\Delta_\perp$ but not its angle, we plot it in the $\arg \vec\Delta_\perp=0,\pi$ plane.
Form factors evaluated at different $(z, \vec\Delta_\perp)$ but at the same $q^2$ could reveal the frame dependence. In particular, we introduce two special frames for detailed consideration.
\begin{itemize}
\item Drell-Yan frame ($z=0$) :  $q^+=0$, $\vec \Delta_\perp = \vec q_\perp$  and $q^2=-\vec\Delta_\perp^2$. This frame is shown as a single thick solid line in each panel of Fig.~\ref{fig:q2_z_delta}. The Drell-Yan frame is conventionally used together with the plus
  current $J^+$ to calculate the electromagnetic form factors. This choice, on the one hand, avoids spurious effects related to the orientation of the null
  hyperplane where the light-front wavefunction is defined and, on the other hand, it suppresses the contributions from the often-neglected pair creation process, at least for pseudoscalar mesons \cite{CARBONELL1998215, DEMELO1998574, BRODSKY1999239, PhysRevD.65.094043, Simula:2002vm, PhysRevD.88.025036}. For the transition form factor, this is only true if zero-mode contributions are neglected. The
  transition form factor obtained in the Drell-Yan frame is significantly restricted in the space-like region, i.e. $q^2 \le 0$. Although one could analytically continuate the form factor to the time-like region by changing $\vec q_\perp$ to $i\vec q_\perp$
  ~\cite{Melikhov:1995xz, Jaus:1996np, Bakker:2003up}, we elect to calculate transition form factors directly from wavefunctions.
\item longitudinal frame ($\vec \Delta_\perp=0$): $q^2=zm_A^2-zm_B^2/(1-z)$. Note that we use the same definition for the
  longitudinal frame as in Ref.~\cite{Li:2017uug, Li:2019kpr}, which is different from those in the literature where $\vec
  q_\perp=0$ is called the longitudinal frame~\cite{Isgur:1988iw, Sawicki:1992qj, BRODSKY1999239, Bakker:2003up}. In this frame, we have access to the
  kinematic region up to $q_{\max}^2=(m_A-m_B)^2$, the point where the final
  meson does not recoil. This maximal value occurs at $z = 1 - m_B/m_A \equiv z_{\text{turn}} $. For a given $q^2$, there are two solutions for $z$, corresponding to either the positive or the negative recoil direction of the final meson relative to the initial meson, namely,
\begin{itemize}
\item longitudinal-I: $z=\bigg[m_A^2-m_B^2+q^2+\sqrt{(m_A^2-m_B^2+q^2)^2-4m_A^2q^2}~\bigg]/(2m_A^2)$. $ z_{\text{turn}}\le z < 1$. This branch joins the second branch at
  $q^2=q^2_{\max}$ with $z=z_{\text{turn}},\vec\Delta_\perp=0$. The time-like region is accessed at $z_{\text{turn}}\le z <  z_{\text{node}}$, and the space-like region is at $ z_{\text{node}}\le z < 1$, where $z_{\text{node}}\equiv 1-m^2_B/m^2_A$. The longitudinal-I frame is shown as thick dotted lines in Fig.~\ref{fig:q2_z_delta}.
\item longitudinal-II:
$z=\bigg[ m_A^2-m_B^2+q^2-\sqrt{(m_A^2-m_B^2+q^2)^2-4m_A^2q^2}~\bigg]/(2m_A^2)$.
$0\le z\le z_{\text{turn}}$. This second branch only exists in the time-like region, and it joins the Drell-Yan frame at $q^2=0$
with $z=0,\vec\Delta_\perp=0$. The longitudinal-II frame is shown as thick dashed lines in Fig.~\ref{fig:q2_z_delta}.
\end{itemize}
\end{itemize}

\subsection{Calculation of the form factors}
\subsubsection{Elastic form factor of the spin-0 meson}\label{sec:EFF_J0}
The elastic form factor of a (pseudo)scalar $\psi_h$ is the charge form factor $F(q^2)$,  defined as
\begin{align}\label{eq:EFF_J0}
  \braket{\psi_h(P')|J^\mu(0)|\psi_h(P)}=(P+P')^\mu F(q^2)\;,
 \end{align}
 as we have derived in Eq.~\eqref{eq:EFF_J0_e}.
 The charge form factor $F(q^2)$ is interpreted as the Fourier transformation of the charge density in the system. For quarkonium, the physical from factor vanishes due to charge conjugation symmetry (see also the discussion on charge conjugation in Sec. \ref{app:LD_spin0}), so what being calculated is actually the fictitious form factor from the quark current, $J^\mu_q$. 
 In the light-front wavefunction representation of the valence Fock sector, the hadron matrix element reads
\begin{align}
  \begin{split}
  \braket{\psi_h(P', m_j')|J_q^\mu(0)|\psi_h(P, m_j)}=
  &\sum_{s, s', \bar s}
    \int_0^1\frac{\diff x}{2x(1-x)}
    \int\frac{\diff^2 k_\perp}{{(2\pi)}^3}
    \frac{1}{x'}
    \psi_{s \bar s /h}^{(m_j)}(\vec k_\perp, x)
    \psi_{s'\bar s/h}^{(m_j') *}
    (\vec k_\perp ', x' )  \\
  &\times
  \bar{u}_{s ' }(x'{P'}^+ , \vec k'_\perp +x'\vec P'_\perp)
    \gamma^\mu u_s(xP^+, \vec k_\perp +x\vec P_\perp)
    \;,
    \end{split}
\end{align}
where $x'=({P'}^+-(1-x)P^+)/{P'}^+$ and $\vec k_\perp '= \vec k_\perp+(1-x) (P^+\vec P'_\perp-{P'}^+\vec P_\perp)/{P'}^+$. This is essentially the same as Eq.~\eqref{eq:jmuq}. We rewrite $x'$ and $\vec k'_\perp$ in terms of the two boost invariants we have defined in Section~\ref{sec:frame}, 
\(z\) and \(\vec \Delta_\perp\), as 
\[
x' = x + z(1-x),\qquad \vec k'_\perp=\vec k_\perp + (1-x)\vec\Delta_\perp
\;.
\]
The transferred momentum square $q^2$ can be written according to Eq.~\eqref{eq:q2_z_delta} with $m_A=m_B=m_h$,
\begin{align}
  q^2=-(z^2m_h^2+\vec\Delta_\perp^2)/(1-z)\;.
\end{align}
Note that $q^2\le 0$.
 
 One could extract the form factor with different current components. The $+$, $\perp$ and $-$ hadron matrix elements should be related through the transverse Lorentz boost specified by the velocity vector $\vec \beta_\perp$,
\begin{align}\label{eq:tr_boost}
  v^+ \to v^+,\quad \vec{v}_\perp \to \vec{v}_\perp +v^+\vec{\beta}_\perp,
  \quad v^- \to v^- + 2\vec \beta_\perp\cdot\vec v_\perp + \vec \beta_\perp^2 v^+
   \;.
\end{align}
The hadron matrix elements are thereby related through,
 \begin{align}\label{eq:tr_boost_hadron_matrix}
   \begin{split}
     \bra{\psi_h({P'}^+,\vec P'_\perp+{P'}^+\vec\beta_\perp)}&\vec J_\perp\ket{\psi_h(P^+,\vec P_\perp+P^+\vec\beta_\perp)}\\
     =&
     \braket{\psi_h(P')|\vec J_\perp|\psi_h(P)}
     + \vec\beta_\perp\braket{\psi_h(P')|J^+|\psi_h(P)}
     \;,\\
     \bra{\psi_h({P'}^+,\vec P'_\perp+{P'}^+\vec\beta_\perp)}&J^-\ket{\psi_h(P^+,\vec P_\perp+P^+\vec\beta_\perp)}\\
     =&  \braket{\psi_h(P')|J^-|\psi_h(P)}
     +2\vec\beta_\perp\cdot \braket{\psi_h(P')|\vec J_\perp|\psi_h(P)}
     + \vec\beta^2_\perp \braket{\psi_h(P')|J^+|\psi_h(P)}
     \;.
     \end{split}
 \end{align}
This relation implies that the form factors extracted from different current components should be equivalent. One can verify it by substituting Eq.~\eqref{eq:EFF_J0} into Eq.~\eqref{eq:tr_boost_hadron_matrix}. We would like to know if this is still true in the valence Fock sector, and write out the form factor with different current components in the valence light-front wavefunction representation. 
\begin{enumerate}
\item the plus current\\
\begin{align}\label{eq:EFF_spin0_Jpl}
  \begin{split}
    F(q^2)\big|_{J^+}=&
    \braket{\psi_h(P')|J_q^+(0)|\psi_h(P)}/(P^++{P'}^+)\\
    =&\sum_{s, \bar s}
    \int_0^1\frac{\diff x}{2x(1-x)}
    \int\frac{\diff^2 k_\perp}{{(2\pi)}^3}
    \frac{1}{x'}
    \psi_{s \bar s /h}(\vec k_\perp, x)
    \psi_{s \bar s/h}^* (\vec k_\perp ', x' )
    2\sqrt{x'{P'}^+xP^+}/(P^++{P'}^+)\\
    =&\sum_{s, \bar s}
    \int_0^1\frac{\diff x}{2x(1-x)}
    \int\frac{\diff^2 k_\perp}{{(2\pi)}^3}
    \frac{2}{2-z}\sqrt{\frac{x(1-z)}{x+z(1-x)}}
    \psi_{s \bar s /h}(\vec k_\perp, x)
    \psi_{s \bar s/h}^* (\vec k_\perp ', x' )
    \;.
  \end{split}
\end{align}

In the second line, the form factor is written as a function of $(z,\vec\Delta_\perp)$, dependence on $P$ or $P'$ is eliminated. The normalization of the form factor at $q^2=0$ follows as the result of  the normalization of the hadron
wavefunction,
\begin{align}
  F(0)\big|_{J^+}=\sum_{s, \bar s}
    \int_0^1\frac{\diff x}{2x(1-x)}
    \int\frac{\diff^2 k_\perp}{{(2\pi)}^3}
    \psi_{s \bar s /h}(\vec k_\perp, x)
    \psi_{s \bar s/h}^* (\vec k_\perp , x)
    =1
    \;.
\end{align}

\item the transverse current\\
Now we turn to the transverse current. Assuming that the rotational symmetry on the transverse plane is
 preserved, using $J^x$ or $J^y$ component or linear combinations of the two should be equivalent. Here we use
 $J^R\equiv J^x+i J^y$ and $J^L\equiv J^x-i J^y$ as the transverse currents. For
 any transverse vector $\vec k_\perp$, which is expressed as $(k^x, k^y)$ in the Cartesian coordinate or $(k_\perp,\theta)$ in the
 polar coordinate, we will write its complex form as $k^R\equiv k^x +i k^y=k_\perp e^{i \theta}$ and $k^L\equiv k^x -i k^y=k_\perp
 e^{-i\theta}$. The elastic form factor extracted from the $J^R$ current reads,
\begin{align}\label{eq:h0_EFF_JR}
  \begin{split}
    F(q^2)\big|_{J^R}=&
    \braket{\psi_h(P')|J_q^R(0)|\psi_h(P)}/(P^R+{P'}^R)\\
    =& F(q^2)\big|_{J^+}
    +\frac{1}{P^R+{P'}^R}
    \sum_{s \bar s}
    \int_0^1\frac{\diff x}{2x(1-x)}
    \int\frac{\diff^2 k_\perp}{{(2\pi)}^3}
    \psi_{s\bar s /h}(\vec k_\perp, x)
    \psi_{s\bar s/h}^* (\vec k_\perp ', x' )
    \\
    &\times 
    \frac{1}{\sqrt{x(1-z)[x+z(1-x)]^3}}
    \Big\{
    [z+2x(1-z)]k^R 
    +\frac{x}{2-z}(2-2x-3z+2xz)\Delta^R
    \Big\}
    \;.
  \end{split}
\end{align}
We have applied the symmetry among different spin components of spin-0 particle $h_0$,
\begin{align}
    \psi_{\uparrow \uparrow/h_0}(\vec k_\perp, x)=\psi_{\downarrow \downarrow/h_0 }(\vec k_\perp,
    x),
    \qquad
  \psi_{\uparrow \downarrow/h_0 }(\vec k_\perp, x)=-\psi_{\downarrow \uparrow/h_0}(\vec k_\perp, x)\;.
\end{align}

We see that $F(q^2)\big|_{J^R}$ and $F(q^2)\big|_{J^+}$ are different by the second term in the last line of Eq.~\eqref{eq:h0_EFF_JR}. Moreover, this second term depends
on $P^R+{P'}^R$ in the $(z,\vec\Delta_\perp)$ parameter space. This indicates that fixing
$(z,\vec\Delta_\perp)$ is not sufficient to unambiguously determine a frame in this case. However, in the Drell-Yan and the longitudinal frames, it can
be proved that this term actually vanishes, leaving $F(q^2)\big|_{J^R}=F(q^2)\big|_{J^+}$.
\begin{align}
  \begin{split}
    F(q^2)\big|_{J^R,\mathrm{DY}}=&
    \braket{\psi_h(P')|J_q^R(0)|\psi_h(P)}/(P^R+{P'}^R)\\
   =&
   F(q^2)\big|_{J^+,\mathrm{DY}}\\
   &+
   \frac{1}{P^R+{P'}^R} \sum_{s \bar s}
   \int_0^1\frac{\diff x}{2x(1-x)}
   \int\frac{\diff^2 k_\perp}{{(2\pi)}^3}
   \psi_{s\bar s /h}(\vec k_\perp, x)
   \psi_{s\bar s/h}^* (\vec k_\perp ', x' )
   \frac{1}{x}
   [2k^R+(1-x)q^R]\\
   =&
   F(q^2)\big|_{J^+,\mathrm{DY}}
    \;.
  \end{split}
\end{align}
The second term vanishes under the transverse integral with $\vec k'_\perp=\vec k_\perp+(1-x)q^R$ in the Drell-Yan frame. Now, in the longitudinal frame:
\begin{align}
  \begin{split}
    F(q^2)\big|_{J^R,\mathrm{long}}=&
    \braket{\psi_h(P')|J_q^R(0)|\psi_h(P)}/(P^R+{P'}^R)\\
    =&F(q^2)\big|_{J^+,\mathrm{long}}+
    \frac{1}{P^R+{P'}^R}
    \sum_{s \bar s}
    \int_0^1\frac{\diff x}{2x(1-x)}
    \int\frac{\diff^2 k_\perp}{{(2\pi)}^3}
    \psi_{s\bar s /h}(\vec k_\perp, x)
    \psi_{s\bar s/h}^* (\vec k_\perp ', x' )
    \\
    &\times 
    \frac{ [z+2x(1-z)]k^R }{\sqrt{x(1-z)[x+z(1-x)]^3}}\\
    =&F(q^2)\big|_{J^+,\mathrm{long}}
    \;.
  \end{split}
\end{align}
Note that $\vec k'_\perp=\vec k_\perp$ in the longitudinal frame, thus the second term vanishes since the angular integral is zero.
As with the $J^+$ current, $F(0)|_{J^R}=1$ is guaranteed by the normalization of the hadron wavefunction. At $q^2=0$, the
terms proportional to $k^R$ in the integral would vanish since the angular integration would be 0.

\item the minus current\\
Using the $J^-$ current,
\begin{align}\label{eq:h0_EFF_Jmn}
  \begin{split}
    F(q^2)\big|_{J^-}=&
    \braket{\psi_h(P')|J_q^-(0)|\psi_h(P)}/(P^-+{P'}^-)\\
    =&\frac{1}{P^2_\perp+m_h^2+(1-z)({P'}^2_\perp+m_h^2)}
    \sum_{s \bar s}
    \int_0^1\frac{\diff x}{2x(1-x)}
    \int\frac{\diff^2 k_\perp}{{(2\pi)}^3}
    \psi_{s\bar s /h}(\vec k_\perp, x)
    \psi_{s\bar s/h}^* (\vec k_\perp ', x' )
    \\
    &\times 
    2\sqrt{ \frac{1-z}{x[x+z(1-x)]}}
    [
    m_q^2+(\vec k_\perp +x\vec P_\perp)\cdot(\vec k'_\perp +x'\vec P'_\perp)
]
    \;.
  \end{split}
\end{align}
In deriving Eq.~\eqref{eq:h0_EFF_Jmn}, the spin flip terms vanish by exact cancellations among different spin components. The normalization of the elastic form
factor ($F(0)=1$) with $J^-$ has a nontrivial requirement on the wavefunctions, and this is referred to as a type of Virial
theorem~\cite{Burkardt:1989wy}. We can see this explicitly in Eq.~\eqref{eq:J0_EFF_Jmn_q0},
\begin{align}\label{eq:J0_EFF_Jmn_q0}
  \begin{split}
    F(0)\big|_{J^-}=&
    \braket{\psi_h(P')|J_q^-(0)|\psi_h(P)}/(P^-+{P'}^-)\\
    =&\frac{1}{2({P}^2_\perp+m_h^2)}
    \sum_{s \bar s}
    \int_0^1\frac{\diff x}{2x(1-x)}
    \int\frac{\diff^2 k_\perp}{{(2\pi)}^3}
    \psi_{s\bar s /h}(\vec k_\perp, x)
    \psi_{s\bar s/h}^* (\vec k_\perp, x)
    \frac{2}{x}
    [
    m_q^2+(\vec k_\perp +x\vec P_\perp)^2]
    \;.
  \end{split}
\end{align}
In the truncated Fock space, the light-front $J^-$ current is not conserved and it violates the
Ward-Takahashi identity~\cite{Marinho:2007zzb,Marinho:2008pe}. The valence Fock sector is not sufficient to extract the elastic form factor with the $J^-$ current. 

The work by H.M. Choi, H.Y. Ryu and C.R. Ji~\cite{Choi:2019wqx} implemented a replacement of the meson mass $m_h$ by the invariant mass $m^2_0=(m_q^2+\vec k_\perp^2)/x + (m_q^2+\vec k_\perp^2)/(1-x)$ in studying the $(\pi^0,\eta,\eta'\to \gamma^*\gamma^*)$ transitions with a manifestly covariant model. Following the format of this treatment, we see that restoring $F(0)=1$ in Eq.~\eqref{eq:J0_EFF_Jmn_q0} would require a replacement of $m_h^2\to (m_q^2+\vec k_\perp^2)/x - (1-x)\vec P^2_\perp$. In the meson rest frame where $\vec P_\perp = \vec 0_\perp$, the expression reduces to $m_h^2\to (m_q^2+\vec k_\perp^2)/x$, suggesting to replace the meson mass by the invariant mass of the quark, or half of the invariant mass of the meson.
\end{enumerate}

To conclude, the $J^+$ and $\vec J_\perp$ current components could guarantee the normalization of the elastic form factor in the valence
Fock sector, but the $J^-$ component could not. Though the elastic form factors extracted from the
$J^+$ and the $\vec J_\perp$ components are expected to be the same through a transverse boost, the valence light-front
wavefunction representation shows that the two are the same only in the Drell-Yan and the longitudinal frames. In a practical
calculation, $J^+$ and the Drell-Yan frame is often preferred, and the main advantage of this choice is that vacuum pair production/
annihilation is suppressed~\cite{Brodsky:1973kb, Brodsky:1980zm, BRODSKY1999239}. A study on the frame dependence of the elastic form factor of pseudoscalars using the $J^+$ current can be found in Ref.~\cite{Li:2017uug}.

In nonrelativistic quantum mechanics, the root-mean-square charge (mass) radius is the expectation value of the
displacement operator that characterizes the charge (mass) distribution of the system. In quantum field theory, no such local
position operator is allowed and, instead, the charge (mass) radius of the hadron is defined from the charge (gravitational) form factor at small momentum transfer:
\begin{align}\label{eq:radii}
  \braket{r_h^2}=\lim_{q^2 \to 0}-6\frac{\partial}{\partial q^2} F (q^2)
  \;.
\end{align}
In the Drell-Yan frame, $q^+=0$ and  $q^2=-{|\vec q_\perp|}^2$.
We can write $\vec q_\perp$ in the polar coordinate $\{q,\theta\}$. With a change of variable, $t=q^2$,
\begin{align}
  \begin{split}
  \nabla^2_{\vec q_\perp} 
  =&\frac{\partial^2}{\partial q^2}+\frac{1}{q}\frac{\partial}{\partial
    q}+\frac{1}{q^2}\frac{\partial^2}{\partial \theta^2}\\
  =& \frac{\partial^2}{\partial t^2}\bigg(\frac{\partial t}{\partial q}\bigg)^2 + \frac{\partial}{\partial t} \bigg(\frac{\partial^2 t}{\partial q^2}\bigg)
  +\frac{1}{\sqrt{t}}\frac{\partial }{\partial t}\frac{\partial t}{\partial q}+\frac{1}{t}\frac{\partial^2}{\partial \theta^2}
  \\
  =&4t\frac{\partial^2}{\partial t^2}+4\frac{\partial}{\partial t}+\frac{1}{t}\frac{\partial^2}{\partial \theta^2}
  \;.
  \end{split}
\end{align}
At the limit of $q^2\to 0$, the first term vanishes. Since the form factor does not have angular dependence, the third term
vanishes as well. It follows that
\begin{align}
  \frac{\partial}{\partial t}\bigg|_{t=0} = \frac{1}{4}\nabla^2_{\vec q_\perp} 
  \;.
\end{align}
We can thereby rewrite the charge radius in Eq.~\eqref{eq:radii} in terms of the two-dimensional Laplacian of the charge form factor,
\begin{align}
  \braket{r_h^2}=-\frac{3}{2} \nabla^2_{\vec q_\perp} F(q^2)\bigg|_{q^2=0}
  \;.
\end{align}

We have already mentioned that the physical form factors of a hadron should receive contributions from each constituent,  $F(q^2)=\sum_f e_f F_f(q^2)$, where $f$ is the
constituent (anti)quark with charge $e_f$. Though for quarkonium, the physical form factor vanishes due to charge conjugation and we
calculate the fictitious form factor contributed from the quark only. For a charged hadron,
such as $\pi^\pm$ and proton, one should consider its physical form factor that sums over the contributions of all constituent partons.
In the following, we will derive the contributions of the quark and the antiquark separately.
As an example, the charge radius of $\pi^+$ sums over the contributions from $u$ and $\bar{d}$.
\begin{align}
  \begin{split}
    \braket{r_{\pi^+}^2}
    =&\mathcal{Q}_u \braket{r_{\pi^+}^2}_{u} 
    + \mathcal{Q}_{\bar{d}} \braket{r_{\pi^+}^2}_{\bar d}
    \\
    =&\frac{3}{2} \sum_{s,\bar s}
    \int_0^1\frac{\diff x}{4\pi}
    \int \diff^2 r_\perp
    \bigg[
      \frac{2}{3} { (1-x)}^2+\frac{1}{3} x^2
      \bigg]
    \vec{r}^2_\perp
    \tilde{\psi}_{s,\bar s/\pi^+ }(\vec{r}_\perp, x)
    \tilde{\psi}^{*}_{s,\bar s/\pi^+ }
    (\vec{r}_\perp , x)
    \;.
  \end{split}
\end{align}
The dimensionless fractional charge of the quark is, $\mathcal{Q}_u=+2/3$ for the up quark and $\mathcal{Q}_{\bar d}=+1/3$ for the anti-down quark.

\subsubsection{Radiative transition between a spin-0 and a spin-1 mesons}\label{sec:TFF_M1}

The electromagnetic (EM) transition between meson states, which occurs via emission of a photon, $\psi_A\to \psi_B \gamma$, offers insights into the internal structure and the dynamics of such systems.
 The magnetic dipole (M1) transition, which takes place between pseudoscalar and vector mesons ($\psi_A, \psi_B = \mathcal V, \mathcal P$ or $\mathcal P, \mathcal V$), has been detected with strong signals~\cite{PDG2018} and stimulates various theoretical investigations~\cite{Brambilla:2005zw, Donald_JPsi, Bc_NR, cc_GI,Pineda:2013lta}.

The Lorentz covariant decomposition for the electromagnetic transition matrix element between a vector meson ($\mathcal{V}$) and a pseudoscalar  ($\mathcal{P}$) is~\cite{Dudek_JPsi}, as we have derived in Sec.~\ref{app:LD_spin01},
\begin{align}\label{eq:Vq2_def}
  I^\mu_{m_j}\equiv  \bra{\mathcal{P}(P)} J^\mu(0) \ket{\mathcal{V}(P',m_j)}
  =\frac{2 V(q^2)}{m_{\mathcal{P}}+m_{\mathcal{V}}}\epsilon^{\mu\alpha\beta\sigma} {P}_\alpha P'_\beta e_{\sigma}(P', m_j)
  \;,
\end{align}
where $q^\mu = {P'}^\mu - P^\mu$ represents the momentum transfer between the two mesons. $V(q^2)$ is the transition form
factor. $m_{\mathcal{P}}$ and $m_{\mathcal{V}}$ are the masses of the pseudoscalar and the vector, respectively. $e_{\sigma}$ is
the polarization vector of the vector meson, and $m_j=0,\pm 1$ is the magnetic projection.
Writing out all possible formulas of extracting the transition form factor $V(q^2)$ from different current components and different $m_j$ states of the vector meson, one would get Table.~\ref{tab:V_component_mj}. 
 To simplify the expression, we take the two variables defined in Sec.~\ref{sec:frame}, \(z\equiv ({P'}^+-P^+)/{P'}^+\) and \(\vec \Delta_\perp\equiv\vec q_\perp-z\vec P'_\perp\). 
 The five independent extractions in a truncated Fock space are indicated by five different colors in Table.~\ref{tab:V_component_mj}.
 \begin{table}
   \caption{The formulas of extracting the transition form factor $V(q^2)$ from different current components and different $m_j$ states of the
     vector meson. The five independent
     extractions in a truncated Fock space are indicated in five different colors: orange, green, red, blue and brown. Table adapted from Table I in Ref.~\cite{Li:2020wrn}. See detailed derivation in Ref.~\cite{Li:2019ijx}.
   }\label{tab:V_component_mj}
   \begin{tabular}{c ccc}
     \hline\hline\\[-3mm]
     $\dfrac{2 V(q^2)}{m_{\mathcal{P}}+m_{\mathcal{V}}}$ 
     &$m_j = 0$
     &$m_j = 1$
     &$m_j = -1$\\[2mm]\\
     \hline\\[-3mm]
     $J^+$    
& -
 &{\color{Orange}\(\dfrac{i\sqrt{2}I^+_1}{{P'}^+\Delta^R}\)}
 &{\color{Orange}\(\dfrac{-i\sqrt{2}I^+_{-1}}{{P'}^+\Delta^L}\)}
\\[3mm]\\
     $J^R$    
& {\color{PineGreen} \(\dfrac{-i I^R_0}{m_{\mathcal{V}}\Delta^R}\)}
 &{\color{NavyBlue}\(\dfrac{i\sqrt{2}I^R_{1}}{{P'}^R\Delta^R}\)}
&{\color{NavyBlue}\(~~~\dfrac{i\sqrt{2}(1-z)I^R_{-1}}{(m_{\mathcal{P}}^2 - (1-z)^2 m_{\mathcal{V}}^2-P^R\Delta^L )}~~~\)}
     \\[3mm]\\
     $J^L$
  &{\color{PineGreen} \(\dfrac{iI^L_0}{m_{\mathcal{V}}\Delta^L}\)}
& {\color{NavyBlue}\(\dfrac{-i\sqrt{2}(1-z)I^L_{1}}{(m_{\mathcal{P}}^2 - (1-z)^2 m_{\mathcal{V}}^2-P^L\Delta^R )}~~~\)}
  &{\color{NavyBlue}\(\dfrac{-i\sqrt{2}I^L_{-1}}{{P'}^L\Delta^L}\)}
     \\[3mm]\\
     $J^-$
     &{\color{Red} \(\dfrac{-iP^+ I^-_0}{m_{\mathcal{V}}( \Delta^R P^L-\Delta^LP^R )}\)}
     &{\color{Sepia} \(\dfrac{-i\sqrt{2}P^+ {P'}^+I^-_{1}}{{P'}^+{P'}^R(m_{\mathcal{P}}^2 - P^L\Delta^R )- P^+P^R m_{\mathcal{V}}^2}\)}
     &{\color{Sepia}\( \dfrac{i\sqrt{2}P^+ {P'}^+I^-_{-1}}{{P'}^+{P'}^L(m_{\mathcal{P}}^2 - P^R\Delta^L )- P^+P^L m_{\mathcal{V}}^2}\)}
     \\[2mm]\\
     \hline\hline
   \end{tabular}
 \end{table}
 
 In the valence Fock sector, the five independent hadron matrix elements overdetermine the transition form factor. 
In practice, the different prescriptions of extracting the same transition form factor could provide a test of violation of the Lorentz symmetry in the calculation. But more importantly, we would like to know if there is a preferred choice such that the result is closer to the true result that would emerge from a full Fock space basis.

Working in the valence Fock sector, we take the impulse approximation, in which the interaction of the external current with the meson is the summation of its coupling to the quark and to the antiquark. The vertex dressing as well as pair creation/annihilation from higher order diagrams are neglected.
The hadron matrix element can be written accordingly as a sum of the quark term and the antiquark term:
\begin{align}
 \bra{\mathcal{P}(P')} J^\mu(0)\ket{\mathcal{V} (P,m_j)}
  =e \mathcal{Q}_f \bra{\mathcal{P}(P')} J_{q}^\mu(0)\ket{\mathcal{V} (P,m_j)}
  -e \mathcal{Q}_f\bra{\mathcal{P}(P')} J_{\bar{q}}^\mu(0)\ket{\mathcal{V} (P,m_j)}
  \;.
\end{align}
Bu restoring the quark charges, the current operator reads $J^\mu(x) = e \sum_f Q_f \overline \psi_f(x) \gamma^\mu \psi_f(x)$ where $\psi_f(x)$ is the quark field
operator with flavor $f$ ($f=u,d,s,c,b,t$).  $J_q$ and $J_{\bar q}$ are the normal ordered pure quark ($b^\dagger b$) and antiquark ($d^\dagger d$) part of $J^\mu$, 
respectively, where $b$ ($d$) is the quark (antiquark) annihilation operator.
The dimensionless fractional charge of the quark is, $\mathcal{Q}_f=\mathcal{Q}_c=+2/3$ for the charm quark and $\mathcal{Q}_f=\mathcal{Q}_b=-1/3$ for the bottom quark. The electric charge $e=\sqrt{4\pi \alpha_{\text{EM}}}$. For quarkonium, due to the charge conjugation symmetry, the antiquark gives the same contribution as the quark to the total hadronic current. So, for our purpose, we calculate the hadron matrix element for the quark part. As such, we
compute $\hat{V}(q^2)$ which is related to $V(q^2)$ by
$V(q^2)=2e \mathcal{Q}_f \hat{V}(q^2)$.

There are five groups of combinations of the current component and the magnetic projection according to Table~\ref{tab:V_component_mj}. 
  \begin{enumerate}
    \item $J^+$ and $m_j=\pm 1$\\
The light-front wavefunction representation of the transition form factor reads,
\begin{align}\label{eq:V_Jpl_mj1}
  \begin{split}
\hat V|_{J^+, m_j=1}(q^2)
=&\frac{i (m_{\mathcal{V}}+m_{\mathcal{P}})}{\sqrt{2}{P'}^+\Delta^R}
\bra{\mathcal{P}(P)} J_q^+ (0) \ket{\mathcal{V}(P',m_j=1)}\\
=&\frac{i\sqrt{2} (m_{\mathcal{V}}+m_{\mathcal{P}})}{ \Delta^R}
    \sum_{s,\bar s}
    \int_0^1\frac{\diff x}{2x(1-x)}
    \int\frac{\diff^2\vec k_\perp}{{(2\pi)}^3}
    \sqrt{\frac{x(1-z)}{x+z(1-x)}}
     \\
    &\times
    \psi_{s \bar s/\mathcal{P}}^*(\vec k_\perp, x)
    \psi_{s \bar s/\mathcal{V}}^{(m_j=1) }
    (\vec k'_\perp, x') 
    \;.
  \end{split}
\end{align}
Note that the $m_j=-1$ state would lead to the same result, considering the symmetry of the $m_j=\pm 1$ light-front wavefunctions. According to Eq.~\eqref{eq:V_Jpl_mj1}, the transition form factor can be evaluated as a function of $z$ and $\Delta_\perp$. 
It is evident from this expression that the overlapped spin components of the two wavefunctions indicate no spin-flip (between spin-triplet and spin-singlet), which may appear counter-intuitive for the M1 transition.

    \item $J^{R/L}$ and $m_j=0$\\
      Using $J^R$ and $J^L$ current components should give the
      same result with the $m_j=0$ state of the vector meson. Here we present the expression derived from $J^R$,
      \begin{align}\label{eq:V_JR_mj0}
        \begin{split}
          &\hat V|_{J^R, m_j=0}(q^2)\\
          =&-i \frac{m_{\mathcal{V}}+m_{\mathcal{P}}}{2m_{\mathcal{V}}\Delta^R}
          \bra{\mathcal{P}(P)} J_q^R (0) \ket{\mathcal{V}(P',m_j=0)}\\
          =
          &-i \frac{m_{\mathcal{V}}+m_{\mathcal{P}}}{2m_{\mathcal{V}}\Delta^R}\sum_{\bar s}
          \int_0^1\frac{\diff x}{2x(1-x)}
          \int\frac{\diff^2\vec k_\perp}{{(2\pi)}^3}
          \\
          &\times \Bigg\{
          \psi_{\uparrow \bar s/\mathcal{P}}^*(\vec k_\perp, x)
          \psi_{\uparrow\bar s/\mathcal{V}}^{(m_j=0) }(\vec k'_\perp, x') 
     \Big\{
          \frac{2\sqrt{x(1-z)}}{\sqrt{[x+z(1-x)]^3}}(k^R -\frac{x}{z}\Delta^R)
          +\frac{2}{z}\sqrt{\frac{x(1-z)}{x+z(1-x)}}q^R
          \Big\}
          \\
          & +\psi_{\uparrow \bar s/\mathcal{P}}^*(\vec k_\perp, x)
          \psi_{\downarrow\bar s/\mathcal{V}}^{(m_j=0) }(\vec k'_\perp, x') 
          \frac{2m_qz}{\sqrt{x(1-z)[x+z(1-x)]^3}}\\
          &+     \psi_{\downarrow \bar s/\mathcal{P}}^*(\vec k_\perp, x)
          \psi_{\downarrow\bar s/\mathcal{V}}^{(m_j=0) }(\vec k'_\perp, x') 
         \Big\{
          \frac{2}{\sqrt{x(1-z)[x+z(1-x)]}}(k^R -\frac{x}{z}\Delta^R)
          +\frac{2}{z}\sqrt{\frac{x(1-z)}{x+z(1-x)}}q^R
          \Big\}
          \Bigg\}
        \end{split}
      \end{align}
      We can further simplify the expression by taking advantage of the symmetries in the light-front wavefunctions,
      \begin{align}\label{eq:LFWF_sym}
        \begin{split}
          &\psi_{\uparrow \uparrow/\mathcal{V}}^{(m_j=0)}(\vec k_\perp, x)=-\psi_{\downarrow \downarrow/\mathcal{V}}^{(m_j=0)*}(\vec k_\perp,
          x),\qquad
          \psi_{\uparrow \uparrow/\mathcal{P} }(\vec k_\perp, x)=\psi_{\downarrow \downarrow/\mathcal{P} }^*(\vec k_\perp,
          x),\\
          &\psi_{\uparrow \downarrow/\mathcal{V}}^{(m_j=0)}(\vec k_\perp, x)=\psi_{\downarrow \uparrow/\mathcal{V}}^{(m_j=0)}(\vec k_\perp, x),
          \qquad
          \psi_{\uparrow \downarrow/\mathcal{P} }(\vec k_\perp, x)=-\psi_{\downarrow \uparrow/\mathcal{P} }(\vec k_\perp, x)\;.
        \end{split}
      \end{align}
      This leads to a partial cancellation of the first and the third terms in Eq.~\eqref{eq:V_JR_mj0} and reduces it to,
      \begin{align}\label{eq:V_JR_mj0_red}
        \begin{split}
          &\hat V|_{J^R, m_j=0}(q^2)\\
          =
          &-i \frac{m_{\mathcal{V}}+m_{\mathcal{P}}}{2m_{\mathcal{V}}\Delta^R}\sum_{\bar s}
          \int_0^1\frac{\diff x}{2x(1-x)}
          \int\frac{\diff^2\vec k_\perp}{{(2\pi)}^3}
          \frac{2}{\sqrt{x(1-z)[x+z(1-x)]^3}}
          \\
          &\times \Big[
          \psi_{\uparrow\bar s/\mathcal{P}}^*(\vec k_\perp, x)
          \psi_{\uparrow\bar s/\mathcal{V}}^{(m_j=0) }(\vec k'_\perp, x') 
          (zk^R -x\Delta^R)
          +\psi_{ \uparrow\bar s/\mathcal{P}}^*(\vec k_\perp, x)
          \psi_{\downarrow \bar s/\mathcal{V}}^{(m_j=0) }(\vec k'_\perp, x') 
          m_qz
          \Big]\\
          =
          &-i \frac{m_{\mathcal{V}}+m_{\mathcal{P}}}{2m_{\mathcal{V}}\Delta^R}
          \int_0^1\frac{\diff x}{2x(1-x)}
          \int\frac{\diff^2\vec k_\perp}{{(2\pi)}^3}
          \frac{2}{\sqrt{x(1-z)[x+z(1-x)]^3}}
          \\
          &\times \Big[
          [
            \frac{1}{2}\psi_{\uparrow\downarrow-\downarrow\uparrow/\mathcal{P}}^*(\vec k_\perp, x)
\psi_{\uparrow\downarrow+\downarrow\uparrow/\mathcal{V}}^{(m_j=0) }(\vec k'_\perp, x')
+
\psi_{\uparrow\uparrow/\mathcal{P}}^*(\vec k_\perp, x)
          \psi_{\uparrow\uparrow/\mathcal{V}}^{(m_j=0) }(\vec k'_\perp, x') 
]
          (zk^R -x\Delta^R)\\
          &+
          \frac{1}{\sqrt{2}}[
          \psi_{ \uparrow\uparrow/\mathcal{P}}^*(\vec k_\perp, x)
          \psi_{\uparrow\downarrow+\downarrow\uparrow/\mathcal{V}}^{(m_j=0) }(\vec k'_\perp, x') 
          +
          \psi_{\uparrow\downarrow-\downarrow\uparrow/\mathcal{P}}^*(\vec k_\perp, x)
          \psi_{\downarrow\downarrow/\mathcal{V}}^{(m_j=0) }(\vec k'_\perp, x') 
          ]
          m_qz
          \Big]\;.
        \end{split}
      \end{align}
In the second equality, we adopt the notations of spin configurations as $\psi_{\uparrow\downarrow\pm \downarrow\uparrow}\equiv (\psi_{\uparrow\downarrow}\pm \psi_{\downarrow\uparrow})/\sqrt{2}$. This would be convenient to study the non-relativistic limit. According to Eq.~\eqref{eq:V_JR_mj0_red}, the transition form factor can be evaluated as a function of $z$ and $\Delta_\perp$. 
\item $J^{R/L}$ and $m_j=\pm 1$\\
According to our discussion, these four choices should give the same result based on the symmetry in the transverse plane. However, this "equivalence" is not very explicit in the
light-front wavefunction representation, and we see two pairs of choices.

The first pair contains these two extractions:  ($J^R$ and $m_j=1$)  and  ($J^L$ and $m_j=-1$).
  \begin{align}\label{eq:V_JR_mj1}
    \begin{split}
      &\hat V|_{J^R, m_j=1}(q^2)\\
      =& \frac{i(m_{\mathcal{V}}+m_{\mathcal{P}})}{\sqrt{2}{P'}^R\Delta^R} \bra{\mathcal{P}(P)} J_q^R (0) \ket{\mathcal{V}(P',m_j = 1)}\\
      =
      &\frac{i(m_{\mathcal{V}}+m_{\mathcal{P}})}{\sqrt{2}{P'}^R\Delta^R}
      \sum_{\bar s}
      \int_0^1\frac{\diff x}{2x(1-x)}
      \int\frac{\diff^2\vec k_\perp}{{(2\pi)}^3}
      \\
      &\times \Bigg\{
      \psi_{\uparrow \bar s/\mathcal{P}}^*(\vec k_\perp, x)
      \psi_{\uparrow\bar s/\mathcal{V}}^{(m_j=1) }(\vec k'_\perp, x') 
      \frac{2}{\sqrt{x(1-z)[x+z(1-x)]}}(k^R + (1-x)\Delta^R + [x+z(1-x)]{P'}^R)
      \\
      & +\psi_{\downarrow \bar s/\mathcal{P}}^*(\vec k_\perp, x)
      \psi_{\uparrow\bar s/\mathcal{V}}^{(m_j=1) }(\vec k'_\perp, x') 
      \frac{2m_qz}{\sqrt{x(1-z)[x+z(1-x)]^3}}\\
      &+     \psi_{\downarrow \bar s/\mathcal{P}}^*(\vec k_\perp, x)
      \psi_{\downarrow\bar s/\mathcal{V}}^{(m_j=1) }(\vec k'_\perp, x') 
      \frac{2\sqrt{x(1-z)}}{\sqrt{[x+z(1-x)]^3}}(k^R + x (1-z){P'}^R - x\Delta^R)
      \Bigg\}
      \;.
    \end{split}
  \end{align}
  Unlike extracting the transition from factor with the first two choices as in Eqs.~\eqref{eq:V_Jpl_mj1} and~\eqref{eq:V_JR_mj0_red}, fixing the values of $z$ and $\Delta_\perp$ could not uniquely determine the transition form factor in Eq.~\eqref{eq:V_JR_mj1}. There is an extra dependence on the transverse momentum, $\vec P_\perp$, or equivalently on $\vec {P'}_\perp$ or $\vec q_\perp$. This implies that the transition form factor extracted this way is not invariant under the transverse boost.

  The second pair contains  ($J^R$ and $m_j=-1$)  and  ($J^L$ and $m_j=1$).
  \begin{align}\label{eq:V_JR_mjm1}
    \begin{split}
      &\hat V|_{J^R, m_j=-1}(q^2)\\
      =&\frac{i\sqrt{2}(1-z)}{m_{\mathcal{P}}^2 - (1-z)^2 m_{\mathcal{V}}^2-P^R\Delta^L } \bra{\mathcal{P}(P)} J_q^R (0) \ket{\mathcal{V}(P',m_j=-1)}\\
      =
      &\frac{i\sqrt{2}(1-z)}{m_{\mathcal{P}}^2 - (1-z)^2 m_{\mathcal{V}}^2-((1-z){P'}^R-\Delta^R)\Delta^L} 
      \sum_{\bar s}
      \int_0^1\frac{\diff x}{2x(1-x)}
      \int\frac{\diff^2\vec k_\perp}{{(2\pi)}^3}
      \\
      &\times \Bigg\{
      \psi_{\uparrow \bar s/\mathcal{P}}^*(\vec k_\perp, x)
      \psi_{\uparrow\bar s/\mathcal{V}}^{(m_j=-1) }(\vec k'_\perp, x') 
      (k^R + (1-x)\Delta^R + [x+z(1-x)]{P'}^R)
      \\
      & +\psi_{\downarrow \bar s/\mathcal{P}}^*(\vec k_\perp, x)
      \psi_{\uparrow\bar s/\mathcal{V}}^{(m_j=-1) }(\vec k'_\perp, x') 
      \frac{2m_qz}{\sqrt{x(1-z)[x+z(1-x)]^3}}\\
      &+     \psi_{\downarrow \bar s/\mathcal{P}}^*(\vec k_\perp, x)
      \psi_{\downarrow\bar s/\mathcal{V}}^{(m_j=-1) }(\vec k'_\perp, x') 
      (k^R + x (1-z){P'}^R - x\Delta^R)
      \Bigg\}
    \end{split}
  \end{align}
This formalism of extracting the transition form factor also has the problematic dependence on ${P'}^R$, and is therefore not used in practice.
      \item $J^-$ and $m_j=0$\\
In this combination, the light-front wavefunction representation of the transition form factor reads,
      \begin{align}
        \begin{split}
            \hat V|_{J^-, m_j=0}(q^2)
          =
          &\frac{-iP^+}{m_{\mathcal{V}}(\Delta^R P^L-\Delta^L P^R)}
          \sum_{\bar s}
          \int_0^1\frac{\diff x}{2x(1-x)}
          \int\frac{\diff^2\vec k_\perp}{{(2\pi)}^3}
          \frac{1}{x'}\frac{2}{\sqrt{x'{P'}^+ xP^+}}\\
          &\times \bigg[
          \psi_{\uparrow \bar s/\mathcal{P}}^*(\vec k_\perp, x)
          \psi_{\uparrow \bar s/\mathcal{V}}^{(m_j=0)}(\vec k'_\perp, x')
          [m_q^2 + ({k'}^R+x' {P'}^R)(k^L + x P^L)]
         \\
          & +\psi_{\uparrow \bar s/\mathcal{P}}^*(\vec k_\perp, x)
          \psi_{\downarrow \bar s/\mathcal{V}}^{(m_j=0) } (\vec k'_\perp, x') 
          m_q(k^L + x P^L -{k'}^L - x' {P'}^L)\\
          & +\psi_{\downarrow \bar s/\mathcal{P}}^*(\vec k_\perp, x)
          \psi_{\uparrow \bar s/\mathcal{V}}^{(m_j=0) } (\vec k'_\perp, x') 
          m_q({k'}^R+x' {P'}^R -k^R - xP^R)\\
          &+     \psi_{\downarrow \bar s/\mathcal{P}}^*(\vec k_\perp, x)
          \psi_{\downarrow \bar s/\mathcal{V}}^{(m_j=0) }(\vec k'_\perp, x') 
           [m_q^2 + ({k'}^L+x' {P'}^L)(k^R + x P^R)]
          \bigg]     \;,
        \end{split}
      \end{align}
We can simplify this expression by applying the symmetries in the light-front wavefunctions in Eq.~\eqref{eq:LFWF_sym}. 
\begin{align}\label{eq:V_Jmn_mj0}
    \begin{split}
        \hat V|_{J^-, m_j=0}(q^2)
      =
      &\frac{-i}{m_{\mathcal{V}}(\Delta^R P^L-\Delta^L P^R)}
      \sum_{\bar s}
      \int_0^1\frac{\diff x}{2x(1-x)}
      \int\frac{\diff^2\vec k_\perp}{{(2\pi)}^3}
     \frac{2\sqrt{1-z}}{\sqrt{x[x+z(1-x)]^3}}\\
      &\times \bigg[
      \psi_{\uparrow \bar s/\mathcal{P}}^*(\vec k_\perp, x)
      \psi_{\uparrow \bar s/\mathcal{V}}^{(m_j=0)}(\vec k'_\perp, x')
      2i[ ({k'}^y+x' {P'}^y)(k^x + x P^x) - ({k'}^x+x' {P'}^x)(k^y + x P^y)]
     \\
      & +\psi_{\uparrow \bar s/\mathcal{P}}^*(\vec k_\perp, x)
      \psi_{\downarrow \bar s/\mathcal{V}}^{(m_j=0) } (\vec k'_\perp, x') 
      (-2 i) m_q(k^y + x P^y -{k'}^y - x' {P'}^y)
      \bigg]     \;,
    \end{split}
  \end{align}

As in Eq.~\eqref{eq:V_JR_mj1}, fixing the values of $z$ and $\Delta_\perp$ could not uniquely determine the transition form factor in Eq.~\eqref{eq:V_JR_mj1}. There is an extra dependence on the transverse momentum of the initial state, $\vec P'_\perp$. This implies that the transition form factor extracted this way is not invariant under the transverse boost. 
  \item $J^-$ and $m_j=\pm 1$\\
With this combination, we see the extra dependence of the transition form factor on the transverse momentum, again,
\begin{align}
    \begin{split}
        \hat V|_{J^-, m_j=1}(q^2)
      =
      &\frac{-i\sqrt{2}P^+ {P'}^+}{{P'}^+{P'}^R(m_{\mathcal{P}}^2 - P^L\Delta^R )- P^+P^R m_{\mathcal{V}}^2}
      \sum_{\bar s}
      \int_0^1\frac{\diff x}{2x(1-x)}
      \int\frac{\diff^2\vec k_\perp}{{(2\pi)}^3}
      \frac{1}{x'}\frac{2}{\sqrt{x'{P'}^+ xP^+}}\\
      &\times \bigg[
      \psi_{\uparrow \bar s/\mathcal{P}}^*(\vec k_\perp, x)
      \psi_{\uparrow \bar s/\mathcal{V}}^{(m_j=1)}(\vec k'_\perp, x')
      [m_q^2 + ({k'}^R+x' {P'}^R)(k^L + x P^L)]
     \\
      & +\psi_{\uparrow \bar s/\mathcal{P}}^*(\vec k_\perp, x)
      \psi_{\downarrow \bar s/\mathcal{V}}^{(m_j=1) } (\vec k'_\perp, x') 
      m_q(k^L + x P^L -{k'}^L - x' {P'}^L)\\
      & +\psi_{\downarrow \bar s/\mathcal{P}}^*(\vec k_\perp, x)
      \psi_{\uparrow \bar s/\mathcal{V}}^{(m_j=1) } (\vec k'_\perp, x') 
      m_q({k'}^R+x' {P'}^R -k^R - xP^R)\\
      &+     \psi_{\downarrow \bar s/\mathcal{P}}^*(\vec k_\perp, x)
      \psi_{\downarrow \bar s/\mathcal{V}}^{(m_j=1) }(\vec k'_\perp, x') 
       [m_q^2 + ({k'}^L+x' {P'}^L)(k^R + x P^R)]
      \bigg]     \;,
    \end{split}
  \end{align}
\end{enumerate}

To summarize, only two combinations of the current component and the magnetic projection of the vector meson could unambiguously extract the transition form factor from the valence hadron matrix element: they are $\hat V|_{J^{R/L}, m_j=0}(q^2)$ in Eq.~\eqref{eq:V_JR_mj0} and $\hat V|_{J^+, m_j=\pm 1}(q^2)$ in Eq.~\eqref{eq:V_Jpl_mj1}. The other choices are not invariant under the transverse boost and are, therefore, not very useful for calculating the transition form factor.
The work in Ref.~\cite{Li:2018uif} compared the two choices and found that $\hat V|_{J^{R/L}, m_j=0}(q^2)$ is preferred, at least for heavy mesons, since it employs the dominant spin components of the light-front wavefunctions and is more robust in practical calculations.
For the study on the frame dependence of the transition form factor, one might find Ref. \cite{Li:2019kpr} interesting.

%% file: Appendix.tex
\section*{Appendix: Conventions}
\subsection{Light-Front coordinates}\label{app:LF_cor}
The contravariant four-vectors of position $x^\mu$ are written as \( x^\mu= (x^+, x^-, x^1, x^2) \), where \(x^+=x^0 + x^3\) is the light-front time,  \(x^-=x^0- x^3\) is the longitudinal coordinate, and \(\vec{x}_\perp=(x^1, x^2)\) are the transverse coordinates. We sometimes write the transverse components with subscript $x$ ($y$) in place of $1$ ($2$), for example \(\vec{r}_\perp=(r^x,r^y)\). For an arbitrary transverse vector $\vec k_\perp$($\vec k_\perp^*$), define its complex representation as $k^R=k^x+i k^y$ ($k^L=k^x - i k^y$).

The covariant vectors are obtained by $x_\mu=g_{\mu\nu}x^\nu$, with the metric tensors $g_{\mu\nu}$ and $g^{\mu\nu}$. The nonzero components of the metric tensors are,
\begin{align}
  g^{+-}=g^{-+}=2, \qquad 
g_{+-}=g_{-+}=\frac{1}{2}, 
\qquad g^{ii}=g_{ii}=-1~(i=1,2) \;.
\end{align}
Scalar products are
\begin{align}
a\cdot b = a^\mu b_\mu=a^+ b_+ + a^-b_-+a^1 b_1 + a^2 b_2=\frac{1}{2}(a^+b^- + a^- b^+)-\vec a_\perp\cdot \vec b_\perp\;.
\end{align}
Derivatives are written as
\begin{align}
  \partial_+=\frac{\partial}{\partial x^+}=\frac{\partial}{2\partial x_-}=\frac{1}{2}\partial^-,\quad
  \partial_-=\frac{\partial}{\partial x^-}=\frac{\partial}{2\partial x_+}=\frac{1}{2}\partial^+
  \;.
\end{align}
We define the integral operators
\begin{align}
   & \frac{1}{\partial^+} f(x^-)=\frac{1}{4}\int_{-\infty}^{+\infty} \epsilon(x^--y^-) f(y^-)\;,\\
   &    \left( \frac{1}{\partial^+}\right)^2 f(x^-)=\frac{1}{8}\int_{-\infty}^{+\infty} |x^--y^-| f(y^-)\;.
\end{align}
Here, the antisymmetric step function
\begin{align}
    \epsilon(x)=\theta(x)-\theta(-x)\;,
    \qquad
    \frac{\partial\epsilon (x)}{\partial x}=2\delta(x)\;.
\end{align}
with the step function $\theta(x) =0 (x<0);1 (x>0)$. It follows that $|x|=x\epsilon(x)$.

\exercise{For the exponential function, check the following relation,
\begin{align}
    \frac{1}{i \partial^+} e^{-ikx}=\frac{1}{k^+} e^{-ikx}\;.
\end{align}
}

The Levi-Civita tensor is 
\begin{align}
  \epsilon^{\mu\nu\rho\sigma}=\frac{1}{\sqrt{|g|}}
\begin{cases}
+1,&\text{if $\mu,\nu,\rho,\sigma$ is an even permutation of $-,+,1,2$}\\
-1,&\text{if $\mu,\nu,\rho,\sigma$ is an odd permutation of $-,+,1,2$}\\
0,&\text{other cases}
\end{cases}
\end{align}
in which $g\equiv \det g_{\mu\nu}=-\frac{1}{2}$.

The full four-dimensional integral is
\begin{align}
  \int \diff^4 x= 
\int \diff x^0\diff x^1\diff x^2\diff x^3
  =\frac{1}{2}\int \diff x^+ \diff x^- \diff^2 x_\perp
=\int \diff^3 x \diff x^+ \;,
\end{align}
where we also define the volume integral as
\begin{align}
  \int \diff^3 x\equiv \int \diff x_+\diff^2 x^\perp=\frac{1}{2}\int \diff x^- \diff^2 x^\perp \;.
\end{align}

In the momentum space, the Lorentz invariant integral is,
\begin{align}
\begin{split}
\int \frac{\diff ^4 p}{{(2\pi)}^4}\theta(p^+)(2\pi)\delta(p^+ p^--\vec p_\perp^2-m^2)
=&\frac{1}{2}\int \frac{\diff p^+ \diff p^- \diff^2 p_\perp}{{(2\pi)}^4}\theta(p^+)(2\pi)\delta(p^+ p^--\vec p_\perp^2-m^2)\\
=&\int \frac{\diff^2 p_\perp \diff p^+}{{(2\pi)}^32p^+}\theta(p^+)
\end{split}
\end{align}
The Fourier transform of a function $f(\vec{r}_\perp)$ and the inverse transform are defined as 
\begin{align}\label{eq:FourierT}
f(\vec{r}_\perp)= \int\frac{\diff^2 p_\perp}{{(2\pi)}^2}e^{i\vec{p}_\perp\cdot\vec{r}_\perp}\tilde f(\vec{p}_\perp),
\qquad
  \tilde f(\vec{p}_\perp)= \int \diff^2 \vec{r}_\perp e^{-i\vec{p}_\perp\cdot\vec{r}_\perp}f(\vec{r}_\perp)
  \;.
\end{align}
The Dirac deltas read
  \begin{align}
    \begin{split}
    \int \diff^2 \vec r_\perp e^{-i \vec p_\perp\cdot \vec r_\perp}
    = {(2\pi)}^2\delta^2(\vec p_\perp),\qquad
      \int \diff^2 \vec p_\perp e^{i \vec p_\perp\cdot \vec r_\perp}={(2\pi)}^2\delta^2(\vec r_\perp)\;.
    \end{split}
  \end{align}
\subsection{$\gamma$ matrices}\label{app:gamma}
The Dirac matrices are four unitary traceless $4 \times 4$ matrices:
\begin{equation}
  \gamma^0=\beta=
  \begin{pmatrix}
    0&-i\\
    i&0
  \end{pmatrix},
  \quad
  \gamma^+=
  \begin{pmatrix}
    0 & 0\\
    2i & 0
  \end{pmatrix},
  \quad
  \gamma^-=
  \begin{pmatrix}
    0&-2i\\
    0&0
  \end{pmatrix},
  \quad
  \gamma^i=
  \begin{pmatrix}
    -i\hat{\sigma}^i&0\\
    0&i\hat{\sigma}^i
  \end{pmatrix}
  \;.
\end{equation}
They are expressed in terms of the $2\times 2$ Pauli matrices,
\begin{align}
  \hat{\sigma}^1=\sigma^2=
  \begin{pmatrix}
    0&-i\\
    i&0
  \end{pmatrix},
       \quad
       \hat{\sigma}^2=-\sigma^1=
       \begin{pmatrix}
         0&-1\\
         -1&0
       \end{pmatrix}
       \;.
\end{align}
Note that $\gamma^3=\gamma^+-\gamma^0$. It is also convenient to define $\gamma^R\equiv\gamma^1+i\gamma^2$ and $\gamma^L\equiv\gamma^1+i\gamma^2$. The chiral matrix is $\gamma^5=i\gamma^0\gamma^1\gamma^2\gamma^3$.
Some useful relations,
\begin{align}\label{eq:gmiplj}
  \gamma^1\gamma^+\gamma^1=\gamma^2\gamma^+\gamma^2=\gamma^+,
  \quad
  \gamma^1\gamma^+\gamma^2=-\gamma^2\gamma^+\gamma^1=i\gamma^+
\end{align}
\begin{align}
 \gamma^0\gamma^\mu={\gamma^\mu}^\dagger\gamma^0,\qquad \{\gamma^\mu,\gamma^\nu \}=2g^{\mu\nu}\bm{I}
\end{align}
\begin{align}
  \alpha^\kappa=\gamma^0\gamma^\kappa,\quad {(\alpha^1)}^2={(\alpha^2)}^2=\bm{I},\quad \alpha^1\alpha^2=-\alpha^1\alpha^2
\end{align}
Combinations of Dirac matrices as projection operators,
\begin{align}
\begin{split}
  \Lambda^\pm=\frac{1}{4}\gamma^\mp\gamma^\pm=\frac{1}{2}\gamma^0\gamma^\pm=\frac{1}{2}(\bm I\pm\alpha^3)
  \;.
\end{split}
\end{align}
They have the following properties,
\begin{align}
  \begin{split}
    &\Lambda^+ + \Lambda^- = \bm I\,,
    \quad
    {(\Lambda^\pm)}^2= \Lambda^\pm\,,
    \quad \Lambda^\pm\Lambda^\mp=0\,,
    \quad {(\Lambda^\pm)}^\dagger=\Lambda^\pm\,, \\
    &\alpha^i\Lambda^\pm=\Lambda^\mp\alpha^i\,,
    \quad  \gamma^0\Lambda^\pm=\Lambda^\mp\gamma^0
    \;.
  \end{split}
  \end{align}

 \subsection{Spin vectors}\label{sec:spinvector}
We use the following spinor representation,
The $u$, $v$ spinors are defined as,
\begin{equation}
  \begin{split}
    &u(p,\lambda=\frac{1}{2})=\frac{1}{\sqrt{p^+}}{(p^+,0,im_q,ip^x-p^y)}^\intercal \;,\\
    &u(p,\lambda=-\frac{1}{2})=\frac{1}{\sqrt{p^+}}{(0,p^+,-ip^x-p^y,im_q)}^\intercal \;,\\
    &\bar{u}(p,\lambda=\frac{1}{2})=\frac{1}{\sqrt{p^+}}(m_q,p^x- i p^y,-ip^+,0)\;,\\
    &\bar{u}(p,\lambda=-\frac{1}{2})=\frac{1}{\sqrt{p^+}}(-p^x - i p^y,m_q,0,-ip^+)\;,
  \end{split}
\end{equation}
and
\begin{equation}\label{eq:lambda_pro}
  \begin{split}
    &v(p,\lambda=\frac{1}{2})=\frac{1}{\sqrt{p^+}}{(p^+,0,-im_q, ip^x-p^y)}^\intercal \;,\\
    &v(p,\lambda=-\frac{1}{2})=\frac{1}{\sqrt{p^+}}{(0,p^+,-ip^x-p^y, -im_q)}^\intercal \;,\\
    &\bar{v}(p,\lambda=\frac{1}{2})=\frac{1}{\sqrt{p^+}}(-m_q, p^x- i p^y,-ip^+,0)\;,\\
    &\bar{v}(p,\lambda=-\frac{1}{2})=\frac{1}{\sqrt{p^+}}(-p^x - i p^y, -m_q,0,-ip^+)\;.
  \end{split}
\end{equation}

Define the spin vector for the massive spin 1 particles with momentum $k^\mu$ and spin projection $\lambda$:
\begin{align}
e(k,\lambda=0)=(\frac{k^+}{m},\frac{{\vec{k}_\perp}^2-m^2}{mk^+},\frac{\vec{k}_\perp}{m})\\
e(k,\lambda=\pm 1)
=(0,\frac{2\bm{\epsilon}^\perp_\lambda\cdot \vec{k}_\perp}{k^+},\bm{\epsilon}^\perp_\lambda)
\end{align}
where $\bm{\epsilon}^\perp_\pm=(1,\pm i)/\sqrt{2}$ and $m$ is the mass of the particle.

The polarization vectors for gluon are defined as 
\begin{align}
e(k,\lambda=\pm 1)
=(0,\frac{2\bm{\epsilon}^\perp_\lambda\cdot \vec{k}_\perp}{k^+},\bm{\epsilon}^\perp_\lambda)
\end{align}
where $\bm{\epsilon}^\perp_\pm=(1,\pm i)/\sqrt{2}$.

Spin vector identities:
\begin{itemize}
\item Proca equation:
\(k_\mu e^\mu(k,\lambda)=0 \;.\)
\item Orthogonality:
\( e^\mu(k,\lambda)e^*_\mu(k,\lambda')=-\delta_{\lambda,\lambda'}
;.\)
\item Crossing symmetry:
\(e^*_\mu(k,\lambda)=e_\mu(k,-\lambda),\qquad
e^\mu(-k,\lambda)={(-1)}^{\lambda+1}e^\mu(k,\lambda)\)
\end{itemize}

\subsection{QCD color space}\label{app:colorT}
The specification of the quark state in the color space is by a three-element column vector $c$, 
\begin{align}
 c=
  \begin{pmatrix}
   1\\
   0\\
   0
  \end{pmatrix}
  \text{~for red,}
\quad
  \begin{pmatrix}
   0\\
   1\\
  0
  \end{pmatrix}
  \text{~for blue,}
\quad
  \begin{pmatrix}
   0\\
  0\\
   1
  \end{pmatrix}
  \text{~for green.}
\end{align}
We use the standard basis for the fundamental representation of SU(3), i.e. the Gell-Mann matrices,
\begin{align}
  \begin{split}
 & T^1=\frac{1}{2}
  \begin{pmatrix}
   0&1&0\\
   1&0&0\\
   0&0&0
  \end{pmatrix}
  \;,\quad
  T^2=\frac{1}{2}
  \begin{pmatrix}
   0&-i&0\\
   i&0&0\\
   0&0&0
  \end{pmatrix}
  \;,\quad
  T^3=\frac{1}{2}
  \begin{pmatrix}
   1&0&0\\
   0&-1&0\\
   0&0&0
  \end{pmatrix}
  \;,\\
 & T^4=\frac{1}{2}
  \begin{pmatrix}
   0&0&1\\
   0&0&0\\
   1&0&0
  \end{pmatrix}
  \;,\quad
  T^5=\frac{1}{2}
  \begin{pmatrix}
   0&0&-i\\
   0&0&0\\
   i&0&0
  \end{pmatrix}
  \;,\quad
  T^6=\frac{1}{2}
  \begin{pmatrix}
   0&0&0\\
   0&0&1\\
   0&1&0
  \end{pmatrix}
  \;,\\
  & T^7=\frac{1}{2}
  \begin{pmatrix}
   0&0&0\\
   0&0&-i\\
   0&i&0
  \end{pmatrix}
  \;,\quad
  T^8=\frac{1}{2\sqrt{3}}
  \begin{pmatrix}
   1&0&0\\
   0&1&0\\
   0&0&-2
  \end{pmatrix}
  \;.
\end{split}
\end{align}

In the matrix notation, $\bm{A}^\mu=T^a A^\mu_a$ with the gluon index $a=1,\ldots,8$. The color matrix element $A^\mu_{cc'}=T^a_{cc'} A^\mu_a$
\begin{align}
  \bm{A}^\mu=\frac{1}{2}
  \begin{pmatrix}
    \dfrac{1}{\sqrt{3}}A^\mu_8+A^\mu_3 & A^\mu_1-iA^\mu_2 &A^\mu_4-iA^\mu_5\\
    A^\mu_1+iA^\mu_2 &\dfrac{1}{\sqrt{3}}A^\mu_8-A^\mu_3 &A^\mu_6-iA^\mu_7\\
    A^\mu_4+iA^\mu_5&A^\mu_6+iA^\mu_7 & -\dfrac{2}{\sqrt{3}}A^\mu_8
  \end{pmatrix}
  \;.
\end{align}

\subsection{Discrete symmetries}\label{app:symmetry} 
Consider a particle state with momentum $p^\mu$ and parity $ \mathrm{P}$,
\begin{align}
  \varmathbb{P}\ket{\phi (p^\mu, \mathrm{P})}
  = \mathrm{P}\ket{\phi ( \mathcal{P}^\mu_\nu p^\nu, \mathrm{P})}\;.
\end{align}
The parity operator is
\begin{align}
  \mathcal{P}^\mu_\nu={(\mathcal{P}^{-1})}^\mu_\nu=
\begin{pmatrix}
+1&\ &\ &\ \\
\ &-1&\ &\ \\
\ &\ &-1&\ \\
\ &\ &\ &-1 
\end{pmatrix}
          \;.
\end{align}
The current operator under the parity transformation is 
\begin{align}
  \varmathbb{P}^{-1} J^\mu \varmathbb{P}=
 \mathcal{P}^\mu_\nu J^\nu
\;.
\end{align}
For the polarization vector,
\begin{align}\label{eq:Pspinvector}
e^\mu(\mathcal{P}\cdot  k, \lambda)=-\mathcal{P}^\mu_\nu e^\nu(k, \lambda)
\;.
\end{align}
Consider a particle state with charge conjugation $ \mathrm{C}$ (if there is one), 
\begin{align}
  \varmathbb{C}\ket{\phi (p^\mu, \mathrm{C})}
  = \mathrm{C}\ket{\phi (p^\mu, \mathrm{C})}\;.
\end{align}
The current operator under the charge conjugation is 
\begin{align}
  \varmathbb{C}^{-1} J^\mu \varmathbb{C}=
  - J^\mu
  \;.
\end{align}